\documentclass[fleqn,usenatbib]{mnras}
\usepackage{newtxtext,newtxmath}
\usepackage[T1]{fontenc}

\usepackage{graphicx}
\usepackage{amsmath}
\usepackage{bm}
\usepackage{xr-hyper}
\usepackage{bm,upgreek}
\usepackage{lineno}
\usepackage{soul}
\usepackage{cancel}
\usepackage[normalem]{ulem}

\def\msunoh{\,h^{-1}\mathrm{M}_\odot}
\def\msun{\,\mathrm{M}_\odot}

\usepackage{hyperref}

\title[kSZ baryon fractions]{Mass dependence of halo baryon fractions from the kinetic Sunyaev--Zeldovich effect}
\author[F. A. Roper et al.]{\newauthor
Finn A. Roper,\thanks{E-mail: finn.roper@ed.ac.uk}
Yan-Chuan Cai,
and John A. Peacock
\\
Institute for Astronomy, University of Edinburgh, Royal Observatory, Blackford Hill, Edinburgh EH9 3HJ, United Kingdom\\
}

\date{Accepted XXX. Received YYY; in original form ZZZ}

\pubyear{\the\year}

\begin{document}
\label{firstpage}
\pagerange{\pageref{firstpage}--\pageref{lastpage}}
\maketitle

\begin{abstract}
We detect the kinetic Sunyaev--Zeldovich imprint of peculiar motions of galaxy groups and clusters, using the photometric DESI Legacy Survey together with cosmic microwave background (CMB) maps from the Atacama Cosmology Telescope (ACT). We develop a comprehensive forward model based on the \textsc{AbacusSummit} cosmological simulations: mock galaxy group catalogues and synthetic kSZ maps are generated, together with a reconstructed peculiar velocity field that allows for photo-$z$ errors, redshift-space distortions, and survey masks. We investigate possible contamination from the cosmic infrared background (CIB), finding that CIB effects are subdominant to the kSZ signal in the relevant ACT frequency channel.
We then predict the kSZ signal expected when stacking CMB temperature maps around groups, taking account of their estimated radial velocities.
Comparing the model with observations, we are able to constrain the total baryon fraction within haloes, as well as their internal gas profiles. 
We find evidence for mass dependence of the halo baryon fraction within the virial radius.
The gas fraction in massive groups is consistent with the universal baryon fraction, but low-mass groups ({$10^{12.5} \lesssim M\,/\msunoh \lesssim 10^{14}$}) are depleted to $0.38 \pm 0.11$ times the universal baryon fraction. 
We find this low virial baryon fraction to be consistent with an extended gas profile, for which the total baryon content reaches the universal value well beyond the virial radius.
This conclusion is consistent with previous analyses using X-ray, kSZ, and weak lensing, and plausibly reflects energetic feedback processes from the galaxies in these haloes. 
\end{abstract}


\section{Introduction} \label{sec:intro}

The cycle of baryons in the Universe is central both to understanding the formation and evolution of galaxies, and to cosmology. In terms of galaxies, the consensus physical picture is that they form and grow in the context of dark matter haloes. The gravitational well of a halo accretes the surrounding gas, of which some is shock heated to form an ionised halo, while some cools to form the stellar component of a galaxy \citep{WhiteRees1978}. In turn, stars can provide energy feedback to the interstellar medium via stellar winds and supernovae. Also, thermal and kinetic feedback can carry energy out from around a central massive black hole to heat up the gas and even transport some of it beyond the virial radius \citep[e.g.][]{Croton2006}. 

These various feedback mechanisms are intertwined with star-forming activity: they shift baryons between phases, and drive the evolution of both stars and gas. This baryonic redistribution can also alter the overall distribution of dark matter within the host halo \citep{vandaalen2011}.
The main observational constraints on models for galaxy formation are commonly taken from integrated stellar properties such as stellar mass functions and luminosity functions. However, the mass of baryons in the gas phase is usually dominant over the mass in stars \citep{McGaugh_etal_10}. A complete physical picture of the baryon cycle therefore demands a detailed understanding of the properties of the gas phase.

As regards cosmology more broadly, it is known that the redistribution of baryons around haloes can affect the strength of gravitational lensing, and the cross-correlations between galaxies and lensing \citep{vandaalen2011}. Strong AGN feedback in particular can suppress the fraction of baryons in haloes, thus altering the clustering of the total mass distribution and its cross-correlation with galaxies and gas \citep{Dave2019}. This was proposed as a possible means of explaining the so-called $S_8$ tension \citep{Amon2022}, where the amplitude of clustering in the late-time Universe, parameterised as $S_8\equiv(\Omega_\mathrm{m}/0.3)^{0.5}\sigma_8$, has been observed for some while to be low compared to the best-fitting cosmology from {\it Planck\/} \citep{KiDS2021, DES2022} -- although the most recent results from \cite{KiDS2025} may change the picture somewhat.

The above issues all relate rather closely to the gas content of haloes of different total masses: 
\begin{align}
f_\mathrm{gas}(M_\mathrm{total}) \equiv M_\mathrm{gas} / M_\mathrm{total}, 
\end{align}
and this quantity is a focus of our present study. Here, we will employ the kinetic Sunyaev--Zeldovich (kSZ) effect to constrain $f_\mathrm{gas}$ in haloes. The kSZ effect arises from the scattering of CMB photons by free electrons that undergo bulk motion.
In general, Compton scattering by a free electron in ionised gas will change the energy of a scattered photon. If a halo is stationary, the relevant velocities are the internal thermal velocities of the gas, which cause the thermal Sunyaev--Zeldovich (tSZ) effect \citep{Sunyaev_Zeldovich_70}. The tSZ effect depends on frequency, and has a null at a frequency of 217\,GHz. However, the electron cloud could have bulk motions along the radial direction. With the expected bulk motion at the order of $1000\,\mathrm{km\,s}^{-1}$ or smaller, the scattering is almost entirely elastic due to the relatively small velocity, i.e. there is almost no exchange of energy, but the velocity will cause a Doppler shift in scattered photons, altering the observed CMB temperature equally at all frequencies. This temperature distortion is known as the kinematic or kinetic Sunyaev--Zeldovich (kSZ) effect \citep{Sunyaev_Zeldovich_80b} and, in the non-relativistic limit, is equal to an integral over the line of sight (LoS):
\begin{align}
    \delta T_\mathrm{kSZ}(\hat{\bm{r}}) = - T_0\, \sigma_\mathrm{T} \int n_\mathrm{e} \left( \frac{\bm{v}}{c} \cdot \hat{\bm{r}} \right) \mathrm{d}r,
    \label{eq:kSZ}
\end{align}
where ${\bm{v} \cdot \hat{\bm{r}} = v_\parallel}$ is the LoS component of the bulk electron peculiar velocity, $\sigma_\mathrm{T}$ is the Thomson scattering cross-section, $T_0$ is the average present-day CMB temperature, and $c$ is the speed of light \citep[see derivations in][]{Phillips_95,Birkinshaw_99}. The negative sign ensures that approaching haloes (i.e.~${v_\parallel < 0}$) imprint positive temperature shifts. All variables here are proper, not comoving, quantities.

This equation can be rewritten solely in terms of projected quantities as
\begin{align}
    \delta T_\mathrm{kSZ}(\bm{\theta}) = - T_0 \, \tau(\bm{\theta}) \frac{v_\parallel}{c},
    \label{eq:kSZ_proj}
\end{align}
which makes it clear that the kSZ effect depends only on optical depth, ${\tau = \sigma_\mathrm{T} \int n_\mathrm{e} \, \mathrm{d}r}$, and LoS velocity ${v_\parallel = \bm{v} \cdot \hat{\bm{r}}}$, where $\bm{\theta}$ is the sky position of the halo. The kSZ signal is therefore proportional to the density of free electrons and the bulk velocity of the gas halo integrated along the line of sight. It is sensitive to all sources of ionised gas -- with no dependence on gas temperature, in contrast with the tSZ effect.

One may imagine that the redshift of kSZ scattering would be important, as if it takes place at high $z$ then the unperturbed CMB temperature, $T$, is larger than today's $T_0$, meaning that the kSZ temperature perturbation, $\delta T$, is also increased. 
However, both $\delta T$ and $T$ are then redshifted to the present by the same expansion factor, and so the observed fractional perturbation to temperature, $\delta T/T$ at $z=0$, is the same as at the time of kSZ scattering.
Note that this is not in conflict with changes in kSZ strength over cosmological time due to redshift-dependences in the masses, absolute LoS velocities, or mean density of electrons of galaxy groups \citep[e.g.][]{Chen2024}.

There are two major obstacles to the use of the kSZ effect as a means of constraining $f_\mathrm{gas}$. One is that any prediction of the kSZ signal requires some knowledge about the distribution of ionised baryons around a given halo; in addition, we must deal with the cancellation effects from different haloes, since the radial velocity of a given halo is equally likely to be positive or negative. We therefore need to provide a sufficiently robust model for the gas distribution within a halo, as well as some means of estimating the peculiar velocity field. We will undertake this exercise using the group catalogue derived by \cite{Yang_etal_21}, using photometric redshifts obtained with the DESI Legacy Survey. We will employ velocity reconstruction based on the estimated mass fluctuations within the survey, in order to separate the group samples into a receding sub-sample and an approaching sub-sample, thus avoiding cancellations when stacking CMB maps centred on the groups.

To date, significant efforts have been made to measure $f_\mathrm{gas}(M)$ using X-ray, tSZ and kSZ data. All these probes rely on photon-electron interactions, but are sensitive to different aspects of the electron properties. The emissivity of X-rays from ionised gas is proportional to $n_\mathrm{e}^2 \, T^{1/2}$, where $n_\mathrm{e}$ and $T$ are the number density and temperature of free electrons respectively. This signal is typically sensitive to the central part of a galaxy cluster, and may be less constraining for the total gas fraction at the outskirts of a halo where $n_\mathrm{e}$ is low. The tSZ effect is proportional to $n_\mathrm{e} \, T$. Both X-ray and tSZ probes suffer from the degeneracy between $n_\mathrm{e}$ and $T$; in addition, X-ray emission is more sensitive to internal inhomogeneities in gas density, on account of its $n_\mathrm{e}^2$ dependence. The kSZ signal is proportional to $n_\mathrm{e} \, v_{\parallel}$, with $v_{\parallel}$ being the bulk peculiar velocity of electrons along the line of sight. In principle, the combination of kSZ with the other probes can break the degeneracy with temperature. Using the above observations, reported constraints on $f_\mathrm{gas}(M)$ exhibit a diversity of results.

\citet{Bigwood_etal_24} use weak lensing (WL) data from the Dark Energy Survey (DES) primarily constraining $S_8$, and kSZ data from SDSS-BOSS,  deriving constraints on gas mass and baryon mass fractions as functions of halo mass. They find slightly lower fractions than X-ray observations from XMM with lensing mass calibration from HSC \citep[reported in][]{Akina_etal_22}, by $\sim 1.6\sigma$. However, the latest results of X-ray observations from eROSITA data in the eFEDS area \citep{Popesso_etal_24} suggest slightly lower gas fractions for low-mass haloes, which may be more consistent with the kSZ+WL analysis.

In recent years, the DESI Legacy Survey dataset has been exploited by several research groups for kSZ, and all have obtained significant detections of the signal \citep{Chen_etal_22, Li_etal_24, Hadzhiyska_etal_24, McCarthy2025, Lai2025, Hotinli2025}. Among them, \citet{Hadzhiyska_etal_24, Hadzhiyska2025} have found a lower than expected kSZ signal amplitude at small scales using reconstructed velocities of the LRGs, interpreted as an indication of strong feedback. The low-amplitude of the kSZ signal from their analysis seems to be consistent with large-scale analysis using velocity reconstructions and cross-correlations \citep{Lai2025, Hotinli2025}.

In this work, we adopt an analysis that is similar to the one in \citet{Hadzhiyska_etal_24, Hadzhiyska2025}, but our approach has a particular focus on the gas content of groups of different masses. We perform velocity reconstruction using the DESI Legacy group catalogue defined in \citet{Yang_etal_21}. The estimated masses and velocities of these groups then yields a predicted kSZ signal, which allow us to set constraints on the mass-dependence of the gas fraction.

This paper is structured as follows.
Sec.~\ref{sec:data} describes the data from observation and simulation that are used in this work.
Sec.~\ref{sec:AbSu_alterations} explains how the mock halo catalogue used is altered in order to forward model the observed group catalogue.
Sec.~\ref{sec:vel_rec} describes how velocities are reconstructed from group catalogues.
Sec.~\ref{sec:kSZ_template} describes the generation of mock kSZ template skies from the mock halo catalogue.
Sec.~\ref{sec:stacking} details the stacking procedure used to condense the observed CMB and mock kSZ templates into kSZ temperature profiles.
Sec.~\ref{sec:fgas} presents the measured gas mass fractions both overall and as a function of halo mass, which are the main results of this work.
Sec.~\ref{sec:discussion} then elaborates on these results: discussing null tests and foreground contamination.
Finally, Sec.~\ref{sec:conclusions} briefly concludes.

Throughout this paper, we assume a flat $\Lambda$CDM cosmology with ${H_0 = 67.66\,\mathrm{km}\,\mathrm{s}^{-1}\,\mathrm{Mpc}^{-1}}$, ${\Omega_\mathrm{b}/\Omega_\mathrm{m} = 0.04897/0.3111 = 0.1574}$, and all other cosmological parameters consistent with \citet{PlanckCollaboration_18_VI} findings using combined TT,TE,EE+lowE+lensing+BAO constraints.

\section{Data} \label{sec:data}

\subsection{DESI Legacy group catalogue} \label{ssec:DESI_groups}

We use a publicly-available galaxy group catalogue derived from the DR9 data release of the DESI Legacy Imaging Survey\footnote{\url{https://www.legacysurvey.org/}} (DESI-LS). This dataset is described in detail by \cite{Dey_etal_19}, but in essence it is an attempt to construct a uniform nearly all-sky galaxy catalogue to act as a basis for target selection in the different components of the DESI spectroscopic survey \citep{DESI_DR1}. The survey provides approximately 18\,000\,deg$^2$ of ground-based $grz$ imaging, combined with 3--5\,$\upmu$m photometry from the WISE satellite. 

\begin{figure}
\begin{centering}
    \includegraphics[width=\columnwidth]{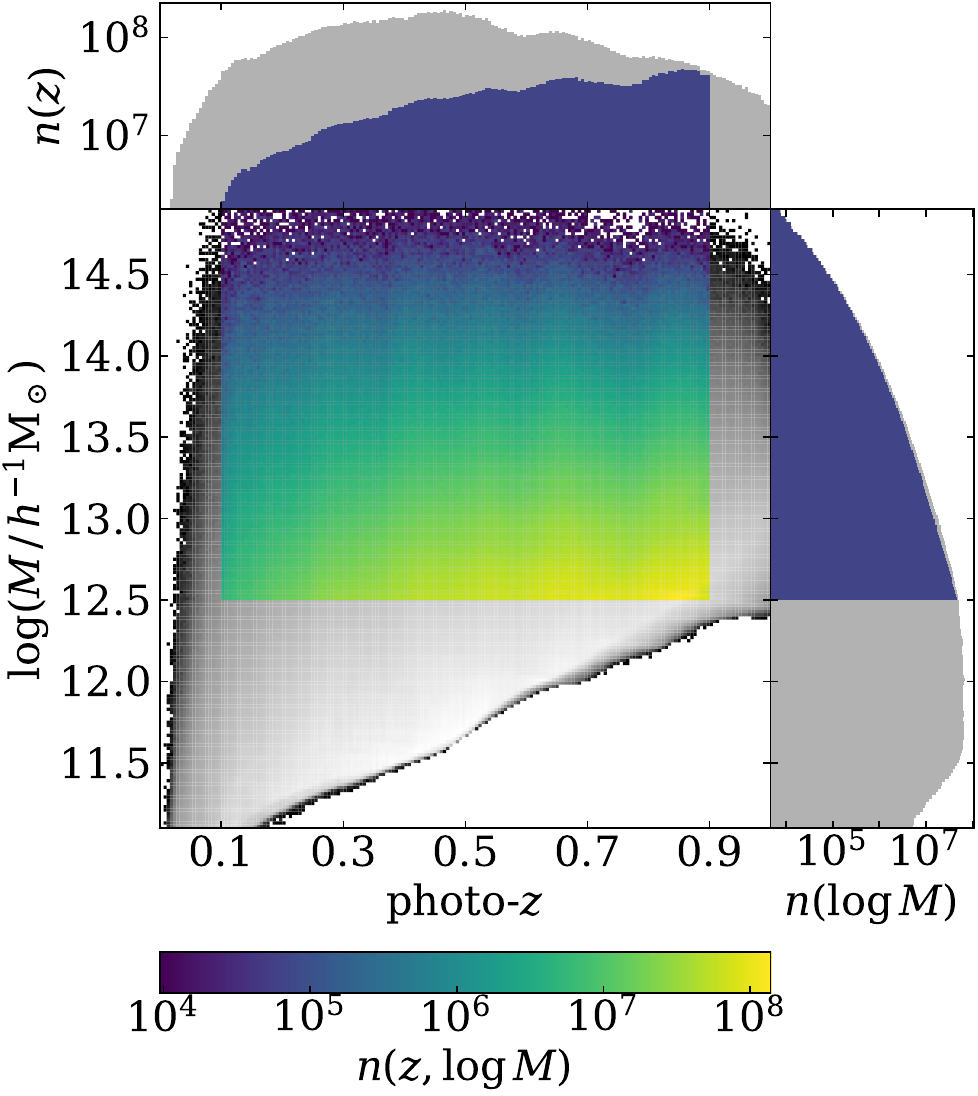}
    \caption{
        Two-dimensional histogram showing groups in the DESI Legacy catalogue in photometric redshift and mass. The full catalogue is shown in greyscale and the catalogue after redshift and mass cuts is shown in colour, with colour bar showing $n(z,\mathrm{log}\,M) \equiv \mathrm{d}^2 N / \mathrm{d}z \, \mathrm{d}\,\mathrm{log}_{10}(M/h^{-1}\mathrm{M}_\odot)$. Both grey and colour scales are logarithmic.
        Minor panels show marginalised redshift [top; $n(z) \equiv \mathrm{d}N/\mathrm{d}z$] and mass [right; $n(\mathrm{log}\,M) \equiv \mathrm{d}N/\mathrm{d}\,\mathrm{log}_{10}(M/h^{-1}\mathrm{M}_\odot)$] distributions. Grey and colour show full and cut DESI-LS catalogues. Note that the cuts in redshift have almost no effect on the marginalised mass distribution.
    }
    \label{fig:hists}
\end{centering}
\end{figure}

This material allows the generation of photometric redshifts with a typical precision of about 0.02 in redshift \citep{Hang2021, Zhou2021}, which are available for $\sim$\,$10^8$ galaxies, depending on precision cuts. \cite{Yang_etal_21} developed a group finder that could employ such approximate redshifts \citep[see also the earlier papers:][]{Yang_etal_05,Yang_etal_07}. This algorithm was applied to the DESI Legacy Imaging Survey DR9 data release\footnote{\url{https://gax.sjtu.edu.cn/data/DESI.html}}. This DR9 catalogue covers $8601$ and $9622\,\mathrm{deg}^2$ in the North and South Galactic Cap regions (NGC and SGC) respectively, and contains a total of 91\,739\,237 `groups' (the majority of which consist of a single isolated galaxy). 
Fig.~\ref{fig:hists} shows the photometric redshift and mass distributions of the full DESI-LS DR9 group catalogue in greyscale and the cut DESI-LS group catalogue in colour, where we restrict the photometric redshift range to ${0.1 < z < 0.9}$ and the mass range to ${M > 12.5 \msunoh}$. It can be seen that the group catalogue is complete down to this minimum mass over the redshift range selected. These selection cuts reduce the number of groups to 19\,648\,383 -- although we then apply further masking after velocity reconstruction to leave 17\,308\,806 groups, almost all of which overlap the relevant CMB sky masks. Henceforth, `the DESI-LS group catalogue' refers to this selection.

In this analysis, we use the halo masses assigned to these groups by \citet{Yang_etal_21}. These estimates were derived by abundance matching the total $z$-band luminosity of the groups' members (for details of the method, see sec.~3.5 of \citealt{Yang_etal_07}). This approach has been found to yield uncertainties of around $0.2 \, \mathrm{dex}$ for ${M \gtrsim 10^{13.5} \msunoh}$ and $0.4\,\mathrm{dex}$ for lower mass haloes \citep{Yang_etal_21}.

\subsection{\textsc{AbacusSummit} halo light-cone mock catalogue} \label{ssec:Abacus_haloes}

In order to understand the relation of this approximate group catalogue to the underlying mass distribution, we need to consider mock data from simulations.
We make use of an all-sky halo light-cone catalogue \citep{Hadzhiyska_etal_22}, which was built using a simulation from the \textsc{AbacusSummit} suite of cosmological $N$-body simulations \citep{Maksimova_etal_21,Garrison_etal_21}; these were designed to meet the cosmological simulation requirements of the DESI survey. The simulation box used is \texttt{AbacusSummit\_huge\_c000\_ph201}, which was run with the fiducial \citet{PlanckCollaboration_18_VI} cosmology (consistent with the cosmology detailed in Sec.~\ref{sec:intro}), a box side length of $7600 \, h^{-1} \mathrm{Mpc}$, and $86\,403^{3}$ particles of mass $5.7 \times 10^{10} \msunoh$.

Haloes in \textsc{AbacusSummit} are identified by the \textsc{CompaSO} algorithm \citep{Hadzhiyska_etal_22b}, which assigns particles into disjoint haloes with spherical overdensities with respect to the mean background density of $\Delta_\mathrm{b} = 200$; within these, the algorithm identifies disjoint subhaloes with $\Delta_\mathrm{b} = 800$. Haloes flagged as unphysical, such as unbound or split haloes, are cleaned by merging them with nearby massive haloes \cite[for details, see][]{Bose_etal_22}. The mass of each halo is simply the total mass of its constituent particles, whilst its position and velocity are computed using only the particles in its largest subhalo.

The halo light-cone mock catalogue is generated by first separating the halo catalogue into multiple redshift epochs of width $0.05$--$0.075$ (for ${z < 1.0}$). Haloes (i.e.\ progenitors and descendants) are matched between epochs using their merger trees then the progenitor history of each halo is (if available) used to find its comoving distance from the origin at its light-cone crossing time, assuming a constant radial velocity between redshift epochs. The halo's mass and position at the light-cone crossing time are then simply linearly interpolated between epochs. See sec.~3 of \citet{Hadzhiyska_etal_22} for details of this procedure.

This catalogue has been found to agree with both theory and simulation snapshots across multiple measures. After a halo occupation distribution has been applied to produce an emission line galaxy (ELG) catalogue, \citet{Hadzhiyska_etal_22} finds the multipoles of the ELG autocorrelation function consistent with those of the full-box snapshot catalogues, and the auto- and cross-power spectra of the ELG and CMB lensing convergence strongly consistent with those predicted from theory. The halo catalogue also performs well in full-sky visual inspection of density and tidal maps and passes multiple consistency checks \citep[see sec.~4 of][]{Hadzhiyska_etal_22}. We also find its halo mass function to be broadly consistent with that of the DESI-LS group catalogue, and both are complete given the redshift and mass cuts shown in Fig.~\ref{fig:hists}. However, we explain in Sec.~\ref{sec:AbSu_alterations} how and why its mass function is altered for our purposes.

\subsection{CMB maps}

The CMB temperature map primarily used\footnote{\texttt{act-planck\_dr6.02\_coadd\_AA\_daynight\_f220\_map\_healpix.fits} found at \url{https://lambda.gsfc.nasa.gov/product/act/act_dr6.02/act_dr6.02_maps_coadd_get.html}} is from data release six of the Atacama Cosmology Telescope (ACT), covering around $19\,000\,\mathrm{deg}^2$ of sky in the $220 \, \mathrm{GHz}$ frequency band. This map coadds \textit{Planck} and AdvancedACT data, making use of the improved resolution (beam angle of $1\,\mathrm{arcmin}$ compared to \textit{Planck}'s $5\,\mathrm{arcmin}$ at this frequency) and high-$\ell$ effectiveness of ACT, while including contributions from \textit{Planck} at low $\ell$, where ground-based observations experience significant atmospheric contamination (see \citealt{Naess_etal_20,Naess_etal_25} for details).

A bandpass centred at around 217\,GHz (nominally 220\,GHz) is chosen as this is close to the `crossover' null frequency at which the tSZ effect disappears. A residual tSZ signal will however still be present in the map, owing to the significant imaging bandwidth and, less significantly, the dependence of crossover frequency on electron cloud temperature \citep[see][respectively]{Lamarre_etal_10, Terzic_etal_24}. 

About 40~per~cent of the sky is masked out of both maps using the \textit{Planck} PR2 HFI galaxy plane and point source masks \citep{PlanckCollaboration_15_I} in order to minimise contamination from local Universe foregrounds. We find no significant changes if equivalent masks provided by the ACT collaboration are used instead.

Other works on the kSZ signal \citep[such as][]{Hotinli2025,Lai2025} use 90~and~150\,GHz bands instead, in order to avoid foreground contamination such as cosmic infrared background (CIB), which is a non-thermal extragalactic foreground identified by the \textit{Planck} team as being associated with high-redshift star-forming galaxies. One disadvantage of this approach is that the larger beam angles at these frequencies begin to dominate the angular size of haloes, causing any investigation of their angular extent to produce more uncertain results. Also, our experiments with using the lower-frequency maps revealed significant tSZ contamination from hot high-mass groups at these frequencies that is not appropriately removed by the methods described in Sec.~\ref{sec:stacking}, causing any values of $f_\mathrm{gas}$ derived to be suspect. The expected tSZ components in these two other maps act as an extra source of noise, which may require a larger sample of stacks to beat down. Therefore, to avoid these larger beam angles and tSZ contamination, we preferred to use only the 220\,GHz band and employ analysis techniques to nullify any simple CIB contamination and quantify any remaining Doppler-shifted CIB (see Secs.~\ref{ssec:stacking_procedure} and \ref{ssec:CIB_contamination}, respectively).

\section{Tailoring mock group catalogues to match observations} \label{sec:AbSu_alterations}

\begin{figure}
\begin{centering}
    \includegraphics[width=\columnwidth]{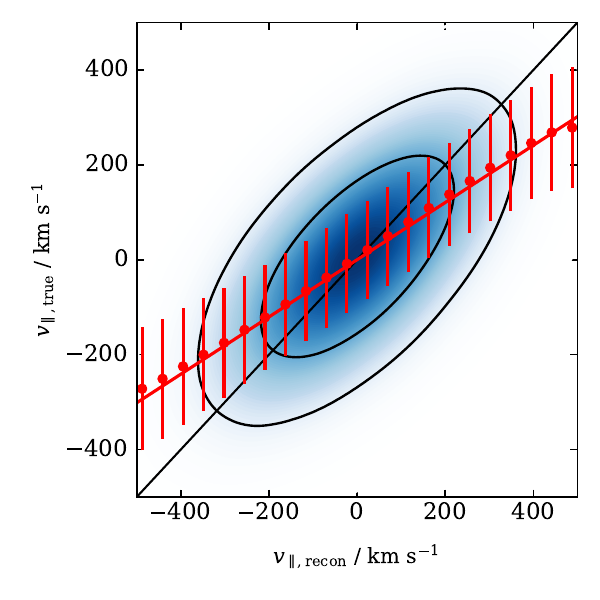}
    \caption{
        Reconstructed grid LoS velocities against true grid LoS velocities for the \textsc{AbacusSummit} halo catalogue, with RSDs and photo-$z$ errors applied to the mock data. Blue shading shows density of haloes in this space, with the contours containing 68 and 95 per~cent of data. Red points and error bars show mean and standard deviation of a Gaussian fitting in bins, with the red line showing a least-squares linear fit. The black line shows {$v_{\parallel,\mathrm{recon}}=v_{\parallel,\mathrm{true}}$}. The halo catalogue is treated identically to the DESI group catalogue, with the same sky masking and redshift distribution. The velocities are reconstructed as described in Sec.~\ref{sec:vel_rec}, and identical grid resolution and smoothing are applied to the true and reconstructed velocities. 
    }
    \label{fig:diag}
\end{centering}
\end{figure}

The observed DESI-LS group catalogue suffers from many observational uncertainties and effects that are known to bias kSZ measurement. As the aim of this work is to model the observed kSZ effect from DESI-LS groups using a mock catalogue, we must also model all these effects such that they reproduce this measurement bias.

Firstly, one potentially biasing inconsistency between DESI-LS and \textsc{AbacusSummit} catalogues as described in Secs.~\ref{ssec:DESI_groups}--\ref{ssec:Abacus_haloes} is their mass functions. We find that the halo mass function produced by the \textsc{CompaSO} algorithm is reasonably consistent with the \citet{Warren_etal_06} mass function used for luminosity abundance matching in the DESI-LS group catalogue. However, the clear difference in halo mass definitions in each do lead to some difference in their mass functions, particularly in the high-mass regime. 
Small deviations in the high-mass tail can potentially lead to discrepancies in predicted kSZ amplitude, since it is natural to weight the contributions of haloes in proportion to mass. 
We therefore match the mass function of the \textsc{AbacusSummit} catalogue to the DESI-LS group catalogue, as shown in Fig.~\ref{fig:hists}. Haloes from simulation and observation are ranked in mass order and DESI-LS masses are mapped across to \textsc{AbacusSummit} haloes in one-to-one fashion. That is, the mass of the most massive observed group is assigned to the most massive simulated halo, the mass of the second-most massive group is given to the second-most massive halo, and so on. This perfectly aligns the already similar mass functions of the two catalogues, whilst minimising the mass change to individual haloes. We note that, as a test, the full analysis was performed both with and without this mass function matching and gave consistent results, emphasising the initial similarity of the mass function of the \textsc{AbacusSummit} halo catalogue to that of the DESI-LS group catalogue.

Secondly, the DESI-LS group catalogue contains redshift-space distortions (RSDs) which, if not corrected for, would cause the absolute halo velocities as predicted by the observed density field to give a systematic underestimate of the true velocities. In concordance with the bulk of this work, we choose to correct for these suppressive effects by forward modelling them using the \textsc{AbacusSummit} halo catalogue. To add these to the mock catalogue, redshift perturbations are added to each \textsc{AbacusSummit} halo individually based on their cosmological redshift and `true' LoS velocities. RSDs can then be corrected for in the same manner in catalogues from observation and simulation, as described in the following section.

Finally, the significant photometric redshift uncertainties present in the DESI-LS group catalogue significantly suppress reconstructed LoS velocities. Again, we model this by adding redshift perturbations individually to \textsc{AbacusSummit} haloes. Halo redshifts are perturbed according to the photometric redshift error distribution given by \citet{Hang2021} for the DESI Legacy Survey. This empirically-derived modified Lorentzian function is described by four redshift-dependent parameters \citep[see secs.~3.2.1 and table~1 of][where we set $x_0^\mathrm{bf}=0$]{Hang2021}, from which we randomly sample a redshift perturbation for each halo. We note that these error distributions refer to photo-$z$s derived by \citet{Hang2021} for a galaxy catalogue based on DESI Legacy DR8, whereas we are modelling a \textit{group} catalogue from DESI Legacy \textit{DR9}, and which was based on a different set of photo-$z$s. But the uncertainty in photo-$z$ estimates is largely set by the width of filter passbands, and we think it is reasonable to assume that the \cite{Hang2021} error distributions will be representative of the errors in the redshifts used by \cite{Yang_etal_21}. The properties of DR8 and DR9 should also be very similar, since the same filters are used in both. Also, the group catalogue we are using has a mean richness of $1.37$, with approximately $84$~per~cent of these `groups' having only one observed galaxy member, so the galaxy error distribution applies directly to these. For a group containing $N$ galaxies, we might naively assume that the width of the error distribution is reduced by a factor $\sqrt{N}$. However, galaxies of the same morphological type in the same group often exhibit similar colours due to their shared environment and formation history, leading to correlated photo-$z$ errors. Furthermore, the primary effect of an underestimation of redshift uncertainty is simply a systematic fractional bias of the true absolute velocities (see the reduction in gradient of the red curve compared to the black in Fig.~\ref{fig:diag}), which will not affect the final results. In this work, reconstructed velocities are only used to separate groups into receding and approaching subsets (where their absolute values are irrelevant) and in weighting each group when stacking (where only an additive, not multiplicative, bias would affect the result; see Sec.~\ref{sec:stacking}). For all the reasons listed above, we therefore choose to neglect any discrepancy between the photometric redshift error distributions given by \citet{Hang2021} and \citet{Yang_etal_21}.

\section{Velocity reconstruction} \label{sec:vel_rec}

The kSZ effect from galaxy groups is linearly dependent on their peculiar velocities, meaning that the net effect over the full sky is zero; in order to detect the effect, we reconstruct the radial velocities of DESI-LS groups. The general approach to this problem uses the linearised continuity equation to estimate the velocity field from the observed (biased) matter density field, while also accounting for the effect of redshift-space distortions and photo-$z$ errors. We now describe the details of this reconstruction method, and show the result of applying it to the \textsc{AbacusSummit} halo catalogue. Our particular aim is forward modelling the degradation in detected kSZ signal that is caused by inaccuracies in the reconstructed velocities. 

\subsection{Mass overdensity field}
\label{ssec:dens_recon}

The following method is used to estimate overdensity fields for both DESI-LS groups and \textsc{AbacusSummit} haloes, though we use the term `haloes' for both set of objects. 

First, we transform haloes to comoving Cartesian coordinates, and build a density field by convolving with a triangular-shaped cloud (TSC) window function and weighting by mass\footnote{Performed using \textsc{Pylians} \citep{Pylians}.}. This mass density field is smoothed in configuration space with a Gaussian kernel of radius $12.5 \,h^{-1}\mathrm{Mpc}$ to reduce the effect of the non-linear fingers-of-God redshift-space distortion \citep[RSD; see][]{Hadzhiyska_etal_23}. 
The larger-scale Kaiser RSD effect is partially mitigated by using a group catalogue, which is more positively biased than the corresponding galaxy catalogue; the remaining distortions are then corrected for as described in Sec.~\ref{ssec:vel_recon}.

To generate a set of unclustered `random' haloes, we use a method similar to that of \citet{Cole_11}. We first choose the random catalogue to be 50 times larger than the source catalogue. Then, to each new group, we assign the redshift and mass of a randomly-selected group from the original catalogue, whilst independently assigning a random angular position within the catalogue's footprint. This ensures that both the number density and mass density redshift distributions of the random catalogue are, once smoothed, equal to those of the original, while still decoupling the line-of-sight and angular distributions.

We then simply calculate the halo overdensity field as 
\begin{align}
    \delta_{\mathrm{h},i} = \frac{\rho_{\mathrm{D},i}}{\rho_{\mathrm{R},i}} - 1,
    \label{eq:overdensities_halo}
\end{align}
where $\rho_\mathrm{D}$ denotes the mass density from the data catalogue, $\rho_\mathrm{R}$ the mass density of the cloned randoms, and $i$ the index of the grid cell.

We note that this procedure to generate the overdensity fields, and the later velocity reconstruction, is always performed separately for the North and South Galactic Cap regions (NGC and SGC) as these are derived from different telescopes and can in principle have different redshift distributions, despite all attempts at making the data uniform. Furthermore, the NGC of the DESI-LS group catalogue is separated into BASS+MzLS and DECaLS regions (along $\mathrm{Dec} = 32^\circ$) when generating random catalogues, as the mass-weighted redshift distributions in these subregions are indeed slightly different. These various random catalogues are then merged, before calculating the overdensity field as normal. This ensures that there is no discontinuity in overdensity between the two regions whilst allowing velocity-inducing interactions between them. 

\subsection{Velocity reconstruction} \label{ssec:vel_recon}

Using this overdensity field, the reconstructed proper peculiar linear velocity field is calculated in Fourier space as follows \citep[see e.g.][following the Fourier convention of e.g. \citealt{Peebles_93}]{Wang_etal_09}:
\begin{align}
\bm{v}_{\bm{k}} = \frac{-\mathrm{i} \, \bm{k}}{k^2} \frac{aHf}{b}\delta_{\bm{k},\mathrm{h}} \, ,
    \label{eq:velocity_field_k}
\end{align}
where a subscript $\bm{k}$ denotes a Fourier-transformed parameter. $H$ is the Hubble parameter, $a$ the normalised scale factor of the Universe, $b$ is the mean linear bias parameter of the surveyed groups, and $f$ is the dimensionless linear growth rate, ${f(\Omega) = \mathrm{d} \ln{D} / \mathrm{d} \ln{a} \simeq \Omega_\mathrm{m}^{0.55}}$. These prefactors are all redshift-dependent. $b$ is calculated in five redshift bins as the mass-weighted average linear bias of haloes, with redshift-dependent biases calculated according to \citet{Tinker_etal_10}\footnote{Calculated using \textsc{cluster-toolkit} \citep{cluster-toolkit} with halo spherical overdensity, $\Delta_\mathrm{b} = 180$.}. This is then interpolated to find $b$ at the redshift of each grid cell. The halo density fluctuation, $\delta_{\rm h}$, is the mass-weighted fractional number density fluctuation, i.e. the matter density contributed by all haloes at a given location, divided by the mean density from these haloes. The latter is smaller than the mean density of the Universe as haloes below the minimum mass are not included, but this undercounting does not affect the analysis.

We correct $\delta_{\bm{k},\mathrm{h}}$ for TSC mass assignment by normalising by the Fourier transform of the TSC window (see \citealt{Hockney_Eastwood_81}; also \citealt{Cui_etal_08}). We also remove velocity modes for which ${k < 4 \times 10^{-3} \, h \, \mathrm{Mpc}^{-1}}$. Along with padding the catalogue region by a minimum extent of ${1 \, h^{-1}\mathrm{Gpc}}$ with ${\delta_{\mathrm{h}}=0}$, this makes effects due to the periodic boundary conditions of the Fourier transform insignificant. 

By inverse fast Fourier transforming we now have a reconstructed velocity field, $\bm{v}$, that is tightly correlated with the true peculiar velocity field \citep{Colombi_Chodorowski_Teyssier_07, Wang_etal_09}, from which the expected peculiar line-of-sight velocities, $v_{\mathrm{\parallel}}$, can be interpolated for each halo. However, this will only be true if the halo positions are known in real space, but this is not so: we position the haloes in 3D assuming a uniform Hubble expansion, but their redshifts are altered by the very peculiar velocities that we seek to estimate, as well as by redshift errors arising from the photo-$z$ estimation process. The following section explains how to allow for these complications.

\subsubsection{Accounting for the impact of RSD and photo-$z$ errors}
Our initial guess for the peculiar velocities has been reconstructed using the redshift space distribution of haloes as if it were real space, 
but we can hope that the result has some correlation with the truth. In that case, we can approximately correct for peculiar velocities and estimate the true cosmological redshifts: 
\begin{align}
    z_\mathrm{corr} = \frac{z_\mathrm{obs} - (v_{\mathrm{\parallel}}/c)}{1 + (v_{\mathrm{\parallel}}/c)}.
    \label{eq:velocity-corrected_redshifts}
\end{align}
The reconstruction method of Secs.~\ref{ssec:dens_recon}--\ref{ssec:vel_recon} is then repeated using these corrected redshifts for each group to obtain new line-of-sight velocities. This process should be iterated until velocities converge, but in practice we find that a single iteration is sufficient -- most probably as a result of the large spatial smoothing kernel. 

Fig.~\ref{fig:diag} shows a comparison between LoS velocities reconstructed using this method and the true LoS velocities from the \textsc{AbacusSummit} light-cone catalogue. There is a high degree of scatter, but the reconstructed velocities are nevertheless significantly correlated with the actual peculiar velocities. This remains true even though the plot compares velocity fields on a mesh, and the velocity of a single halo will have a further scatter around its local mesh value. The challenge is to use this information in a way that reveals the kSZ signal as robustly as possible, without placing excessive reliance on the details of the velocity reconstruction. In practice, we have chosen to do this by using the sign of the reconstructed velocities to classify groups as either approaching or receding. We will use the difference in mean CMB temperature between the approaching and receding subsets as our main measure of the kSZ signal.

If our classification of approaching vs receding was entirely accurate, the observed kSZ signal from the difference of the two subsets would be the same as if we possessed the true velocities, and would be independent of the overall amplitude of the reconstructed velocities. However, the substantial scatter in the photo-$z$'s complicates this picture: the sign of the radial velocity is only estimated reliably for reconstructed velocities above about $400\,\mathrm{km\,s}^{-1}$, and groups with lower velocities are frequently assigned an incorrect sign.
This misclassification biases the observed signal low, to an extent that depends on how we weight the data. As discussed below, we would naturally weight as a function of group momentum: for a weighting that is quadratic in momentum, the average kSZ signal is biased low by about a factor 2; for a linear momentum weighting, the bias is a factor 3. 
In order to interpret any kSZ measurement, we will therefore need to make a detailed forward model of the expected signal that allows for our imperfect approaching/receding classification. We now describe how this was done.

\section{kSZ template} \label{sec:kSZ_template}

In order that we properly forward model the kSZ effect within observational data, we must perform the equivalent filtering, stacking, and analysis procedure on synthetic, or mock, data. As described in Sec.~\ref{ssec:Abacus_haloes}, we are using a halo light-cone catalogue from the \textsc{AbacusSummit} simulations with identical mass, redshift, and sky cuts as applied to its observational counterpart: the DESI group catalogue. We explain below how this group catalogue is used to generate the expected kSZ temperature map on the sky.

In principle the kSZ signal could be predicted from the ionised gas distribution in hydrodynamic simulations; but this would require multiple realisations of the simulation, each providing different values of $f_\mathrm{gas}$ \citep[as in e.g.][]{McCarthyIan_etal_25,Siegel_etal_25}. 
We therefore adopt a more generic approach and generate mock kSZ templates based on the N-body \textsc{AbacusSummit} light-cone described above using an isotropic gas profile model with two free parameters to describe the gas fraction and radial extent. Our semi-analytical approach allows us to explore with flexibility the gas properties around haloes, and set constraints on them.  

The components of the CMB not dependent on group velocity will be an independent source of effective noise that is added on top of the ideal kSZ signal. There is an additional velocity-dependent effect, the Doppler-shifted cosmic infrared background, but this is found to be negligible when compared to the kSZ signal at {$220\,\mathrm{GHz}$} (see Sec. \ref{ssec:CIB_contamination}), and we neglect it in the modelling.

\subsection{Optical depth profile} \label{ssec:tau_profiles}

Following a similar method to that in sec.~3.1 of \citet{Sugiyama_Okumura_Spergel_17}, we assume that the luminosity-weighted centre of the group aligns with the centre of the surrounding electron cloud and model this cloud with a two-dimensional Gaussian profile,
\begin{align}
    \tau(\theta) = \frac{N_{\mathrm{e}} \, \sigma_\mathrm{T}}{D_{\mathrm{A}}^2} \, W_\mathrm{G}(\theta),
    \label{eq:tau_Gaussian_1}
\end{align}
where $N_{\mathrm{e}}$ is the total number of electrons in the halo; $D_\mathrm{A}$ is its angular diameter distance; $\theta = \lvert \bm{\theta} - \bm{\theta}_i \lvert$ is the angle from the halo's centre, $\bm{\theta}_i$; and the normalised Gaussian profile, $W_\mathrm{G}$, has a spatial integral of unity:
\begin{align}
    W_\mathrm{G}(\theta) = \frac{1}{2\pi \, \Sigma^2} \, \mathrm{e}^{-\theta^2 / (2\Sigma^2)},
    \label{eq:Gaussian_profile}
\end{align}
where $\Sigma^2 = \sigma_\mathrm{B}^2 + \sigma_\mathrm{R}^2$ includes both the beam angle of the instrument, $\sigma_\mathrm{B}$, and the angular radius of the gas cloud itself, $\sigma_\mathrm{R}$. We approximate the ACT beam as an isotropic Gaussian with FWHM of $1.0 \, \mathrm{arcmin}$ \citep[see table~1 of][]{Hincks_etal_10}, corresponding to a beam angle of $\sigma_\mathrm{B} = \mathrm{FWHM}/\sqrt{8 \ln{2}}$. 

This simplified Gaussian gas profile is fully described by two parameters: the amplitude, which is fully degenerate with $f_\mathrm{gas}$, and the radial extent, discussed later. Gaussians have been shown in this context to provide very similar results to more realistic gas profiles (e.g.\ a $\beta$-profile: see \citealt{Sugiyama_Okumura_Spergel_17}; see also the discussion in \citealt{Yang_Cai_etal_22}), albeit for CMB maps with lower angular resolution. Moreover, with beam smearing and the fact that we are binning many profiles with a range of angular extents, their stacked average will converge to a Gaussian regardless of the individual physical profiles used.

We model the characteristic angular radius, $\sigma_\mathrm{R}$, as the angular virial radius, $R/D_\mathrm{A}$, multiplied by a dimensionless scaling factor, $\alpha$. For $R$, we use the proper halo radius at which the spherical overdensity is 180 times the background matter density:
\begin{align}
\alpha\equiv \frac{\sigma_\mathrm{R} \, D_\mathrm{A}}{R}; \quad
    R = \left( \frac{3}{4\pi} \frac{M}{\bar\rho(z) \, \Delta} \right)^{1/3} \,,
    \label{eq:radius_from_mass}
\end{align}
where {$\bar\rho(z) = \Omega_\mathrm{m}(z) \, \rho_\mathrm{crit}(z)$}, {$\Delta=180$}, and {$M = M_{180\mathrm{b}}$}. 

The radial scaling parameter, $\alpha$, is introduced to incorporate our lack of knowledge of the concentration of the gas cloud \citep[see e.g.][]{Oppenheimer_etal_25}, together with any profile smearing due to misalignments of the centres of the group's luminosity and gas distributions. In previous analyses \citep[e.g.][]{Schaan_etal_16,Sugiyama_Okumura_Spergel_17} this scaling parameter was neglected due to a dominant beam angle. As a simple guess for $\alpha$, we can note that the radius of a sphere centred on a 3D Gaussian density distribution and enclosing half its mass is approximately $1.538\sigma$, where $\sigma$ is the isotropic standard deviation of this Gaussian; if we use this model for the baryons and assume that they are co-spatial with the total mass, then this baryon half-mass radius would equal half the virial radius (for a truncated isothermal halo), suggesting $\alpha \simeq 0.33$ if baryons and dark matter have the same distribution. But the gas can be more or less extended than the mass, so $\alpha$ can in practice be larger or smaller than this baseline figure.

Over the redshift and mass range analysed in this work, with the above $\alpha=0.33$, we find that $\sigma_\mathrm{R}$ is of the same order of magnitude as the $220 \, \mathrm{GHz}$ ACT $\sigma_\mathrm{B}$. The mean halo mass in the DESI group sample is approximately ${M_{180\mathrm{b}} = 10^{12.5} \msunoh}$. However, because the kSZ signal is proportional to halo mass -- even with the far smaller number of high-mass groups -- we find that the majority of the kSZ signal arises from haloes of mass above $10^{13.5} \msunoh$. With $\alpha=0.33$, for these two masses the beam angle is dominant (i.e.\ ${\sigma_\mathrm{B} > \sigma_\mathrm{R}}$) for redshifts of ${z > 0.27}$ and $>0.76$, respectively.

\subsection{Prior on $\mathbf{\alpha}$}
\label{sec:prior_on_alpha}

The scaling parameter, $\alpha$, allows the baryonic component of a halo to be either more or less extended than its total mass distribution. If $\alpha \ll 1$, then the baryonic mass converges well within the virial radius; but if $\alpha\sim 1$ or larger, then substantial baryon mass will lurk beyond the virial radius. In principle, this possible `bloating' of the baryon distribution is independent of the total quantity of baryons in a halo, but there are good theoretical reasons to expect that the two quantities are related. Suppose we endow a halo with a gas mass that is set by the universal baryon fraction: $M_\mathrm{gas} = (\Omega_\mathrm{b}/\Omega_\mathrm{m})\,M$. If the potential well of the halo is shallow and/or feedback heating of the gas is strong, the gas will naturally become more distended -- but we will still expect to recover a universal baryon fraction if we integrate to large radii. This issue is discussed by \citet{Sorini_etal_22} and \cite*{Ayromlou_Nelsoneta_Pillepich_23}, who compare various hydrodynamical simulations and conclude that a universal baryon fraction is always recovered -- although it may be necessary to integrate to ten times the virial radius.

As illustrated in fig.~7 of \cite{Ayromlou_Nelsoneta_Pillepich_23}, there is general agreement from a range of galaxy formation codes that the impact of feedback should depend on halo mass: massive haloes retain a global baryon fraction, while low-mass haloes eject almost all their baryons beyond the virial radius. Approximately 50~per~cent of baryons are retained within the virial radius for masses around $10^{12.5}\msun$, albeit with a substantial dispersion in this value, depending on the model. We can use this information to develop an approximate prior for $\alpha(M)$, as follows. The radial profiles for dark matter and for baryons will be different in detail, but we assume that the appropriate value of $\alpha$ will be given by the ratio of the half-mass radii (in 3D) for these two components. For the total mass, we assume an NFW profile with a concentration-mass relation $c=(M/10^{18}\msun)^{-0.23}$ \citep{Gastaldello2007}. For the baryons, we assume a density $n_\mathrm{e}(r) \propto r^{-\gamma}\exp(-r/r_\mathrm{b})$, where the power-law slope can be approximated as $\gamma=1.4 - 0.3\log_{10}(M/10^{14}\msun)$ based on gas density profiles of haloes from the \textsc{simba} simulations with fiducial modelling of baryonic feedback (\citealt{Dave2019}; analysed by \citealt{Sorini_etal_24}); the typical baryon fraction within $r_{\rm vir}$ as given in fig.~7 of \cite{Ayromlou_Nelsoneta_Pillepich_23} can then be approximated via $r_\mathrm{b}/r_\mathrm{vir}  = (M/10^{12.8}\msun)^{-0.4}$. Using these model profiles to compute half-mass radii, we then get the final approximation for $\alpha$:
\begin{equation}
\alpha(M) = 0.26\, (M / 10^{14}\msun)^{-0.27}.
\end{equation}

This informative prior for $\alpha$ is not to be taken too seriously: there is a substantial difference between the predictions of different models, and the estimate is based on calculations at $z=0$ and neglects any evolution. 
Rather, we will use this prior as a fiducial case, to illustrate that the gas mass fraction can be measured quite robustly.
We then show that its observational analogue -- the gas mass fraction within the virial radius, $f_\mathrm{gas,vir}$ -- can still be measured robustly when gas extent, i.e. $\alpha$, is not known a priori.
This is achieved by allowing $\alpha$ to be a free parameter that is fitted independently in different mass bins.
Having measured a value of $\alpha$, we can then predict a template profile for the kSZ signal using our mock observations, normalised via the assumption that the halo has a total baryon content that is universal ($f_\mathrm{gas}$, as previously defined, equal to $\Omega_\mathrm{b}/\Omega_\mathrm{m}$). 
We could then scale this template profile to match the CMB data, but this scaling factor will not give a robust estimate of the total baryon mass: much of the baryon content can lie at large radii, where the kSZ signal is small and therefore not typically observed. 
Rather, we will show that the kSZ amplitude gives an estimate of the baryonic mass within the virial radius that is rather insensitive to $\alpha$, and for this reason it is helpful to contrast results using an external prior on $\alpha$ with those deduced by fitting $\alpha$ directly. 

\subsection{Optical depth amplitude} \label{ssec:optical_depth}

\begin{figure*}
\begin{centering}
    \includegraphics[width=2.\columnwidth]{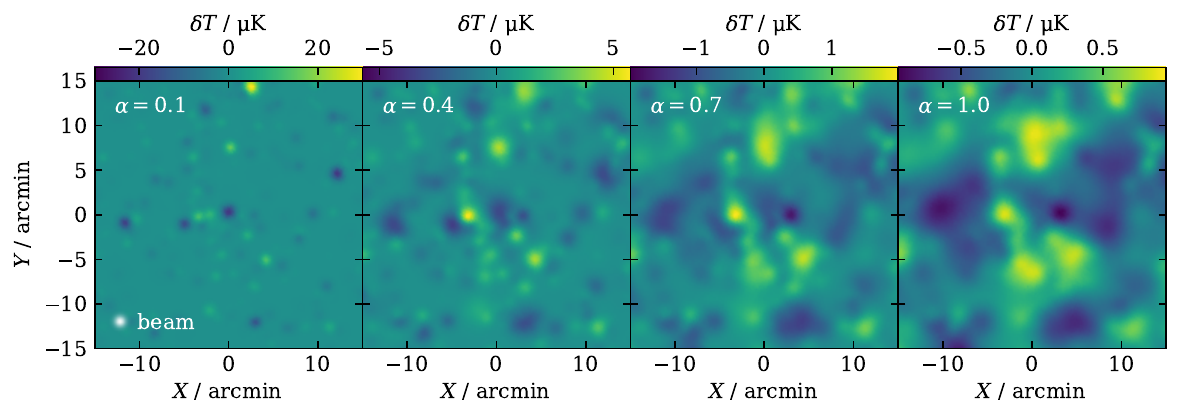}
    \caption{A comparison of CMB temperature fluctuations, $\delta T$, over the same region of kSZ templates with different radial scaling parameters, $\alpha$. These maps are convolved with an isotropic Gaussian beam with FWHM of $1.0\,\mathrm{arcmin}$, shown in white in the left panel. The mock kSZ templates are generated from \textsc{AbacusSummit} halo light-cone as described in Sec.~\ref{sec:kSZ_template}. Lower values of $\alpha$ give sharply peaked halo profiles with smaller angular extent compared to the extended and shallower profiles produced for higher $\alpha$, but different haloes are affected by $\alpha$ to different extents. 
    These effects, expected to occur in real observations, are reproduced in our kSZ template (see Sec.~\ref{ssec:optical_depth} for details).
    Note that the colour scale is not shared between panels for visual clarity. The full templates cover around 40~per~cent of the sky over {$0.1<z<0.9$}, equivalent to the DESI-LS catalogue (see Sec.~\ref{sec:kSZ_template} for details).}
    \label{fig:alpha_gnomes}
\end{centering}
\end{figure*}

\begin{figure*}
\begin{centering}
    \includegraphics[width=1.39\columnwidth]{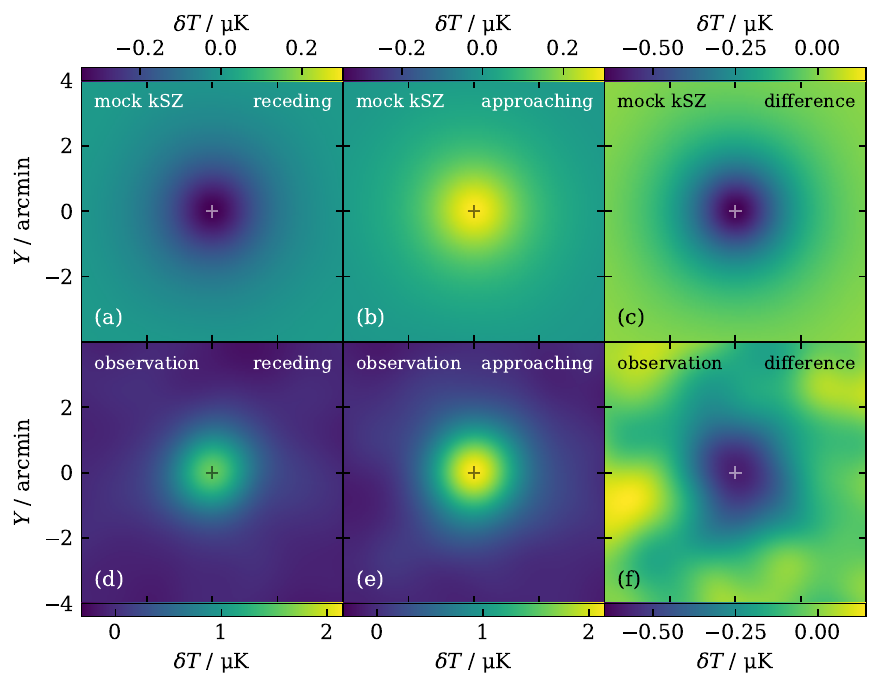}
    \raisebox{0.02\height}{\includegraphics[width=0.61\columnwidth]{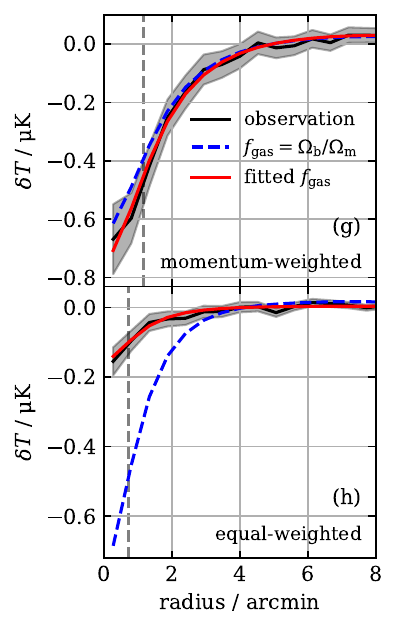}}
    \caption{
        Stacked simulated and observed temperature fluctuations, $\delta T$, showing absolute LoS momentum-weighted averages of regions around haloes.
        Panels (a--c) show stacks of a mock kSZ template (without superimposed intrinsic CMB fluctuations), generated using a halo light-cone catalogue from the \textsc{AbacusSummit} simulations (with {$\alpha=0.6$}; see Sec.~\ref{sec:kSZ_template}). 
        Panels (d--f) show stacks of a filtered {$220\,\mathrm{GHz}$} ACT CMB temperature map over the positions of DESI-LS galaxy groups, smoothed with a $0.5\,\mathrm{arcmin}$ FWHM Gaussian kernel to remove visual noise.
        Panels (a) \& (d), (b) \& (e) show stacks over only the receding and approaching haloes, respectively. Observational stacks are both positive due to possible CIB emission of groups.
        Panels (c) \& (f) show the receding stack minus the approaching stack, thereby cancelling out all temperature fluctuations that are not halo velocity-dependent. The dominant remaining effect being kSZ. 
        Panels (g) \& (h) show radial means of difference stacks with absolute LoS momentum weighting and equal weighting, respectively. Note that the radius extends further than shown in the left panels and shows means of pixel values in annuli of $0.5\,\mathrm{arcmin}$ width.      
        The black curve corresponds to observation [i.e. panel (f) with no smoothing]; the grey vertical dashed line denotes the absolute LoS momentum-weighted average angular halo virial radius, $\sigma_\mathrm{R}$; the blue dashed curve shows prediction from the mock kSZ template assuming $f_\mathrm{gas} = \Omega_\mathrm{b}/\Omega_\mathrm{m}$;
        and the red curve shows the best-fitting profile corresponding to {$f_\mathrm{gas} = (1.2 \pm 0.4) (\Omega_\mathrm{b}/\Omega_\mathrm{m})$} and {$\alpha = 0.59 \pm 0.12$}.   
        Panel~(h) is similar to panel~(g), but with no group weighting, with red curve corresponding to {$f_\mathrm{gas} = (0.2 \pm 0.2) (\Omega_\mathrm{b}/\Omega_\mathrm{m})$} and {$\alpha = 0.4 \pm 0.2$}. See Sec.~\ref{sec:fgas} for details. 
    }
    \label{fig:stack}
\end{centering}
\end{figure*}

For the purpose of template construction, we initially assume a universal gas-mass fraction in haloes and that all electrons are in fully-ionised gas associated with haloes, i.e. we do not model any diffuse component that is not gravitationally bound to any haloes in our catalogues. From these assumptions we obtain $N_{\mathrm{e},i} = f_\mathrm{gas} \, M_i / (\mu_\mathrm{e} \, m_\mathrm{p})$ for a halo of mass $M_i$, where $\mu_\mathrm{e} = 1.17$ is the mean particle weight per electron and $m_\mathrm{p}$ is the proton mass. From our full ionisation assumption, we can take $f_\mathrm{gas}$ to be equal to the universal baryon fraction, $f_\mathrm{b} = \Omega_\mathrm{b} / \Omega_\mathrm{m} = 0.158$, found by \citet{PlanckCollaboration_18_VI} using combined TT,TE,EE+lowE+lensing+BAO constraints. This serves as a fiducial assumption that that is later relaxed in order to fit to the data from observation.

Substituting this into equation~(\ref{eq:tau_Gaussian_1}), and applying the Gaussian convolution in equation~(\ref{eq:Gaussian_profile}), we obtain
\begin{align}
    \tau_i(\theta) = \frac{\sigma_\mathrm{T} \, f_\mathrm{gas} \, M_i}{2\pi \, \mu_\mathrm{e} \, m_\mathrm{p} \, \Sigma_i^2 \, D_{\mathrm{A},i}^2} \, \mathrm{e}^{-\theta^2 / (2\Sigma^2)}.
    \label{eq:tau_Gaussian}
\end{align}
We can then substitute $\tau_i(\theta_i)$ into equation~(\ref{eq:kSZ_proj}), summing over all $N$ haloes:
\begin{align}
    \delta T_\mathrm{kSZ}(\bm{\theta}) = - T_0 \sum_i^N \tau_i(\theta_i) \frac{\bm{v}_{\parallel,i}}{c}.
    \label{eq:kSZ_proj_sum}
\end{align}
To build up the kSZ template, this is then calculated at the position of each \textsc{HEALPix}\footnote{\textsc{HEALPix} (\citealt{HEALPix}; \url{http://healpix.sf.net}); \textsc{healpy} \citep{healpy}.} pixel over the sky. 

An example patch of these kSZ templates is shown in Fig.~\ref{fig:alpha_gnomes} for multiple values of $\alpha$. We see that $\alpha$ strongly affects both the angular extent of haloes and their maximum amplitude. We also see that $\alpha$ affects the angular extent of individual haloes differently, as it only alters $\sigma_\mathrm{R}$ and not $\sigma_\mathrm{B}$ in equation~\ref{eq:Gaussian_profile}. For example, the kSZ signal of the pair of haloes producing a dipole at the centre of the panels are less affected by $\alpha$ than other objects in the region of sky shown, i.e. as the value of $\alpha$ increases from the left to the right, they become more extended, but their increase in extent is not as significant as that of other structures in the field. This causes the dipole to remain fairly constant in amplitude whilst other haloes fade as they become extended; meaning that the dipole appears comparatively brighter as $\alpha$ is increased. This is because the two haloes making the dipole in temperature lie at high redshift, meaning $\sigma_\mathrm{B}$ is more dominant over {$\sigma_\mathrm{R} = \alpha R / D_\mathrm{A}$}. Effective point sources of this kind simply remain approximately the same angular size of the beam no matter the $\alpha$, so that their predicted profile is less sensitive to any change in $\alpha$ than is the profile of a well-resolved halo.

We can therefore anticipate that the observed stacked kSZ signal will vary for haloes of different masses and redshifts. These two factors are entangled: a low-mass halo at low $z$ may contribute significantly to the stacked signal up to relatively large radii if its gas profile is more extended than the ACT beam, but the same halo may be unresolved at high $z$, and therefore contribute only at small angular scales. All such effects are incorporated in our kSZ templates.

In summary, our template model of the kSZ has two free parameters: the gas fraction parameter $f_{\rm gas}$ which regulates the amplitude of the overall kSZ signal, and the $\alpha$ parameter which characterises the extent of the gas density profile. They are correlated, but not completely degenerate, allowing them to be constrained from the data.

\subsection{Matching mock and real observations} 

To use our mocks as a forward model, it is crucial to ensure that the simulated signal is extracted in a realistic manner. This is taken care of by the following procedure.

We generate the ideal kSZ signal on the sky in the fiducial cosmology, following the method described in Sec.~\ref{sec:kSZ_template}, using the `true' positions and velocities of haloes in the \textsc{AbacusSummit} light-cone catalogue described in Sec.~\ref{ssec:Abacus_haloes}. This mock kSZ effect is the exact signal that would be imprinted on the CMB and is free of any observational effects associated with a galaxy survey.

Then we make a reconstructed velocity field, allowing for RSD effects and photo-$z$ errors; incomplete knowledge of the underlying matter field due to the sparse sampling of the density field; survey geometry; and simplifying assumptions in the reconstruction processes. This involves a number of steps: 

(1) We match the halo mass distribution as a function of redshift to the DESI-LS catalogue, with DESI-LS-like photometric redshift errors and RSDs added as described in Sec.~\ref{sec:AbSu_alterations}.

(2) With the halo light-cone catalogue having DESI-LS-like selection effects applied (see Fig.~\ref{fig:hists}), we apply the same sky mask and redshift cut to the simulated data as is applied to the observed data.

(3) We reconstruct the LoS velocities for the simulation haloes as described in Sec.~\ref{sec:vel_rec}, treating the input quantities (positions and masses of either simulated or observed haloes) identically. 

We can then apply the same stacking procedure described in Sec.~\ref{sec:stacking} to real observations and simulations to produce the DESI-stacked CMB map and the \textsc{AbacusSummit}-stacked mock kSZ templates, such as those shown in Fig.~\ref{fig:stack}.

\section{CMB stacking} \label{sec:stacking}

We present here the methods used to overcome the rather low signal-to-noise ratio of the kSZ signature. As we will see, the typical signal for our groups is below $1\,\mathrm{\upmu K}$, which is to be compared with  root-mean-square fluctuations of around $100\,\mathrm{\upmu K}$ in the primordial CMB temperature. This disparity can be reduced by taking account of the different angular scales of the kSZ and intrinsic CMB fluctuations, which is achieved via filtering of the CMB sky in harmonic space. Regions of the CMB sky around galaxy groups are then stacked with weights to maximise the kSZ signal whilst cancelling out primordial fluctuations and other sources of contamination. Crucially, we stack the receding and approaching groups separately to avoid cancellation, using the sign of the reconstructed LoS peculiar velocities. This procedure is applied to both observational and mock datasets.

\subsection{CMB filtering} \label{ssec:filtering}

The dominant source of noise is the primordial CMB, particularly the first few peaks of its power spectrum. These lower-multipole modes cause a `global' offset to the final stacked profiles, where the profile does not fall to zero within expected radii, making it difficult to appropriately compare to mock profiles, where no such offset exists. Whereas, the power spectrum of the kSZ signal is known to peak at high multipoles: the peak of all kSZ templates generated using the method in Sec.~\ref{sec:kSZ_template} is at $\ell>1000$ ($\theta \lesssim 10\,\mathrm{arcmin}$). We therefore design the filter to remove low-multipole noise whilst keeping the majority of the high-multipole signal untouched.
After testing various filters on mock kSZ and CMB templates, we find that the following filter does not significantly impact the amplitude of the expected one-halo kSZ signal, preserves its profile shape, and substantially reduces primordial CMB noise:
\begin{equation}
  f(\ell)=\begin{cases}
    0, & \text{if $\ell<540$}\\
    3x^2 - 2x^3, & \text{if $540\leq\ell\leq1080$}\\
    1, & \text{if $\ell>1080$}
  \end{cases}
\end{equation}
where $x=\ell/540-1$, and $\ell$ of 540 and 1080 correspond to angles of $20$ and $10\,\mathrm{arcmin}$, respectively. This smooth B\'ezier curve transition is included as apodisation to maintain the profile shapes. See fig.~2 of \citet{Tanimura_Zaroubi_Aghanim_21} for the effect of a similar filter on the primordial CMB power spectrum. This filter is therefore applied to the observed CMB and to all mock kSZ templates in this work.

\subsection{Stacking procedure} \label{ssec:stacking_procedure}

For both the mock and real data, we image small regions of the sky centred on haloes (`stamps') using gnomonic projection and average these pixel-by-pixel into `stacks'. We repeat this for the receding and approaching groups separately and take the difference between them ({receding $-$ approaching}). We show this in panels (a--c) \& (d--f) of Fig.~\ref{fig:stack} for mock and real observations, respectively.
The stacking process averages out sources of contamination that are uncorrelated with the positions of haloes, but signals such as CIB and tSZ may still remain in each of the two sub-samples. This is evident in panels (d) \& (e) of Fig.~\ref{fig:stack}, where the stacked results from both the receding and the approaching groups yield positive temperature fluctuations, but we anticipate them to have the opposite sign of temperature if the signal is pure kSZ, as seen in the results from mocks [panels (a) \& (b)]. As long as the statistical properties of the two subsamples are the same, taking the difference of the two stacks should remove contaminants that are in common, leaving the pure kSZ signal. This is evident in the difference between the receding and approaching groups, shown in panels (c) \& (f) of Fig.~\ref{fig:stack}. We can see that the difference of these two stacks converges to zero at large radii (greater than $\sim4\,\mathrm{arcmin}$) -- a reassuring sign that any non-kSZ residuals in each of the stacks must be very similar between the two stacks, such that their difference is consistent with null. We see here that there is a clear velocity-dependent signal remaining, with a negative dip at small radii. The only possible remaining velocity-dependent contaminant is the Doppler-shifted component of CIB emission, which we show in Sec.~\ref{ssec:CIB_contamination} to be sub-dominant compared to kSZ
\citep*[see also][]{Maniyar_Ferraro_Schaan_22}.

These two-dimensional stacks are then transformed into one-dimensional profiles for further analysis. 
Pixel values of stacks are mean averaged in annuli of $0.5\,\mathrm{arcmin}$ width out to a maximum radius of $8.0\,\mathrm{arcmin}$. Covariances are then calculated as described in Sec.~\ref{ssec:errors}. The shaded regions in Figs.~\ref{fig:stack} \& \ref{fig:profiles_M_bins} show 1-$\sigma$ uncertainty derived from the diagonal components of the covariance matrix for the profile from observation (black curves). Uncertainties of profiles from mock observation (red and blue curves) are too small to be seen and so are not plotted.

\subsubsection{Weights}

When stacking, we wish to up-weight stamps with a high kSZ signal-to-noise ratio (SNR) and down-weight those without. 
The kSZ signal expected from each group is linearly dependent on both its halo mass and LoS velocity, i.e.\ on its LoS momentum (see Sec.~\ref{sec:kSZ_template}).
We can then classify any non-kSZ temperature fluctuation as a noise term. We first minimise noise due to foreground contamination by applying a $40$~per~cent galactic plane mask and a $\mathrm{SNR} > 5$ point source Boolean mask \citep{PlanckCollaboration_15_I}. Instrumental noise is accounted for by applying ACT's inverse variance mask \citep{Naess_etal_25}. 
After appropriate sky masking is applied, the noise is then dominated by primordial CMB temperature fluctuations. This Gaussian random field, when filtered as described in Sec.~\ref{ssec:filtering} and averaged over stamps of the angular scale used in this analysis, has a variance that is treated as equal for each stamp. 

Whilst the theoretically optimal weighting for each stamp would be the square of SNR, equivalent to the square of LoS momentum, this gives most of the weight to a few high-velocity groups whose velocities may not be reliable. We instead choose to weight in proportion to SNR, and so we weight each stamp by the absolute value of its group's LoS momentum when stacking over receding and approaching groups. Again we emphasise that the result depends only on the relative magnitude of the reconstructed velocities.

We implement this momentum-weighted stacking on the filtered ACT map and obtain the images shown in panels (d--g) in Fig.~\ref{fig:stack}. This will be the default choice of our weights. We will also explore results with equal weighting for each halo [e.g. panel (h)] for comparison with results in the literature.

\section{Constraining gas properties} \label{sec:fgas}

The results shown in Fig.~\ref{fig:stack} demonstrate that we have successfully detected the kSZ effect using the method described in Sec.~\ref{sec:stacking}. We now use the kSZ templates constructed as in Sec.~\ref{sec:kSZ_template} as a forward model to be compared with the detected signal, and thus set constraints on the properties of gas around haloes.

\subsection{The overall gas fraction} 

\begin{figure}
\begin{centering}
    \includegraphics[width=\columnwidth]{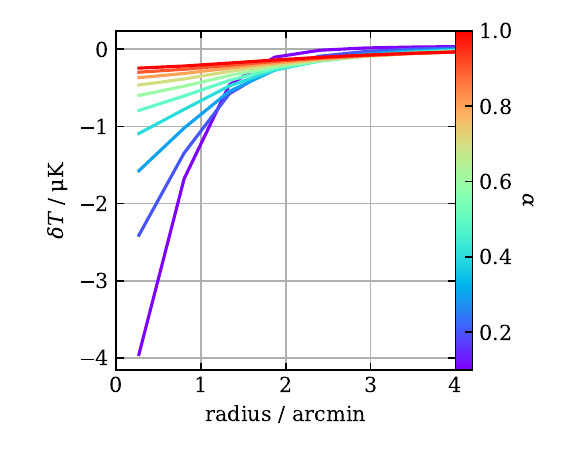}
    \caption{Stacked temperature profiles of the mock kSZ templates, with universal baryon content and varying radial scaling parameter, $\alpha$, where {$\sigma_\mathrm{R}=\alpha R / D_\mathrm{A}$}. Note that the beam angle, $\sigma_\mathrm{B}$ is included in the total Gaussian radius $\Sigma = (\sigma_\mathrm{R}^2 + \sigma_\mathrm{B}^2)^{1/2}$ (see Sec.~\ref{ssec:tau_profiles} for details). The total spatially integrated optical depth is unchanged by $\alpha$, meaning {$\int \delta T \, \theta \, \mathrm{d}\theta$} is conserved, causing sharp slopes and strong temperature decrements for low $\alpha$.}
    \label{fig:rainbow}
\end{centering}
\end{figure}

\begin{figure}
\begin{centering}
    \includegraphics[width=\columnwidth]{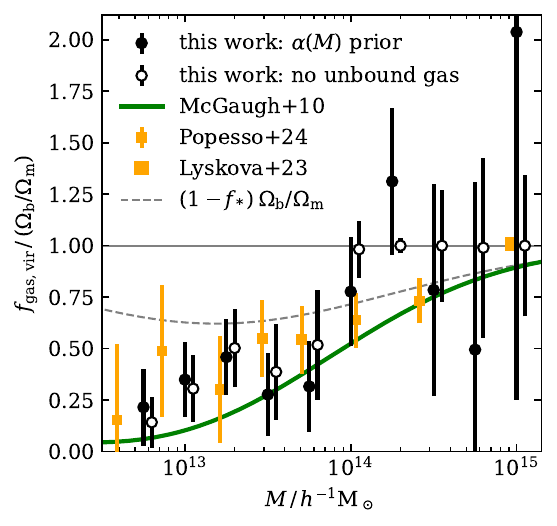}
    \caption{
        Gas fraction within the projected virial radius as a function of halo mass.
        Filled black points show values calculated from fitting $f_\mathrm{gas,vir}$ of stacked mock profiles to stacked profiles from observation, with the radial extent of the gas distribution, $\alpha(M)$, informed by hydrodynamical simulation.
        Unfilled points show values inferred from fitting the radial extent of stacked mock profiles to those from observation, assuming the total halo gas mass fraction (i.e. not within a certain radius) to be equal to the universal baryon mass fraction. From the fitted $\alpha$ values, we analytically infer the fraction of halo gas that lies within the virial radius in projection. They are visually offset in $M$.
        Orange squares show values from eFEDS (small squares; \citealt{Popesso_etal_24}) and eROSITA (large square; \citealt{Lyskova_etal_23}) X-ray data. 
        The green curve is derived from best-fitting formulae given by \citet{Oh_etal_20} based on data compiled by \citet{McGaugh_etal_10} using a generalised baryonic Tully-Fisher relation.
        All data have been transformed to show {$M \equiv M_\mathrm{180b}$} and {$f_\mathrm{gas,vir} \equiv f_\mathrm{gas}(<R_\mathrm{180b})$} assuming a Gaussian gas profile and mass--concentration relation from \citet{Ludlow+16}.
        Mass bins are $0.5 \, \mathrm{dex}$ in width, so adjacent points overlap.
        The universal baryon fraction is shown as a solid grey line, and the same minus the fraction of baryons converted to stars is shown as a dashed grey curve \citep[informed by][]{Oh_etal_20}.
    }
    \label{fig:main_with_integrals}
\end{centering}
\end{figure}

\begin{figure*}
\begin{centering}
    \includegraphics[width=2.\columnwidth]{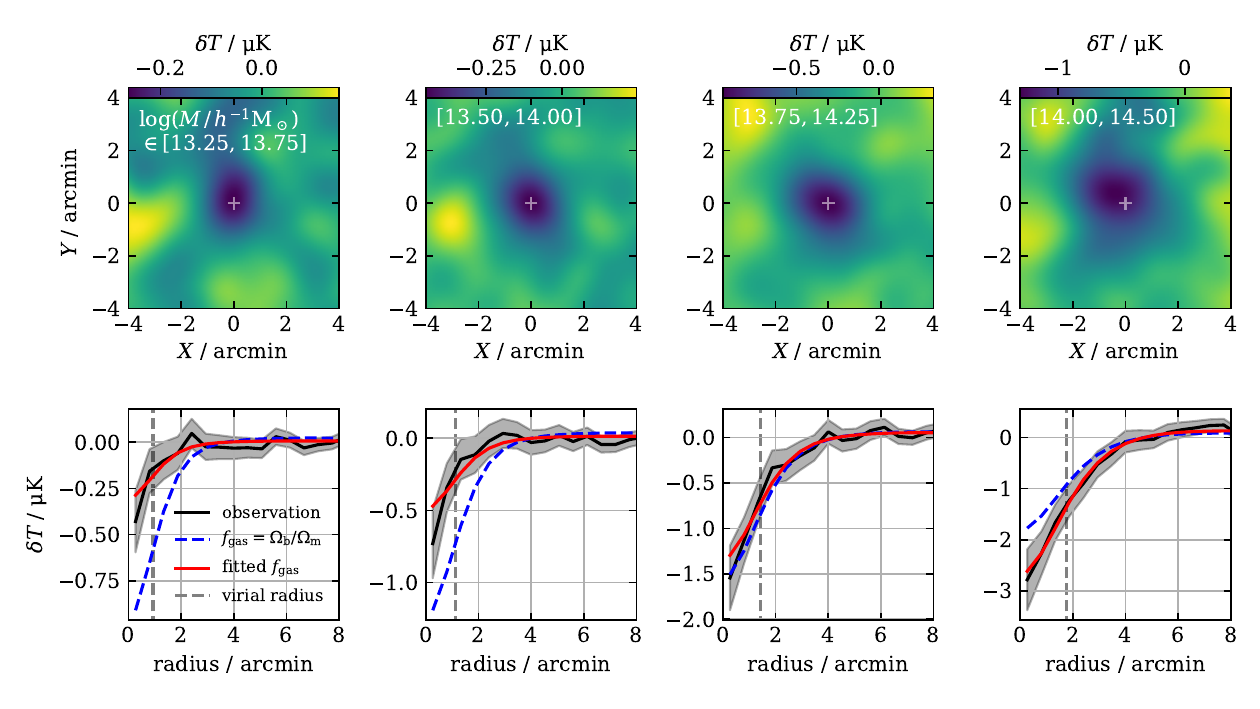}
    \caption{
        Stacks and stacked profiles for groups binned by halo mass, $M$, corresponding to the four central bins of Fig.~\ref{fig:main_with_integrals}. 
        Top row: Absolute LoS momentum-weighted stacks of the difference between receding and approaching groups, with the mass range labelled, similar to the third column of Fig.~\ref{fig:stack}, with the same $0.5\,\mathrm{arcmin}$ FWHM Gaussian smoothing to remove visual noise.
        Bottom row: Profiles corresponding to the above stacks, but with no added smoothing, and an extended radius range. Curves and shaded regions have the same meaning as in panel (g) of Fig.~\ref{fig:stack}.
        Logarithmic $M$ bins increase in rightwards panels. Note that all bins are one decade in width and overlap. As halo mass increases, an increase in both kSZ amplitude and $\sigma_\mathrm{R}$ can clearly be seen, as well as an increase in the ratio between red and blue dashed curves (i.e. $f_\mathrm{gas}$).
    }
    \label{fig:profiles_M_bins}
\end{centering}
\end{figure*}

With the mock and real observational procedure of analysis matched, we can now fit the mock kSZ temperature profiles to observations. From our template, we are left with two degrees of freedom: the gas fraction, $f_{\rm gas}$; and $\alpha$, describing the radial profile of the gas (see Sec. \ref{ssec:tau_profiles}). We will sample these two parameters using Markov chain Monte Carlo (MCMC) sampling with \textsc{emcee} \citep{ForemanMackey_etal_13}.

We generate multiple kSZ templates assuming a wide range of radial gas spatial extents: {$\alpha \in [0.05, 3.0]$} where the angular scale-length of the gas is $\sigma_\mathrm{R} = \alpha R / D_\mathrm{A}$ (see equation.~\ref{eq:radius_from_mass}). In previous kSZ stacking work, the detailed gas profile was often not a concern because the \textit{Planck} beam angle dominated the physical radii of haloes \citep[e.g.][see also discussion in \citealt{Yang_Cai_etal_22}]{Tanimura_Zaroubi_Aghanim_21}. The stacked temperature profiles of a selection of these templates can be seen in Fig.~\ref{fig:rainbow}. We may then fit for values of $\alpha$ between those generated by linearly interpolating between their profiles, allowing us to continuously sample $\alpha$.

Altering $\alpha$ conserves the total spatially-integrated optical depth of haloes, meaning we can then alter the the second fitting parameter, $f_\mathrm{gas}$, independently to scale the amplitude of the mock profile to fit the profile from observation. We note again that all mock templates are initially constructed assuming all haloes to have the universal {$f_\mathrm{gas} = \Omega_\mathrm{b} / \Omega_\mathrm{m} = 0.158$} found by \citet{PlanckCollaboration_18_VI}. The fitted value of $f_\mathrm{gas}$ is then deduced via a linear scaling of these universal profiles with a MCMC search for the minimal $\chi^2$ with respect to the data curve (see Sec.~\ref{ssec:errors} for the calculation of the covariance). 

However, this means that $f_\mathrm{gas}$ is not defined as a gas mass fraction within a fixed radius, as is commonly the case for SZ and X-ray studies. The inclusion of $\alpha$ means that $f_\mathrm{gas}$ measures the `overall' gas mass divided by the total mass of a halo, where `overall' here includes all halo gas with coherent LoS velocity. However, highly attenuated gas in the extreme outskirts of a halo's gas distribution would not contribute to measured $f_\mathrm{gas}$ as this would not conform to a Gaussian profile and any gas at over $8\,\mathrm{arcmin}$ from a halo is not included in the profiles.

\subsubsection{Momentum weighting vs. equal weighting}

The result of the best-fitting profile from stacking of kSZ templates is shown as the red curve in panel (g) of Fig.~\ref{fig:stack}, which finds a $5.4\sigma$ kSZ detection with {$f_\mathrm{gas} = (1.2 \pm 0.4) (\Omega_\mathrm{b} / \Omega_\mathrm{m})$} consistent with the universal baryon fraction and a radial scaling parameter of {$\alpha = 0.59 \pm 0.12$}.

Whilst this is a stack over the full $10^{12.5} < M\,/\msunoh \lesssim 10^{15.5}$ mass range (see Fig.~\ref{fig:hists}), the values of $f_\mathrm{gas}$ and $\alpha$ it produces can be interpreted as indicative of the values at an effective mass the stack is probing. For momentum-weighted stamps, this effective mass probed by kSZ stacking is a kSZ signal- and momentum-weighted average group mass (i.e. momentum-squared-weighted) of $10^{14.5}\msunoh$, i.e. the stacked kSZ sample is dominated by high-mass haloes. As we see in the next subsection, this relatively high averaged halo mass from this weighting scheme is the key to obtaining a gas fraction {\it apparently} consistent with the universal baryon fraction, which could be misleading if it is not interpreted properly.

Using unweighted stamps, however, the stacked sample is dominated by low-mass haloes, which are much more abundant. The average halo mass of the stack falls to the momentum-weighted average group mass of $10^{13.7}\msunoh$, which is nearly an order of magnitude lower than the momentum-squared-weighted case above. This result is shown in panel (h) of Fig.~\ref{fig:stack}. In this case, we obtain a $2.8\sigma$ kSZ measurement with {$f_\mathrm{gas} = (0.2 \pm 0.2) (\Omega_\mathrm{b} / \Omega_\mathrm{m})$} and {$\alpha = 0.4 \pm 0.2$}, significantly lower than the universal baryon fraction (corresponding to the blue dashed curve). At face value, this seems to be more consistent with the relatively low values of $f_\mathrm{gas}$ found in the literature using a similar combination of datasets \citep[e.g.][]{Hadzhiyska_etal_24, Hadzhiyska2025}.

The difference in the best-fitting values of $f_{\rm gas}$ between the two weightings indicates strongly that the gas fraction per halo mass may vary. We now look in detail at the issue of mass-dependent baryonic properties.

\subsection{Gas fraction as a function of halo mass} \label{ssec:results_gas_fracs}

There have been indications from hydrodynamic simulations and observations that the gas fraction around haloes may vary with the host halo mass \citep[see e.g.][and references therein]{Oh_etal_20,Ayromlou_Nelsoneta_Pillepich_23, Bigwood_etal_24,Popesso_etal_24,Wright_etal_24,Oppenheimer_etal_25,Siegel_etal_25}. 
With the high signal-to-noise ratio of the kSZ measurement, and the halo mass information provided from the group catalogue, we can test this suggestion by analysing the gas fraction for haloes of different masses.

We extend the modelling and fitting procedure to kSZ measurements within ten logarithmic mass bins. To increase the signal-to-noise per mass bin, we allow 0.5\,dex for each mass bin spaced every 0.25\,dex, seen in Fig.~\ref{fig:main_with_integrals}, and so halo samples in neighbouring mass bins have some overlap. This will also introduce extra covariance between consecutive data points as plotted, but we are not fitting any model to these data, so the extra covariance is not very relevant.

There are subtleties in this extended fitting that we would like to clarify. The observed kSZ signal is significant at scales smaller than a few arcminutes. These scales are just slightly larger than the viral radii of most of our haloes. This means that our data are more constraining for the gas properties at around the virial radii, or smaller. Therefore, if we observe a deficit of hot gas within the virial radius for haloes of a given mass, this could indicate a low universal baryon fraction, but it could equally well mean that some gas has been expelled from the halo centre to larger radii where our data are weakly constraining. In other words, there are some degeneracies between the two parameters, $f_\mathrm{gas}$ and $\alpha$, within a single mass bin. However, we will find that a combination of these parameters, the baryon fraction within the virial radius, can be estimated robustly. We illustrate this point by using two rather different choices for the priors during the fitting.

First, we choose a prior on $\alpha(M)$ informed by hydrodynamical simulation as discussed in Sec.~\ref{sec:prior_on_alpha}, and fit for the gas fraction within the projected virial radius of each mass bin, $f_\mathrm{gas,vir}$. These are shown as black points with errors in Fig.~\ref{fig:main_with_integrals}. Second, we assume that the total halo gas mass fraction is equal to the universal baryon mass fraction, and allow $\alpha$ to vary and be constrained by the observed profile per mass bin. We then integrate within the projected virial radius of each mass bin to obtain the best-fitting $f_\mathrm{gas,vir}$. This {$f_\mathrm{gas,vir} \equiv f_\mathrm{gas}(<R_\mathrm{180b})$} is distinct from the overall halo $f_\mathrm{gas}$ discussed elsewhere in this work. These are shown as open circles in Fig.~\ref{fig:main_with_integrals}. All values of $f_\mathrm{gas,vir}$ inferred from fitting $\alpha$ and $f_\mathrm{gas}$ lie within one standard deviation of those calculated using the prior on $\alpha$, demonstrating that $f_\mathrm{gas,vir}$ is robust, and not sensitive to the number of degrees of freedom in the model.

Examples of the kSZ measurements per halo mass bin, and the best-fitting profiles versus observations are shown in Fig.~\ref{fig:profiles_M_bins}. We see that our fitted mock kSZ templates (red curves) are able to model well the observed kSZ signal (black curves; see Sec.~\ref{ssec:errors} for detail). We can see that in both cases, $f_\mathrm{gas,vir}$ is consistent with the universal baryon fraction for haloes with masses greater than $\sim 10^{14}\msunoh$, but there is a clear sign of a deficit of gas within the virial radii for the low-mass haloes. 
Indeed, if we use only two mass bins, we find $f_\mathrm{gas,vir}$ equal to {$0.8\pm0.3$} and {$0.38\pm0.11$} times the universal baryon fraction for halo masses above and below $10^{14}\msunoh$, respectively.

The mass dependence of the best-fitting gas fractions explains the ensemble stacked results that we obtained with and without the momentum weighting in the previous subsection. With the momentum weighting, the average mass of the stacked sample is $10^{14.5}\msunoh$, which is significantly higher than $10^{13.7}\msunoh$ for the unweighted case. As we can see in Fig.~\ref{fig:main_with_integrals}, the gas fraction within the virial radius changes abruptly at $\sim 10^{14}\msunoh$, causing the difference for the averaged gas fractions above and below this threshold mass.

We emphasise that the agreement in $f_\mathrm{gas,vir}$ found by the two methods shown in Fig.~\ref{fig:main_with_integrals} shows that the significant decrement in gas within the virial radius at low halo masses (measured directly to obtain the filled points) is still fully consistent with all haloes having a universal baryon fraction when integrated out to many virial radii (as is assumed to obtain the unfilled points, and as is found by e.g. \citealt{Sorini_etal_22,Ayromlou_Nelsoneta_Pillepich_23}). Similarly, we find that using different maximum radii, e.g. $R_\mathrm{200b}$ or $R_\mathrm{500c}$, leads to the same conclusion as the $R_\mathrm{180b}$ used here.

\subsection{Comparison to other results}

Our results are broadly consistent with a recent study using similar datasets of LRGs from the DESI Legacy Survey and their cross-correlations with ACT, where without any mass splitting, and without the momentum weighting, an overall low amplitude of the kSZ signal is found and interpreted as an indication for strong feedback \citep{Hadzhiyska_etal_24}. This was confirmed by another study combining kSZ and CMB lensing \citep{Hadzhiyska2025}, where a particularly low baryon fraction was found at small radii ({$<4\,\mathrm{arcmin}$}) in their stacked profile, but the baryon fraction appeared to be indistinguishable from the universal one at larger radii ({$>4\,\mathrm{arcmin}$}; see figs.~6 \& 10 of their paper). These are broadly consistent with our results of equal weighting. Seen from our results of $f_{\rm gas}$ versus halo mass, it seems likely that the relatively low baryon fraction near the centre of the stacked system is driven mainly by the dominant population of low-mass haloes, in which there is a deficit of baryons. 
Indeed, \citet{Siegel_etal_25} find similar results when analysing these low-mass ({$M\approx10^{13}\msunoh$}) haloes in simulation (see their fig.~6).

The mass-dependence of $f_{\rm gas}$ from our kSZ measurements is also consistent with X-ray measurements from eFEDS \citep{Popesso_etal_24} and eROSITA (\citealt{Lyskova_etal_23}; see also fig.~5 of \citealt{Siegel_etal_25}) in the common range of halo masses, as shown in Fig.~\ref{fig:main_with_integrals}. All these results are consistent within the errors with the best-fitting formula of gas fraction versus halo mass by \citet{Oh_etal_20}, based primarily on rotation curve measurements compiled by \citet{McGaugh_etal_10}. They are also consistent with the results of the baryon census conducted by \citet{Dev2024} from compiling analyses of multiple observations in the literature (see particularly fig.~4 of their paper, and references therein). These results suggest that ionised gas seems to be depleted around relatively low mass haloes, possibly due to feedback mechanisms such as AGN, and the compounding effect that the gravitational potential wells of those low-mass haloes are too weak to retain their hot gas, as suggested by hydrodynamic simulations.

\cite{Oppenheimer_etal_25} compare various model gas profiles: self-similar haloes with {$f_\mathrm{gas} = \Omega_\mathrm{b}/\Omega_\mathrm{m}$} (analogous to the grey dashed line in Fig.~\ref{fig:main_with_integrals}); a model allowing gas decrements at lower masses; and a model with low-mass gas decrements and profile flattening (analogous to the black points in Fig.~\ref{fig:main_with_integrals}). They find that the model with low-mass gas profile flattening is generally the best at matching observations (X-ray, SZ, fast radio bursts, \ion{O}{vi}, and UV). This conclusion appears to be consistent with our results. However, it is not clearly preferred when they use only stacked tSZ and kSZ observations from \citet{Schaan_etal_21}.

Hydrodynamic simulations predict a range of $f_\mathrm{gas}(M)$ curves and those that allow subgrid astrophysics to vary, even more so. Whilst we do not reproduce simulation data in figures in this work, we direct the reader to fig.~9 of \citet{Bigwood_etal_24} and fig.~7 of \citet{Kugel_etal_23} for representative examples. Suffice it to say here that our $f_\mathrm{gas,vir}(M)$ curve (Fig.~\ref{fig:main_with_integrals}) is broadly consistent with many fiducial curves from simulation, but it is in better agreement with simulations that implement stronger feedback, either as their fiducial setting or as stronger-than-fiducial variations. 

\section{Discussion} \label{sec:discussion}

\subsection{Error analysis and null testing} \label{ssec:errors}

The profile covariances for DESI-LS groups used in our MCMC fitting are determined using bootstrapping techniques. To do this, profiles are randomly sampled with replacement from the DESI-LS group catalogue before the stacking procedure detailed in Sec.~\ref{ssec:stacking_procedure} is applied. This random sampling is repeated 2000 times, giving 2000 stacks, and from the corresponding 2000 profiles, the covariance matrix is calculated. These matrices describe covariances in $\delta T$ between radii (e.g. the shaded regions shown in Figs.~\ref{fig:stack} \& \ref{fig:profiles_M_bins} are derived from their diagonal) which allow us to obtain $f_\mathrm{gas}$ and/or $\alpha$ and their associated uncertainties via MCMC. The error bars on filled points in Fig.~\ref{fig:main_with_integrals} show this uncertainty on $f_\mathrm{gas}$, whereas the error bars on unfilled points propagate this uncertainty on $\alpha$ to $f_\mathrm{gas,vir}$. Crucially, these fit $f_\mathrm{gas}$ or $\alpha$ independently, taking fixed or empirical values for the other. Generally, however, the covariances are such that we are able to constrain $f_\mathrm{gas}$ and $\alpha$ independently or simultaneously with reasonable significance. However, we do find a positive correlation between $f_\mathrm{gas}$ and $\alpha$ values in their posteriors. This makes sense, as an increased $\alpha$ will suppress the peak amplitude of the predicted signal (see Fig.~\ref{fig:rainbow}) and so an increased $f_\mathrm{gas}$ will be required to bring this peak back to its previous value, meaning that they are not wholly independent.

Generally, we see that the fitted profiles align very well with the stacked profile from observation -- both overall (Fig.~\ref{fig:stack}) and in certain mass bins (Fig.~\ref{fig:profiles_M_bins}). The SNRs on observed profiles are particularly high for the middling masses shown in Fig.~\ref{fig:profiles_M_bins}. SNRs reduce at both high and low masses due to having fewer and smaller haloes, respectively. We do therefore see degradation in $\Delta \chi^2 = \chi^2_\mathrm{null} - \chi^2_\mathrm{fit}$ of fits at these extremes from its peak of 41 at $10^{14.25} \msunoh$ (the right-most panels of Fig.~\ref{fig:profiles_M_bins}; cf. $\Delta \chi^2 = 58$ and $16$ for momentum- and equal-weighted fits in Fig.~\ref{fig:stack}). However, we see no evidence of mass-dependent behaviours that are not captured by our model; i.e. $\chi^2/N_\mathrm{dof}$ of the fits appears constant as a function of halo mass. $\chi^2/N_\mathrm{dof}$ across the ten mass bins remains at $0.9 \pm 0.5$ (with no clear outliers or trends), close to the expected distribution of $1 \pm \sqrt{2/N_\mathrm{dof}} \approx 1.0 \pm 0.4$, indicating that the observations are well-modelled by the fitted mock kSZ profiles over the full mass range.

To confirm that the filtering and stacking methods used are sufficient to remove all velocity-independent effects from the final result, we perform the following null test. Using the same filtered ACT map, we shuffle the LoS velocities before stacking as in Sec.~\ref{ssec:stacking_procedure} and repeat this 2000 times to generate a distribution of stacked profiles. Using the mean and covariance from this distribution, we find that the absolute LoS momentum-weighted and unweighted profiles are fully consistent with zero signal, with the mean profile of each distribution having $\chi^2/N_\mathrm{dof} = 1.6$ and $0.5$ from null, respectively.

We also compare the covariances produced by bootstrapping and by velocity shuffling and determine that the two methods produce uncertainties on $f_\mathrm{gas}$ fully consistent with one another. With 2000 samplings of each, we find a standard deviation of less than 5~per~cent difference in $f_\mathrm{gas}$ uncertainty with no significant biasing of $f_\mathrm{gas}$ or $\alpha$ values or uncertainties.

\subsection{Cosmic infrared background contamination} \label{ssec:CIB_contamination}

We have investigated potential contaminations from the foreground CIB. Such contaminations can interfere with kSZ measurement in two ways: (1), the possible overlap of the SED between CMB and CIB around 220 GHz (if the CIB spectral energy distribution were approximated by a thermal spectrum, it would be one order of magnitude hotter than the redshifted CMB temperature); (2), possible contributions from the Doppler-shifted CIB. The first effect should cancel in the difference of approaching and receding haloes; but the second is induced by the LoS peculiar motion of CIB emitters, and thus can mimic the kSZ signal. 
We repeat the above CMB stacking analysis, but now using the high-frequency maps at 353, 545, and 857\,GHz provided by {\it Planck} and measure $\delta T$ signals in the difference stacks. Using the empirical fitting $\delta T \propto \nu^\beta$ for frequency $\nu$, we find $\beta=7.3$.
By extrapolating to the lower 220\,GHz frequency range where our kSZ measurements were made, we determine that any interference from the CIB, including the possible Doppler-shifted CIB, is expected to be around 1.5 orders of magnitude below the observed kSZ signal, which is consistent with forecasts \citep[e.g.][]{Maniyar_Ferraro_Schaan_22}. 

\subsection{Miscentring}

One systematic that is not modelled in the kSZ template is possible miscentring between the luminosity-weighted centre of galaxy groups and the centre of their gas distribution. Any miscentring in individual haloes will generate an effective Gaussian broadening of the stacked kSZ profile, and so will increase the fitted $\alpha$.  Astrometric errors on individual galaxies are insignificant, but significant miscentring can occur for galaxy groups due to mergers, galaxy deblending, and, for the DESI-LS catalogue, large photo-$z$ uncertainties affecting group membership. Such effects should average away in massive groups with many members, but may be important in low-mass groups.

We therefore perform a test by modelling miscentring with a Rayleigh distribution of amplitude $\sigma_\mathrm{mis}$ affecting a fraction $f_\mathrm{mis}$ of haloes in the mock halo catalogue \citep[see e.g. equations 3.4--3.7 of][and references therein]{Currie_etal_24}. We then apply the calculated broadening to mock kSZ temperature profiles of known $\alpha$ and $f_\mathrm{gas}$ before fitting using a MCMC to determine the effect on measured $\alpha$ and $f_\mathrm{gas}$. 
Even assuming all groups are affected by miscentring (i.e. {$f_\mathrm{mis}=1$}), to obtain the high values of $\alpha$ we see in low-mass haloes (unfilled points in Fig.~\ref{fig:main_with_integrals}), we require {$\sigma_\mathrm{mis} > 1.2 \, h^{-1}\mathrm{Mpc}$}: over four times the mean halo radius of DESI-LS groups. As the expected miscentring of DESI-LS haloes has not yet been quantified, we compare to $\sigma_\mathrm{mis} = 0.42, 0.32,$~and~$0.27 \,h^{-1}\mathrm{Mpc}$ for cluster catalogues using SDSS, DES, and HSC, respectively \citep[][with $f_\mathrm{mis} = 0.6, 0.2,$~and~$0.1$]{Johnston_etal_07,Shin_etal_19,Ding_etal_25}.
Due to the very high $f_\mathrm{mis}$ and $\sigma_\mathrm{mis}$ required, we find this explanation to be unlikely and conclude that miscentring alone cannot alone explain the high values of $\alpha$ found for low-mass haloes.
We also find that measured $f_\mathrm{gas}$ remains within 10~per~cent of its true value for {$f_\mathrm{mis}=1$} and {$\sigma_\mathrm{mis} < 1.2 \, h^{-1}\mathrm{Mpc}$}, increasing our confidence in the robustness of measured $f_\mathrm{gas}$.

\section{Conclusions} \label{sec:conclusions}

We have made a detection of the kSZ signal by stacking CMB temperature maps from ACT around the locations of groups with mass $>10^{12.5}\msunoh$ found in the DESI Legacy Survey catalogue. Crucially, velocity reconstruction was performed with the photo-$z$ catalogue, in order to separate the receding and approaching groups and avoid cancellation of the kSZ signal. We have made mock observations of the kSZ signal by building realistic mock group catalogues, generating their corresponding kSZ temperature maps, and repeating the same analysis as in real observations. This allows us to use the mock observation as a template for the observed kSZ signal, yielding constraints on the gas density profile and the mass fraction of gas. We find that: \\
\\
$\bullet$ If we give equal weighting to groups independent of their mass, the stacked kSZ signal contributed by our selected DESI-LS group sample is dominated by the abundant low-mass haloes, with an average halo mass of $\sim 10^{13.7}\msunoh$. The amplitude of the observed kSZ signal is lower than would be expected if these groups contained a universal baryon fraction within a few arcminutes from their centres. \\
\\
$\bullet$ When the signal is weighted by the LoS momentum of each group, the stacked kSZ signal is dominated by haloes with large LoS momentum, typically with large halo masses. The resultant average halo mass becomes larger ($\sim 10^{14.5}\msunoh$). In this case, our template provides a good fit to the average kSZ signal if the halo baryon fraction matches the universal value. \\
\\
$\bullet$ Repeating our analyses for haloes of different masses reveals  evidence for gas depletion in low-mass haloes, which allows us to understand why the baryon fraction derived from the mean kSZ signal depends on the mass weighting. \\
\\
$\bullet$ The kSZ measurements do not allow us to determine the total baryon content of haloes, since diffuse baryons lying well beyond the virial radius are hard to detect. But by exploring different assumptions regarding the gas profile, we demonstrate that the baryonic mass within the virial radius can be estimated robustly, and when considering this mass we find that the baryon content in low-mass haloes is depleted.
Furthermore, we find that this picture is consistent with one in which all haloes have a universal total baryon fraction, but where low-mass haloes have far more extended gas profiles, such that the majority of their baryons lie in their outskirts. \\

This empirical suggestion that baryons in low-mass haloes are more extended relative to their dark matter is in consistency with other lines of argument.
Feedback processes have long been seen as a way of expelling gas from haloes, and this energy injection is generally expected to be more effective in low-mass haloes where the potential wells are shallower. 
A range of such effects has been seen in hydrodynamic simulations such as \textsc{eagle}, \textsc{bahamas}, \textsc{IllustrisTNG}, and \textsc{simba} \citep[][respectively]{Schaye2015,McCarthy2017,Springel2018,Dave2019}. Baryon depletion at low halo masses has also been inferred from analysis of X-ray observations \citep[see e.g.][]{Popesso_etal_24}. 
The evidence for a low gas fraction in low-mass haloes from our independent kSZ measurement boosts the level of confidence in the reality of these feedback effects.

However, we have seen that it is harder to make robust measurements of the full radial distribution of baryons as a function of halo mass, indicating that the virial gas fraction alone is insufficient to understand the full effect of feedback. 
With the promised improved quality and quantity of data from the DESI survey in particular, the uncertainties in future kSZ measurements should be substantially reduced. We can therefore hope that future extensions of our current work will be able make a precise discrimination between the range of predictions seen in different
hydrodynamic simulations, leading to an improved understanding of the impact of feedback on the cosmic baryon distribution.

\section*{Acknowledgements}

We thank the anonymous reviewer for detailed and constructive comments which led to significant improvements in this manuscript.
We thank Aritra Gon for valuable discussions and insights. 
YC acknowledges the support of the UK Royal Society through a University Research Fellowship. 
FR acknowledges the support of the UK Science and Technology Facilities Council. 
For the purpose of open access, the author has applied a Creative Commons Attribution (CC BY) license to any Author Accepted Manuscript version arising from this submission. 

This research was partly supported by the Munich Institute for Astro-, Particle and BioPhysics (MIAPbP) which is funded by the Deutsche Forschungsgemeinschaft (DFG, German Research Foundation) under Germany´s Excellence Strategy – EXC-2094 – 390783311.

Many of the results in this paper have been derived using the \textsc{healpy} and \textsc{HEALPix} packages.

The Legacy Surveys consist of three individual and complementary projects: the Dark Energy Camera Legacy Survey (DECaLS; Proposal ID \#2014B-0404; PIs: David Schlegel and Arjun Dey), the Beijing-Arizona Sky Survey (BASS; NOAO Prop. ID \#2015A-0801; PIs: Zhou Xu and Xiaohui Fan), and the Mayall z-band Legacy Survey (MzLS; Prop. ID \#2016A-0453; PI: Arjun Dey). DECaLS, BASS and MzLS together include data obtained, respectively, at the Blanco telescope, Cerro Tololo Inter-American Observatory, NSF’s NOIRLab; the Bok telescope, Steward Observatory, University of Arizona; and the Mayall telescope, Kitt Peak National Observatory, NOIRLab. Pipeline processing and analyses of the data were supported by NOIRLab and the Lawrence Berkeley National Laboratory (LBNL). The Legacy Surveys project is honoured to be permitted to conduct astronomical research on Iolkam Du’ag (Kitt Peak), a mountain with particular significance to the Tohono O’odham Nation.

NOIRLab is operated by the Association of Universities for Research in Astronomy (AURA) under a cooperative agreement with the National Science Foundation. LBNL is managed by the Regents of the University of California under contract to the U.S. Department of Energy.

This project used data obtained with the Dark Energy Camera (DECam), which was constructed by the Dark Energy Survey (DES) collaboration. Funding for the DES Projects has been provided by the U.S. Department of Energy, the U.S. National Science Foundation, the Ministry of Science and Education of Spain, the Science and Technology Facilities Council of the United Kingdom, the Higher Education Funding Council for England, the National Center for Supercomputing Applications at the University of Illinois at Urbana-Champaign, the Kavli Institute of Cosmological Physics at the University of Chicago, Center for Cosmology and Astro-Particle Physics at the Ohio State University, the Mitchell Institute for Fundamental Physics and Astronomy at Texas A\&M University, Financiadora de Estudos e Projetos, Fundacao Carlos Chagas Filho de Amparo, Financiadora de Estudos e Projetos, Fundacao Carlos Chagas Filho de Amparo a Pesquisa do Estado do Rio de Janeiro, Conselho Nacional de Desenvolvimento Cientifico e Tecnologico and the Ministerio da Ciencia, Tecnologia e Inovacao, the Deutsche Forschungsgemeinschaft and the Collaborating Institutions in the Dark Energy Survey. The Collaborating Institutions are Argonne National Laboratory, the University of California at Santa Cruz, the University of Cambridge, Centro de Investigaciones Energeticas, Medioambientales y Tecnologicas-Madrid, the University of Chicago, University College London, the DES-Brazil Consortium, the University of Edinburgh, the Eidgenossische Technische Hochschule (ETH) Zurich, Fermi National Accelerator Laboratory, the University of Illinois at Urbana-Champaign, the Institut de Ciencies de l’Espai (IEEC/CSIC), the Institut de Fisica d’Altes Energies, Lawrence Berkeley National Laboratory, the Ludwig Maximilians Universitat Munchen and the associated Excellence Cluster Universe, the University of Michigan, NSF’s NOIRLab, the University of Nottingham, the Ohio State University, the University of Pennsylvania, the University of Portsmouth, SLAC National Accelerator Laboratory, Stanford University, the University of Sussex, and Texas A\&M University.

BASS is a key project of the Telescope Access Program (TAP), which has been funded by the National Astronomical Observatories of China, the Chinese Academy of Sciences (the Strategic Priority Research Program “The Emergence of Cosmological Structures” Grant \#XDB09000000), and the Special Fund for Astronomy from the Ministry of Finance. The BASS is also supported by the External Cooperation Program of Chinese Academy of Sciences (Grant \#114A11KYSB20160057), and Chinese National Natural Science Foundation (Grant \#12120101003, \#11433005).

The Legacy Survey team makes use of data products from the Near-Earth Object Wide-field Infrared Survey Explorer (NEOWISE), which is a project of the Jet Propulsion Laboratory/California Institute of Technology. NEOWISE is funded by the National Aeronautics and Space Administration.

The Legacy Surveys imaging of the DESI footprint is supported by the Director, Office of Science, Office of High Energy Physics of the U.S. Department of Energy under Contract No. DE-AC02-05CH1123, by the National Energy Research Scientific Computing Center, a DOE Office of Science User Facility under the same contract; and by the U.S. National Science Foundation, Division of Astronomical Sciences under Contract No. AST-0950945 to NOAO.

\section*{Data availability}

The DESI-LS DR9 group catalogue used in this work is available at \url{https://gax.sjtu.edu.cn/data/DESI.html}.
All \textit{Planck} data products used in this work are available at \url{https://pla.esac.esa.int}. 
All ACT products used are available on LAMBDA at \url{https://lambda.gsfc.nasa.gov/product/act/act_dr6.02/act_dr6.02_maps_coadd_get.html}. 
All \textsc{AbacusSummit} data products and associated documentation are available at \url{https://abacusnbody.org}.
All velocity reconstructions, kSZ templates, and stacks are available from the corresponding author upon reasonable request.

\bibliographystyle{mnras}
\bibliography{bib}

@ARTICLE{Gastaldello2007,
       author = {{Gastaldello}, Fabio and {Buote}, David A. and {Humphrey}, Philip J. and {Zappacosta}, Luca and {Bullock}, James S. and {Brighenti}, Fabrizio and {Mathews}, William G.},
        title = "{Probing the Dark Matter and Gas Fraction in Relaxed Galaxy Groups with X-Ray Observations from Chandra and XMM-Newton}",
      journal = {\apj},
         year = 2007,
        month = nov,
       volume = {669},
       number = {1},
        pages = {158-183},
          doi = {10.1086/521519},
archivePrefix = {arXiv},
       eprint = {astro-ph/0610134},
 primaryClass = {astro-ph},
       adsurl = {https://ui.adsabs.harvard.edu/abs/2007ApJ...669..158G}
}

@ARTICLE{Schaye2015,
       author = {{Schaye}, Joop and {Crain}, Robert A. and {Bower}, Richard G. and {Furlong}, Michelle and {Schaller}, Matthieu and {Theuns}, Tom and {Dalla Vecchia}, Claudio and {Frenk}, Carlos S. and {McCarthy}, I.~G. and {Helly}, John C. and {Jenkins}, Adrian and {Rosas-Guevara}, Y.~M. and {White}, Simon D.~M. and {Baes}, Maarten and {Booth}, C.~M. and {Camps}, Peter and {Navarro}, Julio F. and {Qu}, Yan and {Rahmati}, Alireza and {Sawala}, Till and {Thomas}, Peter A. and {Trayford}, James},
        title = "{The EAGLE project: simulating the evolution and assembly of galaxies and their environments}",
      journal = {\mnras},
         year = 2015,
        month = jan,
       volume = {446},
       number = {1},
        pages = {521-554},
          doi = {10.1093/mnras/stu2058},
archivePrefix = {arXiv},
       eprint = {1407.7040},
 primaryClass = {astro-ph.GA},
       adsurl = {https://ui.adsabs.harvard.edu/abs/2015MNRAS.446..521S}
}

@ARTICLE{Dave2019,
       author = {{Dav{\'e}}, Romeel and {Angl{\'e}s-Alc{\'a}zar}, Daniel and {Narayanan}, Desika and {Li}, Qi and {Rafieferantsoa}, Mika H. and {Appleby}, Sarah},
        title = "{SIMBA: Cosmological simulations with black hole growth and feedback}",
      journal = {\mnras},
         year = 2019,
        month = jun,
       volume = {486},
       number = {2},
        pages = {2827-2849},
          doi = {10.1093/mnras/stz937},
archivePrefix = {arXiv},
       eprint = {1901.10203},
 primaryClass = {astro-ph.GA},
       adsurl = {https://ui.adsabs.harvard.edu/abs/2019MNRAS.486.2827D}
}

@ARTICLE{Springel2018,
       author = {{Springel}, Volker and {Pakmor}, R{\"u}diger and {Pillepich}, Annalisa and {Weinberger}, Rainer and {Nelson}, Dylan and {Hernquist}, Lars and {Vogelsberger}, Mark and {Genel}, Shy and {Torrey}, Paul and {Marinacci}, Federico and {Naiman}, Jill},
        title = "{First results from the IllustrisTNG simulations: matter and galaxy clustering}",
      journal = {\mnras},
         year = 2018,
        month = mar,
       volume = {475},
       number = {1},
        pages = {676-698},
          doi = {10.1093/mnras/stx3304},
archivePrefix = {arXiv},
       eprint = {1707.03397},
 primaryClass = {astro-ph.GA},
       adsurl = {https://ui.adsabs.harvard.edu/abs/2018MNRAS.475..676S}
}

@ARTICLE{McCarthy2017,
       author = {{McCarthy}, Ian G. and {Schaye}, Joop and {Bird}, Simeon and {Le Brun}, Amandine M.~C.},
        title = "{The BAHAMAS project: calibrated hydrodynamical simulations for large-scale structure cosmology}",
      journal = {\mnras},
         year = 2017,
        month = mar,
       volume = {465},
       number = {3},
        pages = {2936-2965},
          doi = {10.1093/mnras/stw2792},
archivePrefix = {arXiv},
       eprint = {1603.02702},
 primaryClass = {astro-ph.CO},
       adsurl = {https://ui.adsabs.harvard.edu/abs/2017MNRAS.465.2936M}
}

@ARTICLE{Hadzhiyska2025,
       author = {{Hadzhiyska}, Boryana and {Ferraro}, Simone and {Farren}, Gerrit S. and {Sailer}, Noah and {Zhou}, Rongpu},
        title = "{Missing baryons recovered: A measurement of the gas fraction in galaxies and groups with the kinematic Sunyaev-Zel'dovich effect and CMB lensing}",
      journal = {\prd},
         year = 2025,
        month = dec,
       volume = {112},
       number = {12},
          eid = {123507},
        pages = {123507},
          doi = {10.1103/mdhz-fgj8},
archivePrefix = {arXiv},
       eprint = {2507.14136},
 primaryClass = {astro-ph.CO},
       adsurl = {https://ui.adsabs.harvard.edu/abs/2025PhRvD.112l3507H}
}

@ARTICLE{Dev2024,
       author = {{Dev}, Ajay and {Driver}, Simon P. and {Meyer}, Martin and {Robotham}, Aaron and {Obreschkow}, Danail and {Popesso}, Paola and {Comparat}, Johan},
        title = "{The baryon census and the mass-density of stars, neutral gas, and hot gas as a function of halo mass}",
      journal = {\mnras},
         year = 2024,
        month = dec,
       volume = {535},
       number = {3},
        pages = {2357-2374},
          doi = {10.1093/mnras/stae2485},
archivePrefix = {arXiv},
       eprint = {2411.00456},
 primaryClass = {astro-ph.CO},
       adsurl = {https://ui.adsabs.harvard.edu/abs/2024MNRAS.535.2357D}
}

@ARTICLE{Hotinli2025,
       author = {{Hotinli}, Selim C. and {Smith}, Kendrick M. and {Ferraro}, Simone},
        title = "{Velocity Reconstruction from KSZ: Measuring $f_{NL}$ with ACT and DESILS}",
      journal = {arXiv e-prints},
         year = 2025,
        month = jun,
          eid = {arXiv:2506.21657},
        pages = {arXiv:2506.21657},
archivePrefix = {arXiv},
       eprint = {2506.21657},
 primaryClass = {astro-ph.CO},
       adsurl = {https://ui.adsabs.harvard.edu/abs/2025arXiv250621657H}
}

@ARTICLE{Lai2025,
       author = {{Lai}, Anderson C. and {Kvasiuk}, Yurii and {M\"{u}nchmeyer},  Moritz},
        title = "{KSZ Velocity Reconstruction with ACT and DESI-LS using a Tomographic QML Power Spectrum Estimator}",
      journal = {arXiv e-prints},
         year = 2025,
        month = jun,
          eid = {arXiv:2506.21684},
        pages = {arXiv:2506.21684},
          doi = {10.1088/arXiv:2506.21684},
archivePrefix = {arXiv},
       eprint = {2506.21684},
 primaryClass = {astro-ph.CO},
       adsurl = {https://ui.adsabs.harvard.edu/abs/arXiv:2506.21684}
}

@ARTICLE{McCarthy2025,
       author = {{McCarthy}, Fiona and {Battaglia}, Nicholas and {Bean}, Rachel and {Richard Bond}, J. and {Cai}, Hongbo and {Calabrese}, Erminia and {Coulton}, William R. and {Devlin}, Mark J. and {Dunkley}, Jo and {Ferraro}, Simone and {Gluscevic}, Vera and {Guan}, Yilun and {Colin Hill}, J. and {Johnson}, Matthew C. and {Kusiak}, Aleksandra and {Lagu{\"e}}, Alex and {MacCrann}, Niall and {Madhavacheril}, Mathew S. and {Moodley}, Kavilan and {Naess}, Sigurd and {Qu}, Frank J. and {Ried Guachalla}, Bernardita and {Sehgal}, Neelima and {Sherwin}, Blake D. and {Sif{\'o}n}, Crist{\'o}bal and {Smith}, Kendrick M. and {Staggs}, Suzanne T. and {van Engelen}, Alexander and {Vavagiakis}, Eve M. and {Wollack}, Edward J.},
        title = "{The Atacama Cosmology Telescope: Large-scale velocity reconstruction with the kinematic Sunyaev-Zel'dovich effect and DESI LRGs}",
      journal = {\jcap},
         year = 2025,
        month = may,
       volume = {2025},
       number = {5},
          eid = {057},
        pages = {057},
          doi = {10.1088/1475-7516/2025/05/057},
archivePrefix = {arXiv},
       eprint = {2410.06229},
 primaryClass = {astro-ph.CO},
       adsurl = {https://ui.adsabs.harvard.edu/abs/2025JCAP...05..057M}
}

@ARTICLE{Chen2024,
       author = {{Chen}, Ziyang and {Zhang}, Pengjie},
        title = "{On the capability of high redshift kSZ measurement with galaxy surveys}",
      journal = {\jcap},
         year = 2024,
        month = aug,
       volume = {2024},
       number = {8},
          eid = {053},
        pages = {053},
          doi = {10.1088/1475-7516/2024/08/053},
archivePrefix = {arXiv},
       eprint = {2406.04735},
 primaryClass = {astro-ph.CO},
       adsurl = {https://ui.adsabs.harvard.edu/abs/2024JCAP...08..053C}
}

@ARTICLE{Zhou2021,
       author = {{Zhou}, Rongpu and {Newman}, Jeffrey A. and {Mao}, Yao-Yuan and {Meisner}, Aaron and {Moustakas}, John and {Myers}, Adam D. and {Prakash}, Abhishek and {Zentner}, Andrew R. and {Brooks}, David and {Duan}, Yutong and {Landriau}, Martin and {Levi}, Michael E. and {Prada}, Francisco and {Tarle}, Gregory},
        title = "{The clustering of DESI-like luminous red galaxies using photometric redshifts}",
      journal = {\mnras},
         year = 2021,
        month = mar,
       volume = {501},
       number = {3},
        pages = {3309-3331},
          doi = {10.1093/mnras/staa3764},
archivePrefix = {arXiv},
       eprint = {2001.06018},
 primaryClass = {astro-ph.CO},
       adsurl = {https://ui.adsabs.harvard.edu/abs/2021MNRAS.501.3309Z}
}

@ARTICLE{Hang2021,
       author = {{Hang}, Qianjun and {Alam}, Shadab and {Peacock}, John A. and {Cai}, Yan-Chuan},
        title = "{Galaxy clustering in the DESI Legacy Survey and its imprint on the CMB}",
      journal = {\mnras},
         year = 2021,
        month = feb,
       volume = {501},
       number = {1},
        pages = {1481-1498},
          doi = {10.1093/mnras/staa3738},
archivePrefix = {arXiv},
       eprint = {2010.00466},
 primaryClass = {astro-ph.CO},
       adsurl = {https://ui.adsabs.harvard.edu/abs/2021MNRAS.501.1481H}
}

@ARTICLE{DESI_DR1,
       author = {{DESI Collaboration} and {Abdul Karim}, M. and {Adame}, A.~G. and {Aguado}, D. and {Aguilar}, J. and {Ahlen}, S. and {Alam}, S. and {Aldering}, G. and {Alexander}, D.~M. and {Alfarsy}, R. and {Allen}, L. and {Allende Prieto}, C. and {Alves}, O. and {Anand}, A. and {Andrade}, U. and {Armengaud}, E. and {Avila}, S. and {Aviles}, A. and {Awan}, H. and {Bailey}, S. and {Baleato Lizancos}, A. and {Ballester}, O. and {Bault}, A. and {Bautista}, J. and {Bean}, R. and {Behera}, J. and {BenZvi}, S. and {Beraldo e Silva}, L. and {Bermejo-Climent}, J.~R. and {Beutler}, F. and {Bianchi}, D. and {Blake}, C. and {Blum}, R. and {Bolton}, A.~S. and {Bonici}, M. and {Brieden}, S. and {Brodzeller}, A. and {Brooks}, D. and {Buckley-Geer}, E. and {Burtin}, E. and {Bystr{\"o}m}, A. and {Canning}, R. and {Carnero Rosell}, A. and {Carr}, A. and {Carrilho}, P. and {Casas}, L. and {Castander}, F.~J. and {Cereskaite}, R. and {Cervantes-Cota}, J.~L. and {Chaussidon}, E. and {Chaves-Montero}, J. and {Chen}, S. and {Chen}, X. and {Circosta}, C. and {Claybaugh}, T. and {Cole}, S. and {Cooper}, A.~P. and {Cousinou}, M.-C. and {Cuceu}, A. and {Davis}, T.~M. and {Dawson}, K.~S. and {de Belsunce}, R. and {de la Cruz}, R. and {de la Macorra}, A. and {de Mattia}, A. and {Deiosso}, N. and {Della Costa}, J. and {Demina}, R. and {Demirbozan}, U. and {DeRose}, J. and {Dey}, A. and {Dey}, B. and {Ding}, J. and {Ding}, Z. and {Doel}, P. and {Douglass}, K. and {Dowicz}, M. and {Ebina}, H. and {Edelstein}, J. and {Eisenstein}, D.~J. and {Elbers}, W. and {Emas}, N. and {Escoffier}, S. and {Fagrelius}, P. and {Fan}, X. and {Fanning}, K. and {Favole}, G. and {Fawcett}, V.~A. and {Fern{\'a}ndez-Garc{\'\i}a}, E. and {Ferraro}, S. and {Findlay}, N. and {Font-Ribera}, A. and {Forero-Romero}, J.~E. and {Forero-S{\'a}nchez}, D. and {Frenk}, C.~S. and {G{\"a}nsicke}, B.~T. and {Galbany}, L. and {Garc{\'\i}a-Bellido}, J. and {Garcia-Quintero}, C. and {Garrison}, L.~H. and {Gazta{\~n}aga}, E. and {Gil-Mar{\'\i}n}, H. and {Gloudemans}, A. and {Gnedin}, O.~Y. and {Gontcho A Gontcho}, S. and {Gonzalez}, D. and {Gonzalez-Morales}, A.~X. and {Gonzalez-Perez}, V. and {Gordon}, C. and {Graur}, O. and {Green}, D. and {Gruen}, D. and {Gsponer}, R. and {Guandalin}, C. and {Gutierrez}, G. and {Guy}, J. and {Hahn}, C. and {Han}, J.~J. and {Han}, J. and {He}, S. and {Herrera-Alcantar}, H.~K. and {Heydenreich}, S. and {Honscheid}, K. and {Hou}, J. and {Howlett}, C. and {Huterer}, D. and {Ir{\v{s}}i{\v{c}}}, V. and {Ishak}, M. and {Jacques}, A. and {Jiang}, L. and {Jimenez}, J. and {Jing}, Y.~P. and {Joachimi}, B. and {Joudaki}, S. and {Joyce}, R. and {Jullo}, E. and {Juneau}, S. and {Kara{\c{c}}ayl{\i}}, N.~G. and {Karim}, T. and {Kehoe}, R. and {Kent}, S. and {Khederlarian}, A. and {Kirkby}, D. and {Kisner}, T. and {Kitaura}, F.-S. and {Kizhuprakkat}, N. and {Kong}, H. and {Koposov}, S.~E. and {Kremin}, A. and {Krolewski}, A. and {Lahav}, O. and {Lai}, Y. and {Lamman}, C. and {Lan}, T.-W. and {Landriau}, M. and {Lang}, D. and {Lange}, J.~U. and {Lasker}, J. and {Le Goff}, J.~M. and {Le Guillou}, L. and {Leauthaud}, A. and {Levi}, M.~E. and {Li}, S. and {Li}, T.~S. and {Liu}, W. and {Lodha}, K. and {Lokken}, M. and {Luo}, Y. and {Magneville}, C. and {Manera}, M. and {Manser}, C.~J. and {Margala}, D. and {Martini}, P. and {Maus}, M. and {McCullough}, J. and {McDonald}, P. and {Medina}, G.~E. and {Medina-Varela}, L. and {Meisner}, A. and {Mena-Fern{\'a}ndez}, J. and {Menegas}, A. and {Meneses-Rizo}, J. and {Mezcua}, M. and {Miquel}, R. and {Montero-Camacho}, P. and {Moon}, J. and {Moustakas}, J. and {Mu{\~n}oz-Guti{\'e}rrez}, A. and {Mu noz-Santos}, D. and {Myers}, A.~D. and {Myles}, J. and {Nadathur}, S. and {Najita}, J. and {Napolitano}, L. and {Newman}, J.~A. and {Nikakhtar}, F. and {Nikutta}, R. and {Niz}, G. and {Noriega}, H.~E. and {Nugent}, P.},
        title = "{Data Release 1 of the Dark Energy Spectroscopic Instrument}",
      journal = {\aj},
         year = 2026,
        month = may,
       volume = {171},
       number = {5},
          eid = {285},
        pages = {285},
          doi = {10.3847/1538-3881/ae4c43},
archivePrefix = {arXiv},
       eprint = {2503.14745},
 primaryClass = {astro-ph.CO},
       adsurl = {https://ui.adsabs.harvard.edu/abs/2026AJ....171..285D}
}

@ARTICLE{DES2022,
       author = {{Amon}, A. and {Gruen}, D. and {Troxel}, M.~A. and {MacCrann}, N. and {Dodelson}, S. and {Choi}, A. and {Doux}, C. and {Secco}, L.~F. and {Samuroff}, S. and {Krause}, E. and {Cordero}, J. and {Myles}, J. and {DeRose}, J. and {Wechsler}, R.~H. and {Gatti}, M. and {Navarro-Alsina}, A. and {Bernstein}, G.~M. and {Jain}, B. and {Blazek}, J. and {Alarcon}, A. and {Fert{\'e}}, A. and {Lemos}, P. and {Raveri}, M. and {Campos}, A. and {Prat}, J. and {S{\'a}nchez}, C. and {Jarvis}, M. and {Alves}, O. and {Andrade-Oliveira}, F. and {Baxter}, E. and {Bechtol}, K. and {Becker}, M.~R. and {Bridle}, S.~L. and {Camacho}, H. and {Carnero Rosell}, A. and {Carrasco Kind}, M. and {Cawthon}, R. and {Chang}, C. and {Chen}, R. and {Chintalapati}, P. and {Crocce}, M. and {Davis}, C. and {Diehl}, H.~T. and {Drlica-Wagner}, A. and {Eckert}, K. and {Eifler}, T.~F. and {Elvin-Poole}, J. and {Everett}, S. and {Fang}, X. and {Fosalba}, P. and {Friedrich}, O. and {Gaztanaga}, E. and {Giannini}, G. and {Gruendl}, R.~A. and {Harrison}, I. and {Hartley}, W.~G. and {Herner}, K. and {Huang}, H. and {Huff}, E.~M. and {Huterer}, D. and {Kuropatkin}, N. and {Leget}, P. and {Liddle}, A.~R. and {McCullough}, J. and {Muir}, J. and {Pandey}, S. and {Park}, Y. and {Porredon}, A. and {Refregier}, A. and {Rollins}, R.~P. and {Roodman}, A. and {Rosenfeld}, R. and {Ross}, A.~J. and {Rykoff}, E.~S. and {Sanchez}, J. and {Sevilla-Noarbe}, I. and {Sheldon}, E. and {Shin}, T. and {Troja}, A. and {Tutusaus}, I. and {Tutusaus}, I. and {Varga}, T.~N. and {Weaverdyck}, N. and {Yanny}, B. and {Yin}, B. and {Zhang}, Y. and {Zuntz}, J. and {Aguena}, M. and {Allam}, S. and {Annis}, J. and {Bacon}, D. and {Bertin}, E. and {Bhargava}, S. and {Brooks}, D. and {Buckley-Geer}, E. and {Burke}, D.~L. and {Carretero}, J. and {Costanzi}, M. and {da Costa}, L.~N. and {Pereira}, M.~E.~S. and {De Vicente}, J. and {Desai}, S. and {Dietrich}, J.~P. and {Doel}, P. and {Ferrero}, I. and {Flaugher}, B. and {Frieman}, J. and {Garc{\'\i}a-Bellido}, J. and {Gaztanaga}, E. and {Gerdes}, D.~W. and {Giannantonio}, T. and {Gschwend}, J. and {Gutierrez}, G. and {Hinton}, S.~R. and {Hollowood}, D.~L. and {Honscheid}, K. and {Hoyle}, B. and {James}, D.~J. and {Kron}, R. and {Kuehn}, K. and {Lahav}, O. and {Lima}, M. and {Lin}, H. and {Maia}, M.~A.~G. and {Marshall}, J.~L. and {Martini}, P. and {Melchior}, P. and {Menanteau}, F. and {Miquel}, R. and {Mohr}, J.~J. and {Morgan}, R. and {Ogando}, R.~L.~C. and {Palmese}, A. and {Paz-Chinch{\'o}n}, F. and {Petravick}, D. and {Pieres}, A. and {Romer}, A.~K. and {Sanchez}, E. and {Scarpine}, V. and {Schubnell}, M. and {Serrano}, S. and {Smith}, M. and {Soares-Santos}, M. and {Tarle}, G. and {Thomas}, D. and {To}, C. and {Weller}, J. and {DES Collaboration}},
        title = "{Dark Energy Survey Year 3 results: Cosmology from cosmic shear and robustness to data calibration}",
      journal = {\prd},
         year = 2022,
        month = jan,
       volume = {105},
       number = {2},
          eid = {023514},
        pages = {023514},
          doi = {10.1103/PhysRevD.105.023514},
archivePrefix = {arXiv},
       eprint = {2105.13543},
 primaryClass = {astro-ph.CO},
       adsurl = {https://ui.adsabs.harvard.edu/abs/2022PhRvD.105b3514A}
}

@ARTICLE{KiDS2021,
       author = {{Asgari}, Marika and {Lin}, Chieh-An and {Joachimi}, Benjamin and {Giblin}, Benjamin and {Heymans}, Catherine and {Hildebrandt}, Hendrik and {Kannawadi}, Arun and {St{\"o}lzner}, Benjamin and {Tr{\"o}ster}, Tilman and {van den Busch}, Jan Luca and {Wright}, Angus H. and {Bilicki}, Maciej and {Blake}, Chris and {de Jong}, Jelte and {Dvornik}, Andrej and {Erben}, Thomas and {Getman}, Fedor and {Hoekstra}, Henk and {K{\"o}hlinger}, Fabian and {Kuijken}, Konrad and {Miller}, Lance and {Radovich}, Mario and {Schneider}, Peter and {Shan}, HuanYuan and {Valentijn}, Edwin},
        title = "{KiDS-1000 cosmology: Cosmic shear constraints and comparison between two point statistics}",
      journal = {\aap},
         year = 2021,
        month = jan,
       volume = {645},
          eid = {A104},
        pages = {A104},
          doi = {10.1051/0004-6361/202039070},
archivePrefix = {arXiv},
       eprint = {2007.15633},
 primaryClass = {astro-ph.CO},
       adsurl = {https://ui.adsabs.harvard.edu/abs/2021A&A...645A.104A}
}

@ARTICLE{KiDS2025,
       author = {{Wright}, Angus H. and {St{\"o}lzner}, Benjamin and {Asgari}, Marika and {Bilicki}, Maciej and {Giblin}, Benjamin and {Heymans}, Catherine and {Hildebrandt}, Hendrik and {Hoekstra}, Henk and {Joachimi}, Benjamin and {Kuijken}, Konrad and {Li}, Shun-Sheng and {Reischke}, Robert and {von Wietersheim-Kramsta}, Maximilian and {Yoon}, Mijin and {Burger}, Pierre and {Chisari}, Nora Elisa and {de Jong}, Jelte and {Dvornik}, Andrej and {Georgiou}, Christos and {Harnois-D{\'e}raps}, Joachim and {Jalan}, Priyanka and {William}, Anjitha John and {Joudaki}, Shahab and {Lesci}, Giorgio Francesco and {Linke}, Laila and {Loureiro}, Arthur and {Mahony}, Constance and {Maturi}, Matteo and {Miller}, Lance and {Moscardini}, Lauro and {Napolitano}, Nicola R. and {Porth}, Lucas and {Radovich}, Mario and {Schneider}, Peter and {Tr{\"o}ster}, Tilman and {Valentijn}, Edwin and {Wittje}, Anna and {Yan}, Ziang and {Zhang}, Yun-Hao},
        title = "{KiDS-Legacy: Cosmological constraints from cosmic shear with the complete Kilo-Degree Survey}",
      journal = {\aap},
         year = 2025,
        month = nov,
       volume = {703},
          eid = {A158},
        pages = {A158},
          doi = {10.1051/0004-6361/202554908},
archivePrefix = {arXiv},
       eprint = {2503.19441},
 primaryClass = {astro-ph.CO},
       adsurl = {https://ui.adsabs.harvard.edu/abs/2025A&A...703A.158W}
}

@ARTICLE{Croton2006,
       author = {{Croton}, Darren J. and {Springel}, Volker and {White}, Simon D.~M. and {De Lucia}, G. and {Frenk}, C.~S. and {Gao}, L. and {Jenkins}, A. and {Kauffmann}, G. and {Navarro}, J.~F. and {Yoshida}, N.},
        title = "{The many lives of active galactic nuclei: cooling flows, black holes and the luminosities and colours of galaxies}",
      journal = {\mnras},
         year = 2006,
        month = jan,
       volume = {365},
       number = {1},
        pages = {11-28},
          doi = {10.1111/j.1365-2966.2005.09675.x},
archivePrefix = {arXiv},
       eprint = {astro-ph/0508046},
 primaryClass = {astro-ph},
       adsurl = {https://ui.adsabs.harvard.edu/abs/2006MNRAS.365...11C}
}

@ARTICLE{WhiteRees1978,
       author = {{White}, S.~D.~M. and {Rees}, M.~J.},
        title = "{Core condensation in heavy halos: a two-stage theory for galaxy formation and clustering.}",
      journal = {\mnras},
         year = 1978,
        month = may,
       volume = {183},
        pages = {341-358},
          doi = {10.1093/mnras/183.3.341},
       adsurl = {https://ui.adsabs.harvard.edu/abs/1978MNRAS.183..341W}
}

@ARTICLE{vandaalen2011,
       author = {{van Daalen}, Marcel P. and {Schaye}, Joop and {Booth}, C.~M. and {Dalla Vecchia}, Claudio},
        title = "{The effects of galaxy formation on the matter power spectrum: a challenge for precision cosmology}",
      journal = {\mnras},
         year = 2011,
        month = aug,
       volume = {415},
       number = {4},
        pages = {3649-3665},
          doi = {10.1111/j.1365-2966.2011.18981.x},
archivePrefix = {arXiv},
       eprint = {1104.1174},
 primaryClass = {astro-ph.CO},
       adsurl = {https://ui.adsabs.harvard.edu/abs/2011MNRAS.415.3649V}
}

@ARTICLE{PlanckCollaboration_18_VI,
       author = {{Planck Collaboration} and {Aghanim}, N. and {Akrami}, Y. and {Ashdown}, M. and {Aumont}, J. and {Baccigalupi}, C. and {Ballardini}, M. and {Banday}, A.~J. and {Barreiro}, R.~B. and {Bartolo}, N. and {Basak}, S. and {Battye}, R. and {Benabed}, K. and {Bernard}, J. -P. and {Bersanelli}, M. and {Bielewicz}, P. and {Bock}, J.~J. and {Bond}, J.~R. and {Borrill}, J. and {Bouchet}, F.~R. and {Boulanger}, F. and {Bucher}, M. and {Burigana}, C. and {Butler}, R.~C. and {Calabrese}, E. and {Cardoso}, J. -F. and {Carron}, J. and {Challinor}, A. and {Chiang}, H.~C. and {Chluba}, J. and {Colombo}, L.~P.~L. and {Combet}, C. and {Contreras}, D. and {Crill}, B.~P. and {Cuttaia}, F. and {de Bernardis}, P. and {de Zotti}, G. and {Delabrouille}, J. and {Delouis}, J. -M. and {Di Valentino}, E. and {Diego}, J.~M. and {Dor{\'e}}, O. and {Douspis}, M. and {Ducout}, A. and {Dupac}, X. and {Dusini}, S. and {Efstathiou}, G. and {Elsner}, F. and {En{\ss}lin}, T.~A. and {Eriksen}, H.~K. and {Fantaye}, Y. and {Farhang}, M. and {Fergusson}, J. and {Fernandez-Cobos}, R. and {Finelli}, F. and {Forastieri}, F. and {Frailis}, M. and {Fraisse}, A.~A. and {Franceschi}, E. and {Frolov}, A. and {Galeotta}, S. and {Galli}, S. and {Ganga}, K. and {G{\'e}nova-Santos}, R.~T. and {Gerbino}, M. and {Ghosh}, T. and {Gonz{\'a}lez-Nuevo}, J. and {G{\'o}rski}, K.~M. and {Gratton}, S. and {Gruppuso}, A. and {Gudmundsson}, J.~E. and {Hamann}, J. and {Handley}, W. and {Hansen}, F.~K. and {Herranz}, D. and {Hildebrandt}, S.~R. and {Hivon}, E. and {Huang}, Z. and {Jaffe}, A.~H. and {Jones}, W.~C. and {Karakci}, A. and {Keih{\"a}nen}, E. and {Keskitalo}, R. and {Kiiveri}, K. and {Kim}, J. and {Kisner}, T.~S. and {Knox}, L. and {Krachmalnicoff}, N. and {Kunz}, M. and {Kurki-Suonio}, H. and {Lagache}, G. and {Lamarre}, J. -M. and {Lasenby}, A. and {Lattanzi}, M. and {Lawrence}, C.~R. and {Le Jeune}, M. and {Lemos}, P. and {Lesgourgues}, J. and {Levrier}, F. and {Lewis}, A. and {Liguori}, M. and {Lilje}, P.~B. and {Lilley}, M. and {Lindholm}, V. and {L{\'o}pez-Caniego}, M. and {Lubin}, P.~M. and {Ma}, Y. -Z. and {Mac{\'\i}as-P{\'e}rez}, J.~F. and {Maggio}, G. and {Maino}, D. and {Mandolesi}, N. and {Mangilli}, A. and {Marcos-Caballero}, A. and {Maris}, M. and {Martin}, P.~G. and {Martinelli}, M. and {Mart{\'\i}nez-Gonz{\'a}lez}, E. and {Matarrese}, S. and {Mauri}, N. and {McEwen}, J.~D. and {Meinhold}, P.~R. and {Melchiorri}, A. and {Mennella}, A. and {Migliaccio}, M. and {Millea}, M. and {Mitra}, S. and {Miville-Desch{\^e}nes}, M. -A. and {Molinari}, D. and {Montier}, L. and {Morgante}, G. and {Moss}, A. and {Natoli}, P. and {N{\o}rgaard-Nielsen}, H.~U. and {Pagano}, L. and {Paoletti}, D. and {Partridge}, B. and {Patanchon}, G. and {Peiris}, H.~V. and {Perrotta}, F. and {Pettorino}, V. and {Piacentini}, F. and {Polastri}, L. and {Polenta}, G. and {Puget}, J. -L. and {Rachen}, J.~P. and {Reinecke}, M. and {Remazeilles}, M. and {Renzi}, A. and {Rocha}, G. and {Rosset}, C. and {Roudier}, G. and {Rubi{\~n}o-Mart{\'\i}n}, J.~A. and {Ruiz-Granados}, B. and {Salvati}, L. and {Sandri}, M. and {Savelainen}, M. and {Scott}, D. and {Shellard}, E.~P.~S. and {Sirignano}, C. and {Sirri}, G. and {Spencer}, L.~D. and {Sunyaev}, R. and {Suur-Uski}, A. -S. and {Tauber}, J.~A. and {Tavagnacco}, D. and {Tenti}, M. and {Toffolatti}, L. and {Tomasi}, M. and {Trombetti}, T. and {Valenziano}, L. and {Valiviita}, J. and {Van Tent}, B. and {Vibert}, L. and {Vielva}, P. and {Villa}, F. and {Vittorio}, N. and {Wandelt}, B.~D. and {Wehus}, I.~K. and {White}, M. and {White}, S.~D.~M. and {Zacchei}, A. and {Zonca}, A.},
        title = "{Planck 2018 results. VI. Cosmological parameters}",
      journal = {\aap},
         year = 2020,
        month = sep,
       volume = {641},
          eid = {A6},
        pages = {A6},
          doi = {10.1051/0004-6361/201833910},
archivePrefix = {arXiv},
       eprint = {1807.06209},
 primaryClass = {astro-ph.CO},
       adsurl = {https://ui.adsabs.harvard.edu/abs/2020A&A...641A...6P}
}

@ARTICLE{McGaugh_etal_10,
       author = {{McGaugh}, Stacy S. and {Schombert}, James M. and {de Blok}, W.~J.~G. and {Zagursky}, Matthew J.},
        title = "{The Baryon Content of Cosmic Structures}",
      journal = {\apjl},
         year = 2010,
        month = jan,
       volume = {708},
       number = {1},
        pages = {L14-L17},
          doi = {10.1088/2041-8205/708/1/L14},
archivePrefix = {arXiv},
       eprint = {0911.2700},
 primaryClass = {astro-ph.CO},
       adsurl = {https://ui.adsabs.harvard.edu/abs/2010ApJ...708L..14M}
}

@ARTICLE{Naess_etal_20,
       author = {{Naess}, Sigurd and {Aiola}, Simone and {Austermann}, Jason E. and {Battaglia}, Nick and {Beall}, James A. and {Becker}, Daniel T. and {Bond}, Richard J. and {Calabrese}, Erminia and {Choi}, Steve K. and {Cothard}, Nicholas F. and {Crowley}, Kevin T. and {Darwish}, Omar and {Datta}, Rahul and {Denison}, Edward V. and {Devlin}, Mark and {Duell}, Cody J. and {Duff}, Shannon M. and {Duivenvoorden}, Adriaan J. and {Dunkley}, Jo and {D{\"u}nner}, Rolando and {Fox}, Anna E. and {Gallardo}, Patricio A. and {Halpern}, Mark and {Han}, Dongwon and {Hasselfield}, Matthew and {Hill}, J. Colin and {Hilton}, Gene C. and {Hilton}, Matt and {Hincks}, Adam D. and {Hlo{\v{z}}ek}, Ren{\'e}e and {Ho}, Shuay-Pwu Patty and {Hubmayr}, Johannes and {Huffenberger}, Kevin and {Hughes}, John P. and {Kosowsky}, Arthur B. and {Louis}, Thibaut and {Madhavacheril}, Mathew S. and {McMahon}, Jeff and {Moodley}, Kavilan and {Nati}, Federico and {Nibarger}, John P. and {Niemack}, Michael D. and {Page}, Lyman and {Partridge}, Bruce and {Salatino}, Maria and {Schaan}, Emmanuel and {Schillaci}, Alessandro and {Schmitt}, Benjamin and {Sherwin}, Blake D. and {Sehgal}, Neelima and {Sif{\'o}n}, Crist{\'o}bal and {Spergel}, David and {Staggs}, Suzanne and {Stevens}, Jason and {Storer}, Emilie and {Ullom}, Joel N. and {Vale}, Leila R. and {Van Engelen}, Alexander and {Van Lanen}, Jeff and {Vavagiakis}, Eve M. and {Wollack}, Edward J. and {Xu}, Zhilei},
        title = "{The Atacama Cosmology Telescope: arcminute-resolution maps of 18 000 square degrees of the microwave sky from ACT 2008-2018 data combined with Planck}",
      journal = {\jcap},
         year = 2020,
        month = dec,
       volume = {2020},
       number = {12},
          eid = {046},
        pages = {046},
          doi = {10.1088/1475-7516/2020/12/046},
archivePrefix = {arXiv},
       eprint = {2007.07290},
 primaryClass = {astro-ph.IM},
       adsurl = {https://ui.adsabs.harvard.edu/abs/2020JCAP...12..046N}
}

@article{healpy,
  doi = {10.21105/joss.01298},
  url = {https://doi.org/10.21105/joss.01298},
  year = {2019},
  month = mar,
  publisher = {The Open Journal},
  volume = {4},
  number = {35},
  pages = {1298},
  author = {Andrea Zonca and Leo Singer and Daniel Lenz and Martin Reinecke and Cyrille Rosset and Eric Hivon and Krzysztof Gorski},
  title = {healpy: equal area pixelization and spherical harmonics transforms for data on the sphere in Python},
  journal = {Journal of Open Source Software}
}

@ARTICLE{Bigwood_etal_24,
       author = {{Bigwood}, L. and {Amon}, A. and {Schneider}, A. and {Salcido}, J. and {McCarthy}, I.~G. and {Preston}, C. and {Sanchez}, D. and {Sijacki}, D. and {Schaan}, E. and {Ferraro}, S. and {Battaglia}, N. and {Chen}, A. and {Dodelson}, S. and {Roodman}, A. and {Pieres}, A. and {Fert{\'e}}, A. and {Alarcon}, A. and {Drlica-Wagner}, A. and {Choi}, A. and {Navarro-Alsina}, A. and {Campos}, A. and {Ross}, A.~J. and {Carnero Rosell}, A. and {Yin}, B. and {Yanny}, B. and {S{\'a}nchez}, C. and {Chang}, C. and {Davis}, C. and {Doux}, C. and {Gruen}, D. and {Rykoff}, E.~S. and {Huff}, E.~M. and {Sheldon}, E. and {Tarsitano}, F. and {Andrade-Oliveira}, F. and {Bernstein}, G.~M. and {Giannini}, G. and {Diehl}, H.~T. and {Huang}, H. and {Harrison}, I. and {Sevilla-Noarbe}, I. and {Tutusaus}, I. and {Elvin-Poole}, J. and {McCullough}, J. and {Zuntz}, J. and {Blazek}, J. and {DeRose}, J. and {Cordero}, J. and {Prat}, J. and {Myles}, J. and {Eckert}, K. and {Bechtol}, K. and {Herner}, K. and {Secco}, L.~F. and {Gatti}, M. and {Raveri}, M. and {Kind}, M. Carrasco and {Becker}, M.~R. and {Troxel}, M.~A. and {Jarvis}, M. and {MacCrann}, N. and {Friedrich}, O. and {Alves}, O. and {Leget}, P. -F. and {Chen}, R. and {Rollins}, R.~P. and {Wechsler}, R.~H. and {Gruendl}, R.~A. and {Cawthon}, R. and {Allam}, S. and {Bridle}, S.~L. and {Pandey}, S. and {Everett}, S. and {Shin}, T. and {Hartley}, W.~G. and {Fang}, X. and {Zhang}, Y. and {Aguena}, M. and {Annis}, J. and {Bacon}, D. and {Bertin}, E. and {Bocquet}, S. and {Brooks}, D. and {Carretero}, J. and {Castander}, F.~J. and {da Costa}, L.~N. and {Pereira}, M.~E.~S. and {De Vicente}, J. and {Desai}, S. and {Doel}, P. and {Ferrero}, I. and {Flaugher}, B. and {Frieman}, J. and {Garc{\'\i}a-Bellido}, J. and {Gaztanaga}, E. and {Gutierrez}, G. and {Hinton}, S.~R. and {Hollowood}, D.~L. and {Honscheid}, K. and {Huterer}, D. and {James}, D.~J. and {Kuehn}, K. and {Lahav}, O. and {Lee}, S. and {Marshall}, J.~L. and {Mena-Fern{\'a}ndez}, J. and {Miquel}, R. and {Muir}, J. and {Paterno}, M. and {Plazas Malag{\'o}n}, A.~A. and {Porredon}, A. and {Romer}, A.~K. and {Samuroff}, S. and {Sanchez}, E. and {Sanchez Cid}, D. and {Smith}, M. and {Soares-Santos}, M. and {Suchyta}, E. and {Swanson}, M.~E.~C. and {Tarle}, G. and {To}, C. and {Weaverdyck}, N. and {Weller}, J. and {Wiseman}, P. and {Yamamoto}, M.},
        title = "{Weak lensing combined with the kinetic Sunyaev-Zel'dovich effect: a study of baryonic feedback}",
      journal = {\mnras},
         year = 2024,
        month = oct,
       volume = {534},
       number = {1},
        pages = {655-682},
          doi = {10.1093/mnras/stae2100},
archivePrefix = {arXiv},
       eprint = {2404.06098},
 primaryClass = {astro-ph.CO},
       adsurl = {https://ui.adsabs.harvard.edu/abs/2024MNRAS.534..655B}
}

@ARTICLE{Lamarre_etal_10,
       author = {{Lamarre}, J. -M. and {Puget}, J. -L. and {Ade}, P.~A.~R. and {Bouchet}, F. and {Guyot}, G. and {Lange}, A.~E. and {Pajot}, F. and {Arondel}, A. and {Benabed}, K. and {Beney}, J. -L. and {Beno{\^\i}t}, A. and {Bernard}, J. -Ph. and {Bhatia}, R. and {Blanc}, Y. and {Bock}, J.~J. and {Br{\'e}elle}, E. and {Bradshaw}, T.~W. and {Camus}, P. and {Catalano}, A. and {Charra}, J. and {Charra}, M. and {Church}, S.~E. and {Couchot}, F. and {Coulais}, A. and {Crill}, B.~P. and {Crook}, M.~R. and {Dassas}, K. and {de Bernardis}, P. and {Delabrouille}, J. and {de Marcillac}, P. and {Delouis}, J. -M. and {D{\'e}sert}, F. -X. and {Dumesnil}, C. and {Dupac}, X. and {Efstathiou}, G. and {Eng}, P. and {Evesque}, C. and {Fourmond}, J. -J. and {Ganga}, K. and {Giard}, M. and {Gispert}, R. and {Guglielmi}, L. and {Haissinski}, J. and {Henrot-Versill{\'e}}, S. and {Hivon}, E. and {Holmes}, W.~A. and {Jones}, W.~C. and {Koch}, T.~C. and {Lagard{\`e}re}, H. and {Lami}, P. and {Land{\'e}}, J. and {Leriche}, B. and {Leroy}, C. and {Longval}, Y. and {Mac{\'\i}as-P{\'e}rez}, J.~F. and {Maciaszek}, T. and {Maffei}, B. and {Mansoux}, B. and {Marty}, C. and {Masi}, S. and {Mercier}, C. and {Miville-Desch{\^e}nes}, M. -A. and {Moneti}, A. and {Montier}, L. and {Murphy}, J.~A. and {Narbonne}, J. and {Nexon}, M. and {Paine}, C.~G. and {Pahn}, J. and {Perdereau}, O. and {Piacentini}, F. and {Piat}, M. and {Plaszczynski}, S. and {Pointecouteau}, E. and {Pons}, R. and {Ponthieu}, N. and {Prunet}, S. and {Rambaud}, D. and {Recouvreur}, G. and {Renault}, C. and {Ristorcelli}, I. and {Rosset}, C. and {Santos}, D. and {Savini}, G. and {Serra}, G. and {Stassi}, P. and {Sudiwala}, R.~V. and {Sygnet}, J. -F. and {Tauber}, J.~A. and {Torre}, J. -P. and {Tristram}, M. and {Vibert}, L. and {Woodcraft}, A. and {Yurchenko}, V. and {Yvon}, D.},
        title = "{Planck pre-launch status: The HFI instrument, from specification to actual performance}",
      journal = {\aap},
         year = 2010,
        month = sep,
       volume = {520},
          eid = {A9},
        pages = {A9},
          doi = {10.1051/0004-6361/200912975},
       adsurl = {https://ui.adsabs.harvard.edu/abs/2010A&A...520A...9L}
}

@ARTICLE{HEALPix,
   author = {{G{\'o}rski}, K.~M. and {Hivon}, E. and {Banday}, A.~J. and 
	{Wandelt}, B.~D. and {Hansen}, F.~K. and {Reinecke}, M. and 
	{Bartelmann}, M.},
    title = "{HEALPix: A Framework for High-Resolution Discretization and Fast Analysis of Data Distributed on the Sphere}",
  journal = {\apj},
   eprint = {arXiv:astro-ph/0409513},
     year = 2005,
    month = apr,
   volume = 622,
    pages = {759-771},
      doi = {10.1086/427976},
   adsurl = {http://adsabs.harvard.edu/abs/2005ApJ...622..759G}
}

@ARTICLE{Wang_etal_09,
       author = {{Wang}, Huiyuan and {Mo}, H.~J. and {Jing}, Y.~P. and {Guo}, Yicheng and {van den Bosch}, Frank C. and {Yang}, Xiaohu},
        title = "{Reconstructing the cosmic density field with the distribution of dark matter haloes}",
      journal = {\mnras},
         year = 2009,
        month = mar,
       volume = {394},
       number = {1},
        pages = {398-414},
          doi = {10.1111/j.1365-2966.2008.14301.x},
archivePrefix = {arXiv},
       eprint = {0803.1213},
 primaryClass = {astro-ph},
       adsurl = {https://ui.adsabs.harvard.edu/abs/2009MNRAS.394..398W}
}

@ARTICLE{Yang_etal_07,
       author = {{Yang}, Xiaohu and {Mo}, H.~J. and {van den Bosch}, Frank C. and {Pasquali}, Anna and {Li}, Cheng and {Barden}, Marco},
        title = "{Galaxy Groups in the SDSS DR4. I. The Catalog and Basic Properties}",
      journal = {\apj},
         year = 2007,
        month = dec,
       volume = {671},
       number = {1},
        pages = {153-170},
          doi = {10.1086/522027},
archivePrefix = {arXiv},
       eprint = {0707.4640},
 primaryClass = {astro-ph},
       adsurl = {https://ui.adsabs.harvard.edu/abs/2007ApJ...671..153Y}
}

@ARTICLE{Cui_etal_08,
       author = {{Cui}, Weiguang and {Liu}, Lei and {Yang}, Xiaohu and {Wang}, Yu and {Feng}, Longlong and {Springel}, Volker},
        title = "{An Ideal Mass Assignment Scheme for Measuring the Power Spectrum with Fast Fourier Transforms}",
      journal = {\apj},
         year = 2008,
        month = nov,
       volume = {687},
       number = {2},
        pages = {738-744},
          doi = {10.1086/592079},
archivePrefix = {arXiv},
       eprint = {0804.0070},
 primaryClass = {astro-ph},
       adsurl = {https://ui.adsabs.harvard.edu/abs/2008ApJ...687..738C}
}

@ARTICLE{Dey_etal_19,
       author = {{Dey}, Arjun and {Schlegel}, David J. and {Lang}, Dustin and {Blum}, Robert and {Burleigh}, Kaylan and {Fan}, Xiaohui and {Findlay}, Joseph R. and {Finkbeiner}, Doug and {Herrera}, David and {Juneau}, St{\'e}phanie and {Landriau}, Martin and {Levi}, Michael and {McGreer}, Ian and {Meisner}, Aaron and {Myers}, Adam D. and {Moustakas}, John and {Nugent}, Peter and {Patej}, Anna and {Schlafly}, Edward F. and {Walker}, Alistair R. and {Valdes}, Francisco and {Weaver}, Benjamin A. and {Y{\`e}che}, Christophe and {Zou}, Hu and {Zhou}, Xu and {Abareshi}, Behzad and {Abbott}, T.~M.~C. and {Abolfathi}, Bela and {Aguilera}, C. and {Alam}, Shadab and {Allen}, Lori and {Alvarez}, A. and {Annis}, James and {Ansarinejad}, Behzad and {Aubert}, Marie and {Beechert}, Jacqueline and {Bell}, Eric F. and {BenZvi}, Segev Y. and {Beutler}, Florian and {Bielby}, Richard M. and {Bolton}, Adam S. and {Brice{\~n}o}, C{\'e}sar and {Buckley-Geer}, Elizabeth J. and {Butler}, Karen and {Calamida}, Annalisa and {Carlberg}, Raymond G. and {Carter}, Paul and {Casas}, Ricard and {Castander}, Francisco J. and {Choi}, Yumi and {Comparat}, Johan and {Cukanovaite}, Elena and {Delubac}, Timoth{\'e}e and {DeVries}, Kaitlin and {Dey}, Sharmila and {Dhungana}, Govinda and {Dickinson}, Mark and {Ding}, Zhejie and {Donaldson}, John B. and {Duan}, Yutong and {Duckworth}, Christopher J. and {Eftekharzadeh}, Sarah and {Eisenstein}, Daniel J. and {Etourneau}, Thomas and {Fagrelius}, Parker A. and {Farihi}, Jay and {Fitzpatrick}, Mike and {Font-Ribera}, Andreu and {Fulmer}, Leah and {G{\"a}nsicke}, Boris T. and {Gaztanaga}, Enrique and {George}, Koshy and {Gerdes}, David W. and {Gontcho}, Satya Gontcho A. and {Gorgoni}, Claudio and {Green}, Gregory and {Guy}, Julien and {Harmer}, Diane and {Hernandez}, M. and {Honscheid}, Klaus and {Huang}, Lijuan Wendy and {James}, David J. and {Jannuzi}, Buell T. and {Jiang}, Linhua and {Joyce}, Richard and {Karcher}, Armin and {Karkar}, Sonia and {Kehoe}, Robert and {Kneib}, Jean-Paul and {Kueter-Young}, Andrea and {Lan}, Ting-Wen and {Lauer}, Tod R. and {Le Guillou}, Laurent and {Le Van Suu}, Auguste and {Lee}, Jae Hyeon and {Lesser}, Michael and {Perreault Levasseur}, Laurence and {Li}, Ting S. and {Mann}, Justin L. and {Marshall}, Robert and {Mart{\'\i}nez-V{\'a}zquez}, C.~E. and {Martini}, Paul and {du Mas des Bourboux}, H{\'e}lion and {McManus}, Sean and {Meier}, Tobias Gabriel and {M{\'e}nard}, Brice and {Metcalfe}, Nigel and {Mu{\~n}oz-Guti{\'e}rrez}, Andrea and {Najita}, Joan and {Napier}, Kevin and {Narayan}, Gautham and {Newman}, Jeffrey A. and {Nie}, Jundan and {Nord}, Brian and {Norman}, Dara J. and {Olsen}, Knut A.~G. and {Paat}, Anthony and {Palanque-Delabrouille}, Nathalie and {Peng}, Xiyan and {Poppett}, Claire L. and {Poremba}, Megan R. and {Prakash}, Abhishek and {Rabinowitz}, David and {Raichoor}, Anand and {Rezaie}, Mehdi and {Robertson}, A.~N. and {Roe}, Natalie A. and {Ross}, Ashley J. and {Ross}, Nicholas P. and {Rudnick}, Gregory and {Safonova}, Sasha and {Saha}, Abhijit and {S{\'a}nchez}, F. Javier and {Savary}, Elodie and {Schweiker}, Heidi and {Scott}, Adam and {Seo}, Hee-Jong and {Shan}, Huanyuan and {Silva}, David R. and {Slepian}, Zachary and {Soto}, Christian and {Sprayberry}, David and {Staten}, Ryan and {Stillman}, Coley M. and {Stupak}, Robert J. and {Summers}, David L. and {Sien Tie}, Suk and {Tirado}, H. and {Vargas-Maga{\~n}a}, Mariana and {Vivas}, A. Katherina and {Wechsler}, Risa H. and {Williams}, Doug and {Yang}, Jinyi and {Yang}, Qian and {Yapici}, Tolga and {Zaritsky}, Dennis and {Zenteno}, A. and {Zhang}, Kai and {Zhang}, Tianmeng and {Zhou}, Rongpu and {Zhou}, Zhimin},
        title = "{Overview of the DESI Legacy Imaging Surveys}",
      journal = {\aj},
         year = 2019,
        month = may,
       volume = {157},
       number = {5},
          eid = {168},
        pages = {168},
          doi = {10.3847/1538-3881/ab089d},
archivePrefix = {arXiv},
       eprint = {1804.08657},
 primaryClass = {astro-ph.IM},
       adsurl = {https://ui.adsabs.harvard.edu/abs/2019AJ....157..168D}
}

@ARTICLE{Oh_etal_20,
       author = {{Oh}, Boon Kiat and {Smith}, Britton D. and {Peacock}, John A. and {Khochfar}, Sadegh},
        title = "{Calibration of a star formation and feedback model for cosmological simulations with ENZO}",
      journal = {\mnras},
         year = 2020,
        month = oct,
       volume = {497},
       number = {4},
        pages = {5203-5219},
          doi = {10.1093/mnras/staa2318},
archivePrefix = {arXiv},
       eprint = {2002.02670},
 primaryClass = {astro-ph.GA},
       adsurl = {https://ui.adsabs.harvard.edu/abs/2020MNRAS.497.5203O}
}

@ARTICLE{Hadzhiyska_etal_22,
       author = {{Hadzhiyska}, Boryana and {Garrison}, Lehman H. and {Eisenstein}, Daniel and {Bose}, Sownak},
        title = "{The halo light-cone catalogues of ABACUSSUMMIT}",
      journal = {\mnras},
         year = 2022,
        month = jan,
       volume = {509},
       number = {2},
        pages = {2194-2208},
          doi = {10.1093/mnras/stab3066},
archivePrefix = {arXiv},
       eprint = {2110.11413},
 primaryClass = {astro-ph.CO},
       adsurl = {https://ui.adsabs.harvard.edu/abs/2022MNRAS.509.2194H}
}

@ARTICLE{Chen_etal_22,
       author = {{Chen}, Ziyang and {Zhang}, Pengjie and {Yang}, Xiaohu and {Zheng}, Yi},
        title = "{Detection of pairwise kSZ effect with DESI galaxy clusters and Planck}",
      journal = {\mnras},
         year = 2022,
        month = mar,
       volume = {510},
       number = {4},
        pages = {5916-5928},
          doi = {10.1093/mnras/stab3604},
archivePrefix = {arXiv},
       eprint = {2109.04092},
 primaryClass = {astro-ph.CO},
       adsurl = {https://ui.adsabs.harvard.edu/abs/2022MNRAS.510.5916C}
}

@ARTICLE{Popesso_etal_24,
       author = {{Popesso}, P. and {Biviano}, A. and {Marini}, I. and {Dolag}, K. and {Vladutescu-Zopp}, S. and {Csizi}, B. and {Biffi}, V. and {Lamer}, G. and {Robothan}, A. and {Bravo}, M. and {Lovisari}, L. and {Ettori}, S. and {Angelinelli}, M. and {Driver}, S. and {Toptun}, V. and {Dev}, A. and {Mazengo}, D. and {Merloni}, A. and {Comparat}, J. and {Ponti}, G. and {Mroczkowski}, T. and {Bulbul}, E. and {Grandis}, S. and {Bahar}, E.},
        title = "{The hot gas mass fraction in halos: From Milky Way-like groups to massive clusters}",
      journal = {\aap},
         year = 2026,
        month = mar,
       volume = {707},
          eid = {A362},
        pages = {A362},
          doi = {10.1051/0004-6361/202453256},
archivePrefix = {arXiv},
       eprint = {2411.16555},
 primaryClass = {astro-ph.GA},
       adsurl = {https://ui.adsabs.harvard.edu/abs/2026A&A...707A.362P}
}

@ARTICLE{Li_etal_24,
       author = {{Li}, Shaohong and {Zheng}, Yi and {Chen}, Ziyang and {Xu}, Haojie and {Yang}, Xiaohu},
        title = "{Detection of Pairwise Kinetic Sunyaev{\textendash}Zel'dovich Effect with DESI Galaxy Groups and Planck in Fourier Space}",
      journal = {\apjs},
         year = 2024,
        month = mar,
       volume = {271},
       number = {1},
          eid = {30},
        pages = {30},
          doi = {10.3847/1538-4365/ad1bd8},
archivePrefix = {arXiv},
       eprint = {2401.03507},
 primaryClass = {astro-ph.CO},
       adsurl = {https://ui.adsabs.harvard.edu/abs/2024ApJS..271...30L}
}

@ARTICLE{Lyskova_etal_23,
       author = {{Lyskova}, N. and {Churazov}, E. and {Khabibullin}, I.~I. and {Burenin}, R. and {Starobinsky}, A.~A. and {Sunyaev}, R.},
        title = "{X-ray surface brightness and gas density profiles of galaxy clusters up to 3 {\texttimes} R$_{500c}$ with SRG/eROSITA}",
      journal = {\mnras},
         year = 2023,
        month = oct,
       volume = {525},
       number = {1},
        pages = {898-907},
          doi = {10.1093/mnras/stad2305},
archivePrefix = {arXiv},
       eprint = {2305.07080},
 primaryClass = {astro-ph.CO},
       adsurl = {https://ui.adsabs.harvard.edu/abs/2023MNRAS.525..898L}
}

@ARTICLE{Hadzhiyska_etal_23,
       author = {{Hadzhiyska}, Boryana and {Ferraro}, Simone and {Ried Guachalla}, Bernardita and {Schaan}, Emmanuel},
        title = "{Velocity reconstruction in the era of DESI and Rubin/LSST. II. Realistic samples on the light cone}",
      journal = {\prd},
         year = 2024,
        month = may,
       volume = {109},
       number = {10},
          eid = {103534},
        pages = {103534},
          doi = {10.1103/PhysRevD.109.103534},
archivePrefix = {arXiv},
       eprint = {2312.12434},
 primaryClass = {astro-ph.CO},
       adsurl = {https://ui.adsabs.harvard.edu/abs/2024PhRvD.109j3534H}
}

@ARTICLE{Yang_etal_05,
       author = {{Yang}, Xiaohu and {Mo}, H.~J. and {van den Bosch}, Frank C. and {Jing}, Y.~P.},
        title = "{A halo-based galaxy group finder: calibration and application to the 2dFGRS}",
      journal = {\mnras},
         year = 2005,
        month = feb,
       volume = {356},
       number = {4},
        pages = {1293-1307},
          doi = {10.1111/j.1365-2966.2005.08560.x},
archivePrefix = {arXiv},
       eprint = {astro-ph/0405234},
 primaryClass = {astro-ph},
       adsurl = {https://ui.adsabs.harvard.edu/abs/2005MNRAS.356.1293Y}
}

@ARTICLE{Schaan_etal_21,
       author = {{Schaan}, Emmanuel and {Ferraro}, Simone and {Amodeo}, Stefania and {Battaglia}, Nicholas and {Aiola}, Simone and {Austermann}, Jason E. and {Beall}, James A. and {Bean}, Rachel and {Becker}, Daniel T. and {Bond}, Richard J. and {Calabrese}, Erminia and {Calafut}, Victoria and {Choi}, Steve K. and {Denison}, Edward V. and {Devlin}, Mark J. and {Duff}, Shannon M. and {Duivenvoorden}, Adriaan J. and {Dunkley}, Jo and {D{\"u}nner}, Rolando and {Gallardo}, Patricio A. and {Guan}, Yilun and {Han}, Dongwon and {Hill}, J. Colin and {Hilton}, Gene C. and {Hilton}, Matt and {Hlo{\v{z}}ek}, Ren{\'e}e and {Hubmayr}, Johannes and {Huffenberger}, Kevin M. and {Hughes}, John P. and {Koopman}, Brian J. and {MacInnis}, Amanda and {McMahon}, Jeff and {Madhavacheril}, Mathew S. and {Moodley}, Kavilan and {Mroczkowski}, Tony and {Naess}, Sigurd and {Nati}, Federico and {Newburgh}, Laura B. and {Niemack}, Michael D. and {Page}, Lyman A. and {Partridge}, Bruce and {Salatino}, Maria and {Sehgal}, Neelima and {Schillaci}, Alessandro and {Sif{\'o}n}, Crist{\'o}bal and {Smith}, Kendrick M. and {Spergel}, David N. and {Staggs}, Suzanne and {Storer}, Emilie R. and {Trac}, Hy and {Ullom}, Joel N. and {Van Lanen}, Jeff and {Vale}, Leila R. and {van Engelen}, Alexander and {Maga{\~n}a}, Mariana Vargas and {Vavagiakis}, Eve M. and {Wollack}, Edward J. and {Xu}, Zhilei and {Atacama Cosmology Telescope Collaboration}},
        title = "{Atacama Cosmology Telescope: Combined kinematic and thermal Sunyaev-Zel'dovich measurements from BOSS CMASS and LOWZ halos}",
      journal = {\prd},
         year = 2021,
        month = mar,
       volume = {103},
       number = {6},
          eid = {063513},
        pages = {063513},
          doi = {10.1103/PhysRevD.103.063513},
archivePrefix = {arXiv},
       eprint = {2009.05557},
 primaryClass = {astro-ph.CO},
       adsurl = {https://ui.adsabs.harvard.edu/abs/2021PhRvD.103f3513S}
}

@ARTICLE{Sunyaev_Zeldovich_70,
       author = {{Sunyaev}, R.~A. and {Zeldovich}, Ya. B.},
        title = "{Small-Scale Fluctuations of Relic Radiation}",
      journal = {\apss},
         year = 1970,
        month = apr,
       volume = {7},
       number = {1},
        pages = {3-19},
          doi = {10.1007/BF00653471},
       adsurl = {https://ui.adsabs.harvard.edu/abs/1970Ap&SS...7....3S}
}

@ARTICLE{Schaan_etal_16,
       author = {{Schaan}, Emmanuel and {Ferraro}, Simone and {Vargas-Maga{\~n}a}, Mariana and {Smith}, Kendrick M. and {Ho}, Shirley and {Aiola}, Simone and {Battaglia}, Nicholas and {Bond}, J. Richard and {De Bernardis}, Francesco and {Calabrese}, Erminia and {Cho}, Hsiao-Mei and {Devlin}, Mark J. and {Dunkley}, Joanna and {Gallardo}, Patricio A. and {Hasselfield}, Matthew and {Henderson}, Shawn and {Hill}, J. Colin and {Hincks}, Adam D. and {Hlozek}, Ren{\'e}e and {Hubmayr}, Johannes and {Hughes}, John P. and {Irwin}, Kent D. and {Koopman}, Brian and {Kosowsky}, Arthur and {Li}, Dale and {Louis}, Thibaut and {Lungu}, Marius and {Madhavacheril}, Mathew and {Maurin}, Lo{\"\i}c and {McMahon}, Jeffrey John and {Moodley}, Kavilan and {Naess}, Sigurd and {Nati}, Federico and {Newburgh}, Laura and {Niemack}, Michael D. and {Page}, Lyman A. and {Pappas}, Christine G. and {Partridge}, Bruce and {Schmitt}, Benjamin L. and {Sehgal}, Neelima and {Sherwin}, Blake D. and {Sievers}, Jonathan L. and {Spergel}, David N. and {Staggs}, Suzanne T. and {van Engelen}, Alexander and {Wollack}, Edward J. and {ACTPol Collaboration}},
        title = "{Evidence for the kinematic Sunyaev-Zel'dovich effect with the Atacama Cosmology Telescope and velocity reconstruction from the Baryon Oscillation Spectroscopic Survey}",
      journal = {\prd},
         year = 2016,
        month = apr,
       volume = {93},
       number = {8},
          eid = {082002},
        pages = {082002},
          doi = {10.1103/PhysRevD.93.082002},
archivePrefix = {arXiv},
       eprint = {1510.06442},
 primaryClass = {astro-ph.CO},
       adsurl = {https://ui.adsabs.harvard.edu/abs/2016PhRvD..93h2002S}
}

@ARTICLE{Yang_Cai_etal_22,
       author = {{Yang}, Tianyi and {Cai}, Yan-Chuan and {Cui}, Weiguang and {Dav{\'e}}, Romeel and {Peacock}, John A. and {Sorini}, Daniele},
        title = "{Understanding the relation between thermal Sunyaev-Zeldovich decrement and halo mass using the SIMBA and TNG simulations}",
      journal = {\mnras},
         year = 2022,
        month = nov,
       volume = {516},
       number = {3},
        pages = {4084-4096},
          doi = {10.1093/mnras/stac2505},
archivePrefix = {arXiv},
       eprint = {2202.11430},
 primaryClass = {astro-ph.CO},
       adsurl = {https://ui.adsabs.harvard.edu/abs/2022MNRAS.516.4084Y}
}

@ARTICLE{Tanimura_Zaroubi_Aghanim_21,
       author = {{Tanimura}, Hideki and {Zaroubi}, Saleem and {Aghanim}, Nabila},
        title = "{Direct detection of the kinetic Sunyaev-Zel'dovich effect in galaxy clusters}",
      journal = {\aap},
         year = 2021,
        month = jan,
       volume = {645},
          eid = {A112},
        pages = {A112},
          doi = {10.1051/0004-6361/202038846},
archivePrefix = {arXiv},
       eprint = {2007.02952},
 primaryClass = {astro-ph.CO},
       adsurl = {https://ui.adsabs.harvard.edu/abs/2021A&A...645A.112T}
}

@ARTICLE{Maksimova_etal_21,
       author = {{Maksimova}, Nina A. and {Garrison}, Lehman H. and {Eisenstein}, Daniel J. and {Hadzhiyska}, Boryana and {Bose}, Sownak and {Satterthwaite}, Thomas P.},
        title = "{ABACUSSUMMIT: a massive set of high-accuracy, high-resolution N-body simulations}",
      journal = {\mnras},
         year = 2021,
        month = dec,
       volume = {508},
       number = {3},
        pages = {4017-4037},
          doi = {10.1093/mnras/stab2484},
archivePrefix = {arXiv},
       eprint = {2110.11398},
 primaryClass = {astro-ph.CO},
       adsurl = {https://ui.adsabs.harvard.edu/abs/2021MNRAS.508.4017M}
}

@ARTICLE{Colombi_Chodorowski_Teyssier_07,
       author = {{Colombi}, St{\'e}phane and {Chodorowski}, Micha{\l} J. and {Teyssier}, Romain},
        title = "{Cosmic velocity-gravity relation in redshift space}",
      journal = {\mnras},
         year = 2007,
        month = feb,
       volume = {375},
       number = {1},
        pages = {348-370},
          doi = {10.1111/j.1365-2966.2006.11330.x},
archivePrefix = {arXiv},
       eprint = {0805.1693},
 primaryClass = {astro-ph},
       adsurl = {https://ui.adsabs.harvard.edu/abs/2007MNRAS.375..348C}
}

@ARTICLE{Birkinshaw_99,
       author = {{Birkinshaw}, M.},
        title = "{The Sunyaev-Zel'dovich effect}",
      journal = {\physrep},
         year = 1999,
        month = mar,
       volume = {310},
       number = {2-3},
        pages = {97-195},
          doi = {10.1016/S0370-1573(98)00080-5},
archivePrefix = {arXiv},
       eprint = {astro-ph/9808050},
 primaryClass = {astro-ph},
       adsurl = {https://ui.adsabs.harvard.edu/abs/1999PhR...310...97B}
}

@BOOK{Peebles_93,
       author = {{Peebles}, P.~J.~E.},
        title = "{Principles of Physical Cosmology}",
         year = 1993,
          doi = {10.1515/9780691206721},
    publisher = {Princeton University Press},
       adsurl = {https://ui.adsabs.harvard.edu/abs/1993ppc..book.....P}
}

@ARTICLE{Garrison_etal_21,
       author = {{Garrison}, Lehman H. and {Eisenstein}, Daniel J. and {Ferrer}, Douglas and {Maksimova}, Nina A. and {Pinto}, Philip A.},
        title = "{The ABACUS cosmological N-body code}",
      journal = {\mnras},
         year = 2021,
        month = nov,
       volume = {508},
       number = {1},
        pages = {575-596},
          doi = {10.1093/mnras/stab2482},
archivePrefix = {arXiv},
       eprint = {2110.11392},
 primaryClass = {astro-ph.CO},
       adsurl = {https://ui.adsabs.harvard.edu/abs/2021MNRAS.508..575G}
}

@misc{cluster-toolkit,
       author = {{McClintock}, Tom},
        title = "{Cluster Toolkit: Tools for analyzing galaxy clusters}",
 howpublished = {Astrophysics Source Code Library, record ascl:2209.004},
         year = 2022,
        month = sep,
          eid = {ascl:2209.004},
       adsurl = {https://ui.adsabs.harvard.edu/abs/2022ascl.soft09004M}
}

@MISC{Pylians,
    author = {{Villaescusa-Navarro}, Francisco},
    title = "{Pylians: Python libraries for the analysis of numerical simulations}",
    howpublished = {Astrophysics Source Code Library, record ascl:1811.008},
    year = 2018,
    month = nov,
    eid = {ascl:1811.008},
    archivePrefix = {ascl},
    adsurl = {https://ui.adsabs.harvard.edu/abs/2018ascl.soft11008V}
}

@ARTICLE{Hincks_etal_10,
       author = {{Hincks}, A.~D. and {Acquaviva}, V. and {Ade}, P.~A.~R. and {Aguirre}, P. and {Amiri}, M. and {Appel}, J.~W. and {Barrientos}, L.~F. and {Battistelli}, E.~S. and {Bond}, J.~R. and {Brown}, B. and {Burger}, B. and {Chervenak}, J. and {Das}, S. and {Devlin}, M.~J. and {Dicker}, S.~R. and {Doriese}, W.~B. and {Dunkley}, J. and {D{\"u}nner}, R. and {Essinger-Hileman}, T. and {Fisher}, R.~P. and {Fowler}, J.~W. and {Hajian}, A. and {Halpern}, M. and {Hasselfield}, M. and {Hern{\'a}ndez-Monteagudo}, C. and {Hilton}, G.~C. and {Hilton}, M. and {Hlozek}, R. and {Huffenberger}, K.~M. and {Hughes}, D.~H. and {Hughes}, J.~P. and {Infante}, L. and {Irwin}, K.~D. and {Jimenez}, R. and {Juin}, J.~B. and {Kaul}, M. and {Klein}, J. and {Kosowsky}, A. and {Lau}, J.~M. and {Limon}, M. and {Lin}, Y. -T. and {Lupton}, R.~H. and {Marriage}, T.~A. and {Marsden}, D. and {Martocci}, K. and {Mauskopf}, P. and {Menanteau}, F. and {Moodley}, K. and {Moseley}, H. and {Netterfield}, C.~B. and {Niemack}, M.~D. and {Nolta}, M.~R. and {Page}, L.~A. and {Parker}, L. and {Partridge}, B. and {Quintana}, H. and {Reid}, B. and {Sehgal}, N. and {Sievers}, J. and {Spergel}, D.~N. and {Staggs}, S.~T. and {Stryzak}, O. and {Swetz}, D.~S. and {Switzer}, E.~R. and {Thornton}, R. and {Trac}, H. and {Tucker}, C. and {Verde}, L. and {Warne}, R. and {Wilson}, G. and {Wollack}, E. and {Zhao}, Y.},
        title = "{The Atacama Cosmology Telescope (ACT): Beam Profiles and First SZ Cluster Maps}",
      journal = {\apjs},
         year = 2010,
        month = dec,
       volume = {191},
       number = {2},
        pages = {423-438},
          doi = {10.1088/0067-0049/191/2/423},
archivePrefix = {arXiv},
       eprint = {0907.0461},
 primaryClass = {astro-ph.CO},
       adsurl = {https://ui.adsabs.harvard.edu/abs/2010ApJS..191..423H}
}

@ARTICLE{Phillips_95,
       author = {{Phillips}, Peter R.},
        title = "{Calculation of the Kinetic Sunyaev-Zeldovich Effect from the Boltzmann Equation}",
      journal = {\apj},
         year = 1995,
        month = dec,
       volume = {455},
        pages = {419},
          doi = {10.1086/176589},
       adsurl = {https://ui.adsabs.harvard.edu/abs/1995ApJ...455..419P}
}

@ARTICLE{Maniyar_Ferraro_Schaan_22,
       author = {{Maniyar}, Abhishek S. and {Ferraro}, Simone and {Schaan}, Emmanuel},
        title = "{Doppler boosted dust emission and CIB-galaxy cross-correlations: a new probe of cosmology and astrophysics}",
      journal = {\prl},
         year = 2023,
       volume = {103},
       number = {4},
          eid = {41001},
        pages = {41001},
          doi = {10.1103/PhysRevLett.130.041001},
       adsurl = {https://journals.aps.org/prl/abstract/10.1103/PhysRevLett.130.041001},
}

@ARTICLE{Tinker_etal_10,
       author = {{Tinker}, Jeremy L. and {Robertson}, Brant E. and {Kravtsov}, Andrey V. and {Klypin}, Anatoly and {Warren}, Michael S. and {Yepes}, Gustavo and {Gottl{\"o}ber}, Stefan},
        title = "{The Large-scale Bias of Dark Matter Halos: Numerical Calibration and Model Tests}",
      journal = {\apj},
         year = 2010,
        month = dec,
       volume = {724},
       number = {2},
        pages = {878-886},
          doi = {10.1088/0004-637X/724/2/878},
archivePrefix = {arXiv},
       eprint = {1001.3162},
 primaryClass = {astro-ph.CO},
       adsurl = {https://ui.adsabs.harvard.edu/abs/2010ApJ...724..878T}
}

@ARTICLE{Terzic_etal_24,
       author = {{Terzi{\'c}}, Bal{\v{s}}a and {Krafft}, Geoffrey A. and {Clark}, William and {Deur}, Alexandre and {Rogers}, Emerson and {Sakiotis}, Ioannis and {Velasco}, Brandon},
        title = "{Fully relativistic derivation of the thermal Sunyaev-Zel'dovich effect}",
      journal = {Physics Letters B},
         year = 2024,
        month = dec,
       volume = {859},
          eid = {139119},
        pages = {139119},
          doi = {10.1016/j.physletb.2024.139119},
archivePrefix = {arXiv},
       eprint = {2405.02127},
 primaryClass = {astro-ph.HE},
       adsurl = {https://ui.adsabs.harvard.edu/abs/2024PhLB..85939119T}
}

@ARTICLE{Yang_etal_21,
       author = {{Yang}, Xiaohu and {Xu}, Haojie and {He}, Min and {Gu}, Yizhou and {Katsianis}, Antonios and {Meng}, Jiacheng and {Shi}, Feng and {Zou}, Hu and {Zhang}, Youcai and {Liu}, Chengze and {Wang}, Zhaoyu and {Dong}, Fuyu and {Lu}, Yi and {Li}, Qingyang and {Chen}, Yangyao and {Wang}, Huiyuan and {Mo}, Houjun and {Fu}, Jian and {Guo}, Hong and {Leauthaud}, Alexie and {Luo}, Yu and {Zhang}, Jun and {Zu}, Ying},
        title = "{An Extended Halo-based Group/Cluster Finder: Application to the DESI Legacy Imaging Surveys DR8}",
      journal = {\apj},
         year = 2021,
        month = mar,
       volume = {909},
       number = {2},
          eid = {143},
        pages = {143},
          doi = {10.3847/1538-4357/abddb2},
archivePrefix = {arXiv},
       eprint = {2012.14998},
 primaryClass = {astro-ph.GA},
       adsurl = {https://ui.adsabs.harvard.edu/abs/2021ApJ...909..143Y}
}

@ARTICLE{Sugiyama_Okumura_Spergel_17,
       author = {{Sugiyama}, Naonori S. and {Okumura}, Teppei and {Spergel}, David N.},
        title = "{A direct measure of free electron gas via the kinematic Sunyaev-Zel'dovich effect in Fourier-space analysis}",
      journal = {\mnras},
         year = 2018,
        month = apr,
       volume = {475},
       number = {3},
        pages = {3764-3785},
          doi = {10.1093/mnras/stx3362},
archivePrefix = {arXiv},
       eprint = {1705.07449},
 primaryClass = {astro-ph.CO},
       adsurl = {https://ui.adsabs.harvard.edu/abs/2018MNRAS.475.3764S}
}

@BOOK{Hockney_Eastwood_81,
       author = {{Hockney}, R.~W. and {Eastwood}, J.~W.},
        title = "{Computer Simulation Using Particles}",
         year = 1981,
       adsurl = {https://ui.adsabs.harvard.edu/abs/1981csup.book.....H},
      publisher = "McGraw-Hill",
      address   = "New York"
}

@ARTICLE{Hadzhiyska_etal_24,
       author = {{Hadzhiyska}, B. and {Ferraro}, S. and {Ried Guachalla}, B. and {Schaan}, E. and {Aguilar}, J. and {Ahlen}, S. and {Battaglia}, N. and {Bond}, J.~R. and {Brooks}, D. and {Calabrese}, E. and {Choi}, S.~K. and {Claybaugh}, T. and {Coulton}, W.~R. and {Dawson}, K. and {Devlin}, M. and {Dey}, B. and {Doel}, P. and {Duivenvoorden}, A.~J. and {Dunkley}, J. and {Farren}, G.~S. and {Font-Ribera}, A. and {Forero-Romero}, J.~E. and {Gallardo}, P.~A. and {Gazta{\~n}aga}, E. and {Gontcho Gontcho}, S. and {Gralla}, M. and {Le Guillou}, L. and {Gutierrez}, G. and {Guy}, J. and {Hill}, J.~C. and {Hlo{\v{z}}ek}, R. and {Honscheid}, K. and {Juneau}, S. and {Kehoe}, R. and {Kisner}, T. and {Kremin}, A. and {Landriau}, M. and {Liu}, R.~H. and {Louis}, T. and {MacCrann}, N. and {de Macorra}, A. and {Madhavacheril}, M. and {Manera}, M. and {Meisner}, A. and {Miquel}, R. and {Moodley}, K. and {Moustakas}, J. and {Mroczkowski}, T. and {Naess}, S. and {Newman}, J. and {Niemack}, M.~D. and {Niz}, G. and {Page}, L. and {Palanque-Delabrouille}, N. and {Partridge}, B. and {Percival}, W.~J. and {Prada}, F. and {Qu}, F.~J. and {Rossi}, G. and {Sanchez}, E. and {Schlegel}, D. and {Schubnell}, M. and {Sherwin}, B. and {Sehgal}, N. and {Seo}, H. and {Sif{\'o}n}, C. and {Spergel}, D. and {Sprayberry}, D. and {Staggs}, S. and {Tarl{\'e}}, G. and {Vargas}, C. and {Vavagiakis}, E.~M. and {Weaver}, B.~A. and {Wollack}, E.~J. and {Zhou}, R. and {Zou}, H.},
        title = "{Evidence for large baryonic feedback at low and intermediate redshifts from kinematic Sunyaev-Zel'dovich observations with ACT and DESI photometric galaxies}",
      journal = {\prd},
         year = 2025,
        month = oct,
       volume = {112},
       number = {8},
          eid = {083509},
        pages = {083509},
          doi = {10.1103/kclp-x5j1},
archivePrefix = {arXiv},
       eprint = {2407.07152},
 primaryClass = {astro-ph.CO},
       adsurl = {https://ui.adsabs.harvard.edu/abs/2025PhRvD.112h3509H}
}

@ARTICLE{Cole_11,
       author = {{Cole}, Shaun},
        title = "{Maximum likelihood random galaxy catalogues and luminosity function estimation}",
      journal = {\mnras},
         year = 2011,
        month = sep,
       volume = {416},
       number = {1},
        pages = {739-746},
          doi = {10.1111/j.1365-2966.2011.19093.x},
archivePrefix = {arXiv},
       eprint = {1104.0009},
 primaryClass = {astro-ph.CO},
       adsurl = {https://ui.adsabs.harvard.edu/abs/2011MNRAS.416..739C}
}

@ARTICLE{Akina_etal_22,
       author = {{Akino}, Daichi and {Eckert}, Dominique and {Okabe}, Nobuhiro and {Sereno}, Mauro and {Umetsu}, Keiichi and {Oguri}, Masamune and {Gastaldello}, Fabio and {Chiu}, I. -Non and {Ettori}, Stefano and {Evrard}, August E. and {Farahi}, Arya and {Maughan}, Ben and {Pierre}, Marguerite and {Ricci}, Marina and {Valtchanov}, Ivan and {McCarthy}, Ian and {McGee}, Sean and {Miyazaki}, Satoshi and {Nishizawa}, Atsushi J. and {Tanaka}, Masayuki},
        title = "{HSC-XXL: Baryon budget of the 136 XXL groups and clusters}",
      journal = {\pasj},
         year = 2022,
        month = feb,
       volume = {74},
       number = {1},
        pages = {175-208},
          doi = {10.1093/pasj/psab115},
archivePrefix = {arXiv},
       eprint = {2111.10080},
 primaryClass = {astro-ph.CO},
       adsurl = {https://ui.adsabs.harvard.edu/abs/2022PASJ...74..175A}
}

@ARTICLE{Sunyaev_Zeldovich_80b,
       author = {{Sunyaev}, R.~A. and {Zeldovich}, Ia. B.},
        title = "{Microwave background radiation as a probe of the contemporary structure and history of the universe}",
      journal = {\araa},
         year = 1980,
        month = jan,
       volume = {18},
        pages = {537-560},
          doi = {10.1146/annurev.aa.18.090180.002541},
       adsurl = {https://ui.adsabs.harvard.edu/abs/1980ARA&A..18..537S}
}

@ARTICLE{Hadzhiyska_etal_22b,
       author = {{Hadzhiyska}, Boryana and {Eisenstein}, Daniel and {Bose}, Sownak and {Garrison}, Lehman H. and {Maksimova}, Nina},
        title = "{COMPASO: A new halo finder for competitive assignment to spherical overdensities}",
      journal = {\mnras},
         year = 2022,
        month = jan,
       volume = {509},
       number = {1},
        pages = {501-521},
          doi = {10.1093/mnras/stab2980},
archivePrefix = {arXiv},
       eprint = {2110.11408},
 primaryClass = {astro-ph.CO},
       adsurl = {https://ui.adsabs.harvard.edu/abs/2022MNRAS.509..501H}
}

@ARTICLE{Amon2022,
       author = {{Amon}, Alexandra and {Efstathiou}, George},
        title = "{A non-linear solution to the S$_{8}$ tension?}",
      journal = {\mnras},
         year = 2022,
        month = nov,
       volume = {516},
       number = {4},
        pages = {5355-5366},
          doi = {10.1093/mnras/stac2429},
archivePrefix = {arXiv},
       eprint = {2206.11794},
 primaryClass = {astro-ph.CO},
       adsurl = {https://ui.adsabs.harvard.edu/abs/2022MNRAS.516.5355A}
}

@ARTICLE{Bose_etal_22,
       author = {{Bose}, Sownak and {Eisenstein}, Daniel J. and {Hadzhiyska}, Boryana and {Garrison}, Lehman H. and {Yuan}, Sihan},
        title = "{Constructing high-fidelity halo merger trees in ABACUSSUMMIT}",
      journal = {\mnras},
         year = 2022,
        month = may,
       volume = {512},
       number = {1},
        pages = {837-854},
          doi = {10.1093/mnras/stac555},
archivePrefix = {arXiv},
       eprint = {2110.11409},
 primaryClass = {astro-ph.CO},
       adsurl = {https://ui.adsabs.harvard.edu/abs/2022MNRAS.512..837B}
}

@article{PlanckCollaboration_15_I,
  title={Planck 2015 results-I. Overview of products and scientific results},
  author={{Planck Collaboration} and Adam, R{\'e}mi and Ade, Peter AR and Aghanim, N and Akrami, Y and Alves, MIR and Arg{\"u}eso, F and Arnaud, M and Arroja, F and Ashdown, Mark and Aumont, J and others},
  journal = {\aap},
  volume={594},
  pages={A1},
  year={2016},
  publisher={EDP sciences}
}

@ARTICLE{Oppenheimer_etal_25,
       author = {{Oppenheimer}, Benjamin D. and {Voit}, G. Mark and {Bah{\'e}}, Yannick M. and {Battaglia}, Nicolas and {Bregman}, Joel and {Burchett}, Joseph N. and {Eckert}, Dominique and {Faerman}, Yakov and {Gibson}, Justus and {Hummels}, Cameron and {Medlock}, Isabel and {Nagai}, Daisuke and {Putman}, Mary and {Qu}, Zhijie and {Sun}, Ming and {Werk}, Jessica K. and {Zhang}, Yi},
        title = "{Introducing the Descriptive Parametric Model: gaseous profiles for galaxies, groups, and clusters}",
      journal = {\mnras},
         year = 2025,
        month = nov,
       volume = {543},
       number = {3},
        pages = {2649-2669},
          doi = {10.1093/mnras/staf1581},
archivePrefix = {arXiv},
       eprint = {2505.14782},
 primaryClass = {astro-ph.GA},
       adsurl = {https://ui.adsabs.harvard.edu/abs/2025MNRAS.543.2649O}
}

@ARTICLE{Naess_etal_25,
       author = {{Naess}, Sigurd and {Guan}, Yilun and {Duivenvoorden}, Adriaan J. and {Hasselfield}, Matthew and {Wang}, Yuhan and {Abril-Cabezas}, Irene and {Addison}, Graeme E. and {Ade}, Peter A.~R. and {Aiola}, Simone and {Alford}, Tommy and {Alonso}, David and {Amiri}, Mandana and {An}, Rui and {Atkins}, Zachary and {Austermann}, Jason E. and {Barbavara}, Eleonora and {Battaglia}, Nicholas and {Battistelli}, Elia Stefano and {Beall}, James A. and {Bean}, Rachel and {Beheshti}, Ali and {Beringue}, Benjamin and {Bhandarkar}, Tanay and {Biermann}, Emily and {Bolliet}, Boris and {Bond}, J. Richard and {Calabrese}, Erminia and {Capalbo}, Valentina and {Carrero}, Felipe and {Chen}, Stephen and {Chesmore}, Grace and {Cho}, Hsiao-mei and {Choi}, Steve K. and {Clark}, Susan E. and {Rosado}, Rodrigo Cordova and {Cothard}, Nicholas F. and {Coughlin}, Kevin and {Coulton}, William and {Crichton}, Devin and {Crowley}, Kevin T. and {Devlin}, Mark J. and {Dicker}, Simon and {Duell}, Cody J. and {Duff}, Shannon M. and {Dunkley}, Jo and {Dunner}, Rolando and {Embil Villagra}, Carmen and {Fankhanel}, Max and {Farren}, Gerrit S. and {Ferraro}, Simone and {Foster}, Allen and {Freundt}, Rodrigo and {Fuzia}, Brittany and {Gallardo}, Patricio A. and {Garrido}, Xavier and {Giardiello}, Serena and {Gill}, Ajay and {Givans}, Jahmour and {Gluscevic}, Vera and {Golec}, Joseph E. and {Gong}, Yulin and {Halpern}, Mark and {Harrison}, Ian and {Healy}, Erin and {Henderson}, Shawn and {Hensley}, Brandon and {Herv{\'\i}as-Caimapo}, Carlos and {Hill}, J. Colin and {Hilton}, Gene C. and {Hilton}, Matt and {Hincks}, Adam D. and {Hlo{\v{z}}ek}, Ren{\'e}e and {Ho}, Shuay-Pwu Patty and {Hood}, John and {Hornecker}, Erika and {Huber}, Zachary B. and {Hubmayr}, Johannes and {Huffenberger}, Kevin M. and {Hughes}, John P. and {Ikape}, Margaret and {Irwin}, Kent and {Isopi}, Giovanni and {Jense}, Hidde T. and {Joshi}, Neha and {Keller}, Ben and {Kim}, Joshua and {Knowles}, Kenda and {Koopman}, Brian J. and {Kosowsky}, Arthur and {Kramer}, Darby and {Kusiak}, Aleksandra and {La Posta}, Adrien and {Lagu{\"e}}, Alex and {Lakey}, Victoria and {Lee}, Eunseong and {Li}, Yaqiong and {Li}, Zack and {Limon}, Michele and {Lokken}, Martine and {Louis}, Thibaut and {Lungu}, Marius and {MacCrann}, Niall and {MacInnis}, Amanda and {Madhavacheril}, Mathew S. and {Maldonado}, Diego and {Maldonado}, Felipe and {Mallaby-Kay}, Maya and {Marques}, Gabriela A. and {van Marrewijk}, Joshiwa and {McCarthy}, Fiona and {McMahon}, Jeff and {Mehta}, Yogesh and {Menanteau}, Felipe and {Moodley}, Kavilan and {Morris}, Thomas W. and {Mroczkowski}, Tony and {Namikawa}, Toshiya and {Nati}, Federico and {Nerval}, Simran K. and {Newburgh}, Laura and {Nicola}, Andrina and {Niemack}, Michael D. and {Nolta}, Michael R. and {Orlowski-Scherer}, John and {Page}, Lyman A. and {Pandey}, Shivam and {Partridge}, Bruce and {Perez Sarmiento}, Karen and {Prince}, Heather and {Puddu}, Roberto and {Qu}, Frank J. and {Ragavan}, Damien C. and {Ried Guachalla}, Bernardita and {Rogers}, Keir K. and {Rojas}, Felipe and {Sakuma}, Tai and {Schaan}, Emmanuel and {Schmitt}, Benjamin L. and {Sehgal}, Neelima and {Shaikh}, Shabbir and {Sherwin}, Blake D. and {Sierra}, Carlos and {Sievers}, Jon and {Sif{\'o}n}, Crist{\'o}bal and {Simon}, Sara and {Sonka}, Rita and {London}, Alexander Spencer and {Spergel}, David N. and {Staggs}, Suzanne T. and {Storer}, Emilie and {Surrao}, Kristen and {Switzer}, Eric R. and {Tampier}, Niklas and {Thornton}, Robert and {Trac}, Hy and {Tucker}, Carole and {Ullom}, Joel and {Vale}, Leila R. and {Van Engelen}, Alexander and {Van Lanen}, Jeff and {Vargas}, Cristian and {Vavagiakis}, Eve M. and {Wagoner}, Kasey and {Wenzl}, Lukas and {Wollack}, Edward J. and {Zheng}, Kaiwen and {The Atacama Cosmology Telescope collaboration}},
        title = "{The Atacama Cosmology Telescope: DR6 maps}",
      journal = {\jcap},
         year = 2025,
        month = nov,
       volume = {2025},
       number = {11},
          eid = {061},
        pages = {061},
          doi = {10.1088/1475-7516/2025/11/061},
archivePrefix = {arXiv},
       eprint = {2503.14451},
 primaryClass = {astro-ph.CO},
       adsurl = {https://ui.adsabs.harvard.edu/abs/2025JCAP...11..061N}
}

@ARTICLE{Warren_etal_06,
       author = {{Warren}, Michael S. and {Abazajian}, Kevork and {Holz}, Daniel E. and {Teodoro}, Lu{\'\i}s},
        title = "{Precision Determination of the Mass Function of Dark Matter Halos}",
      journal = {\apj},
         year = 2006,
        month = aug,
       volume = {646},
       number = {2},
        pages = {881-885},
          doi = {10.1086/504962},
archivePrefix = {arXiv},
       eprint = {astro-ph/0506395},
 primaryClass = {astro-ph},
       adsurl = {https://ui.adsabs.harvard.edu/abs/2006ApJ...646..881W}
}

@ARTICLE{Wright_etal_24,
       author = {{Wright}, Ruby J. and {Somerville}, Rachel S. and {Lagos}, Claudia del P. and {Schaller}, Matthieu and {Dav{\'e}}, Romeel and {Angl{\'e}s-Alc{\'a}zar}, Daniel and {Genel}, Shy},
        title = "{The baryon cycle in modern cosmological hydrodynamical simulations}",
      journal = {\mnras},
         year = 2024,
        month = aug,
       volume = {532},
       number = {3},
        pages = {3417-3440},
          doi = {10.1093/mnras/stae1688},
archivePrefix = {arXiv},
       eprint = {2402.08408},
 primaryClass = {astro-ph.GA},
       adsurl = {https://ui.adsabs.harvard.edu/abs/2024MNRAS.532.3417W}
}

@ARTICLE{Sorini_etal_24,
       author = {{Sorini}, Daniele and {Bose}, Sownak and {Dav{\'e}}, Romeel and {Angl{\'e}s-Alc{\'a}zar}, Daniel},
        title = "{The impact of feedback on the evolution of gas density profiles from galaxies to clusters: a universal fitting formula from the Simba suite of simulations}",
      journal = {The Open Journal of Astrophysics},
         year = 2024,
        month = dec,
       volume = {7},
          eid = {115},
        pages = {115},
          doi = {10.33232/001c.126621},
archivePrefix = {arXiv},
       eprint = {2409.05815},
 primaryClass = {astro-ph.GA},
       adsurl = {https://ui.adsabs.harvard.edu/abs/2024OJAp....7E.115S}
}

@ARTICLE{Currie_etal_24,
       author = {{Currie}, Matthew and {Miller}, Kyle and {Shin}, Tae-hyeon and {Baxter}, Eric and {Jain}, Bhuvnesh},
        title = "{Galaxy cluster profiles: a Gaussian mixture model approach to halo miscentering}",
      journal = {\jcap},
         year = 2025,
        month = jul,
       volume = {2025},
       number = {7},
          eid = {072},
        pages = {072},
          doi = {10.1088/1475-7516/2025/07/072},
archivePrefix = {arXiv},
       eprint = {2408.10371},
 primaryClass = {astro-ph.IM},
       adsurl = {https://ui.adsabs.harvard.edu/abs/2025JCAP...07..072C}
}

@ARTICLE{Ding_etal_25,
       author = {{Ding}, Jupiter and {Dalal}, Roohi and {Sunayama}, Tomomi and {Strauss}, Michael A. and {Oguri}, Masamune and {Okabe}, Nobuhiro and {Hilton}, Matt and {Monteiro-Oliveira}, Rog{\'e}rio and {Sif{\'o}n}, Crist{\'o}bal and {Staggs}, Suzanne T.},
        title = "{Miscentring of optical galaxy clusters based on Sunyaev-Zeldovich counterparts}",
      journal = {\mnras},
         year = 2025,
        month = jan,
       volume = {536},
       number = {1},
        pages = {572-591},
          doi = {10.1093/mnras/stae2601},
archivePrefix = {arXiv},
       eprint = {2411.12120},
 primaryClass = {astro-ph.CO},
       adsurl = {https://ui.adsabs.harvard.edu/abs/2025MNRAS.536..572D}
}

@ARTICLE{Shin_etal_19,
       author = {{Shin}, T. and {Adhikari}, S. and {Baxter}, E.~J. and {Chang}, C. and {Jain}, B. and {Battaglia}, N. and {Bleem}, L. and {Bocquet}, S. and {DeRose}, J. and {Gruen}, D. and {Hilton}, M. and {Kravtsov}, A. and {McClintock}, T. and {Rozo}, E. and {Rykoff}, E.~S. and {Varga}, T.~N. and {Wechsler}, R.~H. and {Wu}, H. and {Zhang}, Z. and {Aiola}, S. and {Allam}, S. and {Bechtol}, K. and {Benson}, B.~A. and {Bertin}, E. and {Bond}, J.~R. and {Brodwin}, M. and {Brooks}, D. and {Buckley-Geer}, E. and {Burke}, D.~L. and {Carlstrom}, J.~E. and {Carnero Rosell}, A. and {Carrasco Kind}, M. and {Carretero}, J. and {Castander}, F.~J. and {Choi}, S.~K. and {Cunha}, C.~E. and {Crawford}, T.~M. and {da Costa}, L.~N. and {De Vicente}, J. and {Desai}, S. and {Devlin}, M.~J. and {Dietrich}, J.~P. and {Doel}, P. and {Dunkley}, J. and {Eifler}, T.~F. and {Evrard}, A.~E. and {Flaugher}, B. and {Fosalba}, P. and {Gallardo}, P.~A. and {Garc{\'\i}a-Bellido}, J. and {Gaztanaga}, E. and {Gerdes}, D.~W. and {Gralla}, M. and {Gruendl}, R.~A. and {Gschwend}, J. and {Gupta}, N. and {Gutierrez}, G. and {Hartley}, W.~G. and {Hill}, J.~C. and {Ho}, S.~P. and {Hollowood}, D.~L. and {Honscheid}, K. and {Hoyle}, B. and {Huffenberger}, K. and {Hughes}, J.~P. and {James}, D.~J. and {Jeltema}, T. and {Kim}, A.~G. and {Krause}, E. and {Kuehn}, K. and {Lahav}, O. and {Lima}, M. and {Madhavacheril}, M.~S. and {Maia}, M.~A.~G. and {Marshall}, J.~L. and {Maurin}, L. and {McMahon}, J. and {Menanteau}, F. and {Miller}, C.~J. and {Miquel}, R. and {Mohr}, J.~J. and {Naess}, S. and {Nati}, F. and {Newburgh}, L. and {Niemack}, M.~D. and {Ogando}, R.~L.~C. and {Page}, L.~A. and {Partridge}, B. and {Patil}, S. and {Plazas}, A.~A. and {Rapetti}, D. and {Reichardt}, C.~L. and {Romer}, A.~K. and {Sanchez}, E. and {Scarpine}, V. and {Schindler}, R. and {Serrano}, S. and {Smith}, M. and {Smith}, R.~C. and {Soares-Santos}, M. and {Sobreira}, F. and {Staggs}, S.~T. and {Stark}, A. and {Stein}, G. and {Suchyta}, E. and {Swanson}, M.~E.~C. and {Tarle}, G. and {Thomas}, D. and {van Engelen}, A. and {Wollack}, E.~J. and {Xu}, Z.},
        title = "{Measurement of the splashback feature around SZ-selected Galaxy clusters with DES, SPT, and ACT}",
      journal = {\mnras},
         year = 2019,
        month = aug,
       volume = {487},
       number = {2},
        pages = {2900-2918},
          doi = {10.1093/mnras/stz1434},
archivePrefix = {arXiv},
       eprint = {1811.06081},
 primaryClass = {astro-ph.CO},
       adsurl = {https://ui.adsabs.harvard.edu/abs/2019MNRAS.487.2900S}
}

@ARTICLE{Johnston_etal_07,
       author = {{Johnston}, David E. and {Sheldon}, Erin S. and {Wechsler}, Risa H. and {Rozo}, Eduardo and {Koester}, Benjamin P. and {Frieman}, Joshua A. and {McKay}, Timothy A. and {Evrard}, August E. and {Becker}, Matthew R. and {Annis}, James},
        title = "{Cross-correlation Weak Lensing of SDSS galaxy Clusters II: Cluster Density Profiles and the Mass--Richness Relation}",
      journal = {arXiv e-prints},
         year = 2007,
        month = sep,
          eid = {arXiv:0709.1159},
        pages = {arXiv:0709.1159},
          doi = {10.48550/arXiv.0709.1159},
archivePrefix = {arXiv},
       eprint = {0709.1159},
 primaryClass = {astro-ph},
       adsurl = {https://ui.adsabs.harvard.edu/abs/2007arXiv0709.1159J},
}

@ARTICLE{McCarthyIan_etal_25,
       author = {{McCarthy}, Ian G. and {Amon}, Alexandra and {Schaye}, Joop and {Schaan}, Emmanuel and {Angulo}, Raul E. and {Salcido}, Jaime and {Schaller}, Matthieu and {Bigwood}, Leah and {Elbers}, Willem and {Kugel}, Roi and {Helly}, John C. and {Forouhar Moreno}, Victor J. and {Frenk}, Carlos S. and {McGibbon}, Robert J. and {Ondaro-Mallea}, Lurdes and {van Daalen}, Marcel P.},
        title = "{FLAMINGO: combining kinetic SZ effect and galaxy{\textendash}galaxy lensing measurements to gauge the impact of feedback on large-scale structure}",
      journal = {\mnras},
         year = 2025,
        month = jun,
       volume = {540},
       number = {1},
        pages = {143-163},
          doi = {10.1093/mnras/staf731},
archivePrefix = {arXiv},
       eprint = {2410.19905},
 primaryClass = {astro-ph.CO},
       adsurl = {https://ui.adsabs.harvard.edu/abs/2025MNRAS.540..143M}
}

@ARTICLE{Kugel_etal_23,
       author = {{Kugel}, Roi and {Schaye}, Joop and {Schaller}, Matthieu and {Helly}, John C. and {Braspenning}, Joey and {Elbers}, Willem and {Frenk}, Carlos S. and {McCarthy}, Ian G. and {Kwan}, Juliana and {Salcido}, Jaime and {van Daalen}, Marcel P. and {Vandenbroucke}, Bert and {Bah{\'e}}, Yannick M. and {Borrow}, Josh and {Chaikin}, Evgenii and {Hu{\v{s}}ko}, Filip and {Jenkins}, Adrian and {Lacey}, Cedric G. and {Nobels}, Folkert S.~J. and {Vernon}, Ian},
        title = "{FLAMINGO: calibrating large cosmological hydrodynamical simulations with machine learning}",
      journal = {\mnras},
         year = 2023,
        month = dec,
       volume = {526},
       number = {4},
        pages = {6103-6127},
          doi = {10.1093/mnras/stad2540},
archivePrefix = {arXiv},
       eprint = {2306.05492},
 primaryClass = {astro-ph.CO},
       adsurl = {https://ui.adsabs.harvard.edu/abs/2023MNRAS.526.6103K}
}

@ARTICLE{ForemanMackey_etal_13,
       author = {{Foreman-Mackey}, Daniel and {Hogg}, David W. and {Lang}, Dustin and {Goodman}, Jonathan},
        title = "{emcee: The MCMC Hammer}",
      journal = {\pasp},
         year = 2013,
        month = mar,
       volume = {125},
       number = {925},
        pages = {306},
          doi = {10.1086/670067},
archivePrefix = {arXiv},
       eprint = {1202.3665},
 primaryClass = {astro-ph.IM},
       adsurl = {https://ui.adsabs.harvard.edu/abs/2013PASP..125..306F}
}

@ARTICLE{Ayromlou_Nelsoneta_Pillepich_23,
       author = {{Ayromlou}, Mohammadreza and {Nelson}, Dylan and {Pillepich}, Annalisa},
        title = "{Feedback reshapes the baryon distribution within haloes, in halo outskirts, and beyond: the closure radius from dwarfs to massive clusters}",
      journal = {\mnras},
         year = 2023,
        month = oct,
       volume = {524},
       number = {4},
        pages = {5391-5410},
          doi = {10.1093/mnras/stad2046},
archivePrefix = {arXiv},
       eprint = {2211.07659},
 primaryClass = {astro-ph.GA},
       adsurl = {https://ui.adsabs.harvard.edu/abs/2023MNRAS.524.5391A}
}

@article{Ludlow+16,
       author = {{Ludlow}, Aaron D. and {Bose}, Sownak and {Angulo}, Ra{\'u}l E. and {Wang}, Lan and {Hellwing}, Wojciech A. and {Navarro}, Julio F. and {Cole}, Shaun and {Frenk}, Carlos S.},
        title = "{The mass-concentration-redshift relation of cold and warm dark matter haloes}",
      journal = {\mnras},
         year = 2016,
        month = aug,
       volume = {460},
       number = {2},
        pages = {1214-1232},
          doi = {10.1093/mnras/stw1046},
archivePrefix = {arXiv},
       eprint = {1601.02624},
 primaryClass = {astro-ph.CO},
       adsurl = {https://ui.adsabs.harvard.edu/abs/2016MNRAS.460.1214L}
}

@ARTICLE{Siegel_etal_25,
       author = {{Siegel}, Jared C. and {Amon}, Alexandra and {McCarthy}, Ian G. and {Bigwood}, Leah and {Yamamoto}, Masaya and {Bulbul}, Esra and {Greene}, Jenny E. and {McCullough}, Jamie and {Schaller}, Matthieu and {Schaye}, Joop},
        title = "{Joint X-Ray, Kinetic Sunyaev─Zeldovich, and Weak Lensing Measurements: Toward a Consensus Picture of Efficient Gas Expulsion from Groups and Clusters}",
      journal = {\apj},
         year = 2026,
        month = jun,
       volume = {1003},
       number = {2},
          eid = {151},
        pages = {151},
          doi = {10.3847/1538-4357/ae5dc2},
archivePrefix = {arXiv},
       eprint = {2509.10455},
 primaryClass = {astro-ph.CO},
       adsurl = {https://ui.adsabs.harvard.edu/abs/2026ApJ..1003..151S}
}

@ARTICLE{Sorini_etal_22,
       author = {{Sorini}, Daniele and {Dav{\'e}}, Romeel and {Cui}, Weiguang and {Appleby}, Sarah},
        title = "{How baryons affect haloes and large-scale structure: a unified picture from the SIMBA simulation}",
      journal = {\mnras},
         year = 2022,
        month = oct,
       volume = {516},
       number = {1},
        pages = {883-906},
          doi = {10.1093/mnras/stac2214},
archivePrefix = {arXiv},
       eprint = {2111.13708},
 primaryClass = {astro-ph.GA},
       adsurl = {https://ui.adsabs.harvard.edu/abs/2022MNRAS.516..883S}
}

\label{lastpage}
\end{document}